\documentclass[twocolumn,showpacs,preprintnumbers,amsmath,amssymb,prb]{revtex4-1}

\usepackage{graphicx}% Include figure files
\usepackage{dcolumn}% Align table columns on decimal point
\usepackage[tight]{subfigure}
\usepackage{amsmath}
\usepackage{simplewick}
\usepackage{amsfonts}
\usepackage{bbm} % blackboard font for unit matrix
\usepackage{mathrsfs}
\usepackage{verbatim}
\usepackage{float}
\usepackage[usenames,dvipsnames]{color}
\usepackage{bm} % bold greek math
\usepackage{cases}  % for "numcases" environment
\usepackage{mathtools} % for "dcases" environment

\usepackage{etoolbox} % make user-defined commands robust: (re)newrobustcmd

\usepackage{hyperref} % clickable links in PDF: INCLUDE hyperref LAST!!!
\usepackage[normalem]{ulem} % for strikeout "sout"

\renewcommand\Im{\operatorname{Im}}

\newrobustcmd{\unit}{\mathbbm{1}}   %\MakeRobustCommand\unit
\newrobustcmd{\bra}[1]{\langle #1|}   %\MakeRobustCommand\bra
\newrobustcmd{\ket}[1]{|#1 \rangle}   %\MakeRobustCommand\ket
\newrobustcmd{\braket}[1]{\langle #1 \rangle}   %\MakeRobustCommand\braket

\newrobustcmd{\Tr}[1]{\underset{#1}{\mathsf{Tr}}}   %\MakeRobustCommand\Tr
\newrobustcmd{\tr}{\mathsf{Tr}}   %\MakeRobustCommand\tr
\newrobustcmd{\nohat}[1]{#1}   %\MakeRobustCommand\nohat
\newrobustcmd{\sdagger}{{\dagger}}   %\MakeRobustCommand\sdagger

\newrobustcmd{\vecg}[1]{{\bm #1}}   %\MakeRobustCommand\vecg
\renewrobustcmd{\vec}[1]{{\mathbf{#1}}}   %\MakeRobustCommand\vecg

\newrobustcmd{\timeint}[4]{ \underset{#1 \geq #2 \geq #3 \geq #4}{\int d #2 d #3 } }
\newrobustcmd{\timeintfour}[6]{ \underset{#1 \geq #2 \geq #3 \geq #4 \geq #5 \geq #6}{\int d #2 d #3 d #4 d #5} }  

\newrobustcmd{\tot}{\mathrm{tot}}   %\MakeRobustCommand\tot
\newrobustcmd{\R}{\mathrm{R}}   %\MakeRobustCommand\R

\newrobustcmd{\Cite}[1]{Ref.~\onlinecite{#1}}   %\MakeRobustCommand\Cite
\newrobustcmd{\Cites}[1]{Refs.~\onlinecite{#1}}   %\MakeRobustCommand\Cites
\newrobustcmd{\Tab}[1]{Table~\ref{#1}}   %\MakeRobustCommand\Tab
\newrobustcmd{\Fig}[1]{Fig.~\ref{#1}}   %\MakeRobustCommand\Fig
\newrobustcmd{\Eq}[1]{Eq.~(\ref{#1})}   %\MakeRobustCommand\Eq
\newrobustcmd{\Eqb}[1]{Equation~(\ref{#1})}   %\MakeRobustCommand\Eqb
\newrobustcmd{\Eqs}[1]{Eqs.~(\ref{#1})}   %\MakeRobustCommand\Eqs
\newrobustcmd{\eq}[1]{(\ref{#1})}   %\MakeRobustCommand\eq
\newrobustcmd{\App}[1]{App.~\ref{#1}}   %\MakeRobustCommand\App
\newrobustcmd{\Sec}[1]{Sec.~\ref{#1}}   %\MakeRobustCommand\Sec
\renewrobustcmd{\sec}[1]{(\ref{#1})}   %\MakeRobustCommand\sec

\definecolor{myred}{rgb}{0.9,0,0}
\definecolor{myblue}{rgb}{0,0,0.5}
\definecolor{mygreen}{rgb}{0,0.6,0}
\newrobustcmd{\hl}[1]{\textcolor{blue}{#1}}  % ONLY for referee A
\newrobustcmd{\HL}[1]{\textcolor{blue}{#1}} % ONLY for referee A
\newrobustcmd{\me}[1]{\textcolor{blue}{#1}}
\newrobustcmd{\schoeller}[1]{\textcolor{blue}{#1}}

\newrobustcmd{\refB}[1]{\textcolor{myred}{#1}} % ONLY for referee B
\newrobustcmd{\refAB}[1]{\textcolor{mygreen}{#1}} % BOTH referee A and B

\newrobustcmd{\cut}[1]{\sout{\textbf{\textcolor{Gray}{#1}}}}  

    \renewrobustcmd{\hl}[1]{\textcolor{black}{#1}}  % ONLY for referee A
  \renewrobustcmd{\HL}[1]{\textcolor{black}{#1}} % ONLY for referee A
  \renewrobustcmd{\refB}[1]{\textcolor{black}{#1}} % ONLY for referee B
   \renewrobustcmd{\refAB}[1]{\textcolor{black}{#1}} % BOTH referee A and B
  \renewrobustcmd{\schoeller}[1]{\textcolor{black}{#1}} % BOTH referee A and B
  \renewrobustcmd{\me}[1]{\textcolor{black}{#1}}
\renewrobustcmd{\cut}[1]{}

\robustify{\sum}
\robustify{\int}
\robustify{\nonumber}

\begin{document}

\title{
  Time-dependent quantum transport:
  causal superfermions,
  \hl{exact} fermion-parity protected decay \hl{mode},
  and Pauli exclusion principle for mixed quantum states
}
\author{R. B. Saptsov$^{(1,2)}$}
\author{M. R. Wegewijs$^{(1,2,3)}$}

\affiliation{
  (1) Peter Gr{\"u}nberg Institut,
      Forschungszentrum J{\"u}lich, 52425 J{\"u}lich,  Germany \\
  (2) JARA- Fundamentals of Future Information Technology\\
  (3) Institute for Theory of Statistical Physics,
      RWTH Aachen, 52056 Aachen,  Germany
}
\date{\today}
\begin{abstract}
We extend the recently developed \emph{causal} superfermion approach to the real-time \hl{diagrammatic} transport theory to time-dependent decay problems.
Its usefulness is illustrated for the Anderson model of a quantum dot with tunneling rates depending on
spin due to \cut{the} ferromagnetic electrodes and / or spin polarization of the tunnel junction.
\hl{This approach naturally leads to an exact result  for
one of the time-dependent decay modes for any value of the Coulomb interaction compatible with the wideband limit.
\schoeller{We generalize these results to multilevel Anderson models and indicate constraints they impose on renormalization-group schemes in order to recover the exact noninteracting limit.}
}

%
%
% CAUSAL SUPERFERMIONS
% 
\hl{(i) We first} set up a second quantization scheme in the space of \emph{density operators}
\cut{in the Liouville-Fock space} constructing ``causal'' field
superoperators using the fundamental physical principles of causality
/ probability conservation and \cut{the} fermion-parity superselection
(univalence).  The time-dependent perturbation series for the
time-evolution is renormalized by explicitly performing the wideband
limit on the superoperator level. \hl{As a result, the occurrence of \cut{causal field} {destruction} and {creation}
superoperators are shown to be tightly linked to the physical short- and
long-time reservoir correlations, respectively.}
\hl{This} effective theory takes as a reference a \emph{damped} local system, which may also provide an
interesting starting point for numerical calculations of memory
kernels in real time.

%
%
% FERMION PARITY
%
\hl{(ii) A} remarkable feature of this approach is the natural appearance of a
\emph{fermion-parity protected} decay mode
\hl{which can be measured using a setup proposed earlier [Phys. Rev. B 85, 075301 (2012)].
This mode} can be calculated exactly in the fully Markovian, infinite-temperature
limit by leading order perturbation theory, \hl{but surprisingly} persists unaltered for
finite \hl{temperature, for any} interaction and tunneling spin polarization.

%
%
% SUPER PAULI PRINCIPLE
%
\hl{(iii) Finally, we show how a
{Liouville-space} analog of the \emph{Pauli principle}  directly leads to an exact expression in the noninteracting limit
%We obtain a closed expression
for the \cut{density-operator} time evolution,
\HL{extending previous works by}
starting from an arbitrary initial \hl{mixed} state
 including spin- and pairing coherences and two-particle correlations stored on the quantum dot.
}
This \hl{exact} result is obtained \hl{already} in finite-order renormalized perturbation theory,
\hl{which \me{surprisingly}} is not quadratic but \emph{quartic} in the 
\hl{field superoperators, despite the absence of Coulomb interaction.}
The latter fact we relate to the time-evolution \me{of} the two-particle component of the \hl{mixed state, which is just the fermion-parity operator, \me{a cornerstone of the formalism.}}
% This clarifies why in a previous real-time renormalization (RG) scheme
% one- and \emph{two}-loop corrections are required to recover the exact %noninteracting result for the Anderson model.
\hl{\me{We illustrate how the} super-Pauli principle also simplifies problems with nonzero Coulomb interaction.}
%
%For spin-independent tunneling the decay rates include
%besides the expected rates $\tilde{\Gamma}$
 %and $2\tilde{\Gamma}$ also such rates as $3\tilde{\Gamma}$ and $4\tilde{\Gamma}$,
%\hl{where $\tilde{\Gamma}$ is the tunnel rate for ???}

\end{abstract}

\maketitle
\section{Introduction\label{sec:intro}}
\subsection{Experimental motivation}
% Quantum dynamics of open systems
Quantum dynamics of open systems is of interest in various research fields, ranging from transport through meso- and nanoscopic systems,
quantum information processing, and quantum optics to physical chemistry and biology.
\hl{Typically, the object of investigation is some smaller part of a larger system, e.g., a single molecule attached to macroscopically large contacts, which act as reservoirs and impose strong nonequilibrium boundary conditions.
In the field of quantum transport a high degree of control has been achieved over \emph{fermionic} subsystems, \HL{such as} few-electron quantum dots coupled to \HL{various kinds of electrodes} (e.g, metals, ferromagnets, or superconductors). %topological insulators ?
This control relies mostly on the strong electrostatic effects,
which for very small systems makes the theoretical description challenging.
%due to the strong electrostatic interaction local to the subsystem.
% (4) Time-dependence and quantum dots
This progress has enabled detailed studies of not only stationary but also of time-dependent transport phenomena~\cite{Kouwenhoven94a,Switkes99,Fujisawa03rev,Elzerman04,Hanson05,Gabelli06,Feve07,Amasha08,Hanson07rev}
 down to the scale of atomic quantum dots.~\cite{Zwanenburg13,Roche13}
% (1D) Interaction effects & time-dependence
% (9A) Time-dependent oscillations
Interaction effects in the time domain have been investigated early on, such as the SET oscillations in the weak tunnel coupling regime,~\cite{Likharev99rev} and continue to be of interest.~\cite{Gross10}
\HL{Quantum} fluctuations between such a strongly correlated dot and the electrodes, \HL{lead to additional effects, such as} level renormalization,  inelastic tunneling effects, and Kondo physics in stationary transport and their nontrivial \HL{signatures} in the time-domain \HL{have also attracted} interest.
% (8) Kondo
A problem that received quite some attention,
is the time-dependent response of \HL{a quantum dot} in the Kondo regime.~\cite{Zarchin08, Delattre09, Yamauchi11, Basset12}
Theoretically, it has been studied using various models and methods.~\cite{Anders05, Hackl09a, Lobaskin05, Kehrein, Goker07, Jung12, Pletyukhov10, Cohen13a, Weiss08}
% (1B)
For instance, when starting from a Kondo model description,~\cite{Anders06,Hackl09a,Lobaskin05} the real-time diagrammatic approach,~\cite{Pletyukhov10} which is at the focus of this paper, provides deep analytical insight 
into the renormalization of exchange interactions as well as the renormalization of the various dissipative effects that ultimately destroy the Kondo effect.
On the other hand, recent numerical studies starting from an Anderson model~\cite{Anders05,Kehrein,Weiss08,Cohen13a,Cohen14}
have investigated the development of the Kondo effect in time, in particular, the much debated splitting of the Kondo peaks.~\cite{Cohen13c}
% get some references from Cohen paper and read them
Application of the  real-time diagrammatic approach to the Anderson model at $T=0$ is of high interest \HL{as it can provide analytical insight,} especially regarding the time-evolution towards stationarity.
% (1B), (3B)
%but is very challenging due to additional complications presented by this model.
Outside the Kondo regime we recently reported some progress in this direction in the stationary limit~\cite{Saptsov12a} and noted an interesting relation to the time-evolution decay modes that were studied before in the weak / moderate tunnel coupling limit.~\cite{Contreras12}
% (6A) Fermion parity effect
\me{In~\Cite{Contreras12},} motivated by experimental progress on single-electron sources~\cite{Feve07, Gabelli06}, new measurement setups were suggested to probe the relaxation rates of a quantum dot~\cite{Splettstoesser10} using a quantum point contact (QPC).
\HL{This study and a more recent one~\cite{Schulenborg14a}} focused on the effect of the Coulomb interaction and surprisingly found that certain multiparticle correlators show a remarkable robustness with respect to most details of the setup (see below), in particular to the interaction.
As argued there, this absence of interaction corrections is really an effect that can be measured.
A key result of the present paper, \HL{expressed by  \Eq{rhotxi},} is that this conclusion holds beyond various of the approximations made in \Cite{Contreras12,Schulenborg14a}.
\HL{\me{This result can be written as follows:}
%To relate this result to experiments we write it here as follows:
\begin{align}
  \braket{(n_\uparrow-\tfrac 1 2 )(n_\downarrow-\tfrac 1 2)}(t) = 
%  \braket{\tfrac{1}{2}n(n-1)}(t) =
  e^{-\Gamma t} \braket{(n_\uparrow-\tfrac 1 2) (n_\downarrow-\tfrac 1 2)}(0) + \ldots
  \label{keyresult}
\end{align}
Here, $n_\sigma$,  $\sigma=\downarrow,\uparrow$ are the spin-resolved occupations
of the dot.
The (equal-time) two-particle correlation function \eq{keyresult} contains a term that decays strictly Markovian with rate $\Gamma$.
This function appears as a coefficient in the expansion of the mixed state of the quantum dot.
It has been shown~\cite{Contreras12,Schulenborg14a} that the experimental observation of the decay of the mixed 
\emph{state} is possible:
one can optimally choose the parameters that determine the initial and final state of the time-dependent decay
such that on a well-separated time scale the current through a detector coupled to the quantum dot directly probes this part of the decay \eq{keyresult}.
This has been worked out in detail for a quantum-point contact detector in~\Cite{Contreras12} and was recently extended to a quantum-dot detector~\cite{Schulenborg14a}.
It is therefore of interest to calculate the full mixed state dynamics, and not just focus on the current through the quantum dot itself which does not reveal this effect.
\me{This is undertaken here:
motivated by the above} experimental connection, we investigate the mixed-state dynamics and
our conclusions strengthen the experimental importance of this effect:}
First, the decay is \HL{exactly exponential in the wideband limit},
i.e., the Markovian assumption made in \Cites{Contreras12,Schulenborg14a}
continues to hold, \HL{but only for this special mode}.
Second, this form of the decay is valid for any tunnel coupling, including possible spin-dependence:
still, $\Gamma$ in \Eq{keyresult} is simply \HL{given by} the sum of the Golden Rule expressions for the spin-resolved tunnel rates of the various junctions $r=L,R$:
$\Gamma = \sum_{r\sigma} \Gamma_{r\sigma}$, cf. \Eq{spectdens}.
% (9C)
Finally, we show that any more realistic quantum dot model taking into account  multiple orbitals labeled by $l$ has such a ``protected'' mode, the decay rate being
$\Gamma = \sum_{rl\sigma} \Gamma_{rl \sigma}$.
Notably, this is independent of the experimental details of the quantum dot
as long as \HL{its} energy scales are much below the electrode band width.
This can include more complex forms of the Coulomb interaction -- including all \HL{local two-particle} matrix elements, not just the charging part -- or spin-orbit interaction, etc.
The only crucial assumption is that the tunnel coupling is bilinear in the electron operators, a basic starting point of virtually all modeling of quantum transport \HL{through strongly interacting systems}.
In fact, even the simplifying assumption of collinear spin dependence of the tunneling made in this work, turns out not to be crucial.~\cite{Saptsov14b}
% (9C)
The interesting question is raised as to which physical principle can be responsible for this remarkable effect.
}

\hl{
The theoretical importance of the key result \eq{keyresult} lies in the fact that it arises naturally in the real-time framework
% (9C)
-- by mere formulation, without real calculation --
\HL{when using a particular kind of superfermion approach.
This particular approach arose in the context of stationary state transport problems~\cite{Saptsov12a}
and further below we give an overview of other superfermion constructions.}
The experimental relevance of the striking result \eq{keyresult} thus
 \emph{physically motivates} a reformulation of the general real-time framework.
% (1)
Perhaps, the impact of this should be compared with that of second quantization in standard quantum mechanics and field theory,
which by itself presents no new physical theory or prediction.
That approach, however, had a big impact by making the general framework more intuitively accessible
(e.g., by introducing field operators to represent quasi-particles),
simplifying calculations to such an extent that their results become intuitively clear
and often revealing their physical origin (e.g, particle exchange).
%s, which otherwise would be masked by (overly) complicated manipulations and expressions.
%
Such a second quantization scheme is  well-established for closed quantum systems but still under active study for open systems \HL{(see below)}.
%, as outlined in the following.
Only recently, this idea has been combined with the real-time diagrammatic theory targeting stationary transport.~\cite{Saptsov12a}
By itself, the real-time diagrammatic theory is already a very successful framework for the calculation of transport properties of nanoscale, strongly interacting systems,\cite{Andergassen10rev}  allowing various levels of approximations to be systematically formulated and worked out, 
both analytically~\cite{Schoeller94,koenig-hs-schoen97,Kern13}
 and numerically,~\cite{Kubala06,Leijnse08a,Koller10} which have found application to \HL{transport experiments}~\cite{Schleser05,Huettel09,Zyazin10,Eliasen10,Stevanato12,Klochan13}.
% (1)
Any general progress in \HL{simplifying or clarifying the general structure of this theory} is therefore \HL{ultimately} of experimental relevance since more accurate approximations come within reach.
For example, as mentioned above, the Anderson model and its generalizations present technical obstacles for gaining analytical insight into the low $T$ nonequilibrium physics.
By combining it with a superfermion technique, we were able to make
%set up a full one plus two-loop real-time renormalization group calculation at $T=0$,~\cite{Saptsov12a} allowing
 detailed predictions~\cite{Saptsov12a} at $T=0$ for measurable stationary $dI/dV$ maps, covering large parameter regimes \me{for strong interactions.} This includes level renormalization effects, energy-dependent broadening, charge-fluctuation \HL{renormalization of} cotunneling peaks.
% and intrinsically generates dissipative energy scales that cut off the RG flow.~\cite{Schoeller09a}
The restriction of the approximations \HL{(only one plus two-loop renormalization-group diagrams),} however, precluded a study of the Kondo regime \HL{for the Anderson model}.
%This concerned the stationary limit only, and
Clearly, addressing the time-dependent problem \HL{for this model} presents an even greater challenge.
% (8A)
Therefore the superfermion technique \cut{introduced there} deserves further attention and development before such attempts are to be made.
}

%\HL{
% an intermediate result.
%We have also obtained an exact time-evolution of the strongly interacting quantum dot in the limit of infinite temperature
%$T\rightarrow +\infty$.~\footnote{Note that this result does not rely on the equality of the tunnel rates $\Gamma_{r\sigma}$ for different reservoirs as one can 
%incorrectly conclude from the statements of \Cite{Oguri13a}: in the present work we use reservoir-resolved tunnel rates, which can be, in general, different
%for different reservoirs, $\Gamma_{r\sigma}\neq \Gamma_{r'\sigma}$ for $r\neq r'$. The results of \Cite{Saptsov12a} also do not rely on this assumption.}
%}

\hl{
% (1B)
\HL{Besides the aforementioned general indirect importance to experiments
and the \HL{concrete} nonperturbative predictions \eq{keyresult}}
the present paper also reports
 an extensive discussion of the time-dependence in the effectively noninteracting limit $U \ll \Gamma$.
\HL{In contrast to previous works, we include} spin coherence, electron-pair coherence (superconducting correlations~\cite{Governale08}) and two-particle correlations \HL{in the initial state of the quantum dot.}
In addition to various theoretical motivations mentioned in the following,
this \HL{limit} is also of experimental relevance.
% For instance, single molecule devices may exhibit transport dominated by a single orbital with very strong broadening.~\cite{THIJSSEN}
% Moreover, 
For example,
the mentioned highly controllable single-electron sources~\cite{Feve07, Gabelli06} can be understood very well in such a picture.
%Besides solving such problems efficiently, the formalism presented here also directly gives deep insight into the possible effects of interaction corrections (cf. \Eq{keyresult})
% as well as providing a new starting point for their systematic calculation.
}

\subsection{Theoretical motivation}
\hl{
The above mentioned experimental progress thus motivates theoretical developments
\HL{and 
in} this paper we invest in a reexamination of the fundamental starting points of transport theory
 and show that they can be exploited more explicitly.
As we now outline, this leads to the key physical principle underlying \Eq{keyresult}.}
% Theory
% The object of investigation is some smaller part of a larger system, e.g., a single molecule attached to macroscopically
% large contacts, which act as reservoirs and impose nonequilibrium boundary conditions.
To describe %the subsystem 
a quantum dot
in the presence of the reservoirs one uses a mixed-state theoretical description.
The mixed quantum state is described by the reduced density operator and can be conveniently considered as an element of a linear space of operators,
referred to as Liouville space.
The time evolution of the state is quite generally described by a kinetic or quantum master equation,
whose time-nonlocal kernel or self-energy is a superoperator on this Liouville space.
This picture is formally quite analogous to that of quantum mechanics of closed systems described in a Hilbert space.
However, the Liouville-space self-energy describes dissipative / non-Hamiltonian dynamics, including non-Markovian memory effects.

Technically,  the dynamics in Liouville space is more complicated because one needs to keep track of the evolution of state vectors (kets)  as well as their adjoints (bras):
in the language of Green's functions, the evolution on two Keldysh contours must be described.
As a result, the usual concepts of quasiparticles corresponding to quantum field operators breaks down.
For open \emph{fermion} systems the anticommutation sign presents additional problems in Liouville space.~\cite{Schmutz,Mukamel08,Schoeller09a}

%Super fermions
To address such problems, Schmutz~\cite{Schmutz} introduced \emph{superfermions}, i.e.,
analogs of quantum field operators that act on the many-particle Liouville space and obey a similar algebra.
It was shown that these, in fact, generate the Liouville space starting from some vacuum supervector
and can thus be used to construct mixed state density operators.
Following this analogy, insights from quantum field theory in Hilbert space could then also be applied to density-operator approaches to nonequilibrium systems.~\cite{Prosen08,Kosov09,Dzhioev}
In these works superfermions were applied mostly to {Markovian} quantum dynamics as described by a given self-energy or time-evolution kernel
and were found to simplify the diagonalization of Lindblad time-evolution generators, in particular,
 finding their stationary eigenvectors.~\cite{Prosen08}

% Causal superfermions are even better
In a recent work,~\Cite{Saptsov12a}, we have extended the application of superfermion techniques to the \emph{derivation} of the reduced dynamics from a 
system - bath approach within the framework of the general real-time transport theory~\cite{Koenig96prl,Koenig96b} formulated in Liouville-space.~\cite{Schoeller09a}
This does not rely on Born and/or Markov approximations.
Moreover, in contrast to the previous superfermion approaches,~\cite{Schmutz,Mukamel08,Kosov,Dzhioev,Prosen08} the special superfermions that are involved
simultaneously incorporate the structure imposed by causality,~\cite{Kamenev09,Jakobs10a} related to probability conservation,~\cite{Schoeller09a}
as well as the fermion-parity superselection rule of quantum mechanics.~\cite{Wick52}
The fermion-parity was already included by Schmutz,~\cite{Schmutz} but turns out to play a far more prominent role.
This causal structure furthermore exploits a Liouville-space analog of the ``Keldysh rotation''~\cite{Kamenev09}, \HL{well-known from Green's function approaches.}
Although these particular \HL{\emph{causal}} superoperators were  introduced earlier,~\cite{Schoeller09a}
their role as quantum fields in Liouville space was not recognized or taken advantage of.
% T=oo importance
In this formulation of the real-time approach, the unit operator plays the special role of the vacuum state 
in Liouville-Fock space of the reduced system.
However, physically this operator describes the \emph{infinite-temperature} mixed quantum state of the reduced system with maximal von Neumann entropy.
It was realized that there is a corresponding natural decomposition of the self-energy into an infinite-temperature part and nontrivial \emph{finite-temperature} corrections.
The \HL{causal} superfermion operators are constructed in such a way to maximally simplify and emphasize this fundamental structure of the perturbation theory 
for self-energy kernels, for time-evolution, and for arbitrary observables.
This is a general feature of open fermionic quantum systems
 which other superfermion formulations do not explicitly reveal.
The causal superfermions, furthermore, translate other fundamental properties of the underlying Hilbert-Fock space fields in a particularly clear way,~\cite{Saptsov12a}
such as their irreducible transformation under spin and particle-hole symmetry transformations,
as well as fluctuation-dissipation relations (related to the Liouville-space Wick theorem~\cite{Schoeller09a}).

\begin{figure}[t]
  \includegraphics[width=0.9\columnwidth]{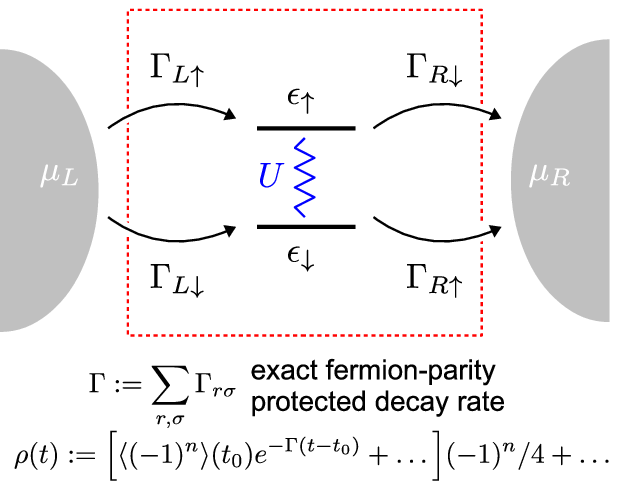}
  \caption{
    Anderson model with spin-dependent tunneling of electrons from reservoirs $r=L,R$
    into orbitals with energy $\epsilon_\sigma$ and local Coulomb interaction $U$.
    The dependence of the tunneling on the spin $\sigma=\uparrow,\downarrow$ can arise either from the tunnel barriers
    or from a spin-polarization of the density of states in the electrodes (e.g. ferromagnets)
    or from both.
    For simplicity the magnetic field $\mathbf{B}=B \mathbf{e}_z$ that causes the Zeeman splitting $\epsilon_\uparrow -\epsilon_\downarrow = B$
    and the axes of spin-polarization of the tunneling are assumed to coincide.
    The sum of all tunnel rates $\Gamma=\sum_{r,\sigma} \Gamma_{r\sigma}$ that connect the quantum dot to the electrodes
    turns out to be an \emph{exact} decay rate in the interacting ($U \neq 0$), nonequilibrium Anderson model.
    The corresponding decay mode is the fermion-parity operator $(-1)^n$,
    a central quantity in the construction of the causal superfermions.
  }
  \label{fig:sketch}
\end{figure}

% In this paper (1)
One of the aims of this paper is to highlight simple applications of causal superfermions
and illustrate the physical insight they convey into nonequilibrium transport through an Anderson quantum dot, sketched in \Fig{fig:sketch}.
\HL{These} particular quantum-field superoperators were introduced in the \HL{admittedly} rather complicated context of \Cite{Saptsov12a},
which constituted one of its major technical applications:
Only by exploiting the properties of the causal superfermions the two-loop real-time renormalization-group (RG) calculation
of the $T=0$ transport could be kept tractable,
even when using the minimalistic Anderson model for the quantum dot.
This may convey the incorrect impression that superfermions are not useful in simpler calculations
or that they even rely on the advanced RG machinery.
Indeed,  in \Cite{Saptsov12a} we already outlined how various aspects of the Liouville-space real-time approach
 are further clarified, independent of the RG formulated ``on top'' of it.
\footnote{
  For instance, it was shown that
  the proof of the Liouville-space Wick theorem for fermions~\cite{Schoeller09a} simplifies greatly when using causal superfermions, reducing to the standard proof by Gaudin.~\cite{Gaudin60}
  Moreover, the causal structure~\cite{Kamenev09} of the reservoir Green's functions in the wideband limit was shown to lead
  to an exponential reduction of the number of diagrams contributing to the self-energy.
  Also, an algebraic term-by-term proof of the bandwidth-independence of the perturbation theory was suggested,
  relating it to the completeness of the basis used for the reduced system.
}
These more formal insights have already found useful applications to real-time calculations
in several  works~\cite{Reckermann10thesis,Hell11thesis,Gergs13thesis,Hell13a} dealing with simpler problems and / or approximations.
The superfermion approach also allowed an \emph{exact} result to be found that is more specific to the Anderson model:~\cite{Saptsov12a}
two complex-valued eigenvalues of the exact self-energy superoperator lie symmetric with respect to an average value depending on known, bare parameters.
This implies a nonperturbative sum rule for the level positions \emph{and} broadenings of nonequilibrium excitations of the quantum dot in the presence of coupling to the reservoirs and interaction.
It was indeed \HL{noted} earlier in real-time perturbation theory~\cite{Splettstoesser12a} and more recently in a Liouville-space Green function study.~\cite{Oguri13a}

\emph{Time-dependence and fermion-parity.}
Another exact result obtained using the causal superfermions  provided more concrete physical insights into another previous work:
we showed that generically the \emph{exact} effective self-energy has an eigenvalue that is protected by the fermion-parity superselection rule of 
quantum mechanics.
This eigenvalue corresponds to the experimentally measurable decay mode,
\HL{the key result \eq{keyresult} mentioned in the previous section.}
Surprisingly, this decay mode depends only on the sum of all tunnel rate constants but not on any of the remaining parameters,
\emph{including the Coulomb interaction} $U$.
Using the superfermion approach this result, first obtained perturbatively in \Cite{Contreras12},
could be shown~\cite{Saptsov12a} to hold nonperturbatively in the tunnel coupling.
However, our study, \Cite{Saptsov12a}, did not consider spin-dependent tunneling, in contrast to \Cite{Contreras12},
a restriction that we lift in this paper.
Moreover, this result can be easily generalized for an arbitrary number of spin orbitals and, in fact, is independent of the details
of the interaction \HL{on the quantum dot (i.e., the concrete form of the quantum dot model Hamiltonian):} only the wideband limit and
the bilinear form of the tunnel Hamiltonian matter.
%
% In this paper (2)
%
This striking result motivates another aim of this paper,  \HL{namely,} to illustrate the usefulness of causal superfermions for time-dependent decay problems,
rather than the stationary state problems at the focus of \Cite{Saptsov12a}.
The first part of the paper is concerned with formulating the general time-dependent perturbation theory using causal superfermions,
and discusses several insights offered into interacting problems.
For instance, we show that the exact time-evolution superoperator has an effective expansion involving only causal ``\emph{creation} superoperators``
with intermediate propagators {which} are exponentially \emph{damped} in time, i.e., dissipative.
This \hl{physical} picture emerges when we integrate out Markovian correlations,
 leaving only that part of the bath dynamics that leads to the nontrivial, low-temperature phenomena in the Anderson model.
This applies generally, i.e., also to interacting systems, and may be an interesting starting point for direct numerical simulation 
schemes of reduced dynamics,~\cite{Cohen11, Cohen13a, Cohen13b, Weiss08,Segal10, Segal11, Segal13} since it eliminates some ``Markovian overhead'' from the start.
\hl{Throughout the paper we highlight such connections of admittedly formal expressions to physical insights,
which is important for effectively applying the technique to \me{more} complex problems \me{that have motivated this work.}}

\emph{Noninteracting limit and super-Pauli principle.}
To \HL{most} clearly highlight  applications of causal superfermions,
  the second part of this paper focuses on the simplest case,
  the noninteracting limit  ($U=0$) of the Anderson model.
Importantly, we include the spin -- distinguishing it from the spinless noninteracting resonant level model (NRLM) --
and we also include a  magnetic field and spin dependence of the tunneling.
\\
%
% Big spaces
%
First, our study provides a clear illustration of how the causal field superoperators allow
 one to deal with large fermionic Liouville spaces encountered in transport problems.
(Already for the very simplistic single-level Anderson model the dimension of this space equals 16.)
\hl{We show by direct calculation that in the noninteracting limit the perturbation theory
-- after a simple skeleton resummation
\cut{(renormalized perturbation theory)} that exploits the wideband limit (a \emph{discrete} renormalization step~\cite{Schoeller09a}) -- 
naturally terminates at the second loop order.
Moreover, the lack of interactions on the quantum dot results in an additional simplification,
namely that the two-loop part of the \cut{full} time-evolution superoperator factorizes.
As we show, this is less clear when considering the two-loop self-energy in Laplace space
\hl{as is often done when focusing on stationary state properties.\cite{Kubala06,Leijnse08a,Koller10}}
Most importantly, we show that this termination and factorization directly follow from the fundamental anticommutation relations of the causal fermionic superoperators.
The %truncation
termination 
is in fact a consequence of the \emph{super-Pauli principle}, the Liouville-space analog of the corresponding principle in the Hilbert-space of 
the quantum mechanics of closed systems,
\HL{which additionally relies on the independent principle of fermion-parity superselection (this principle is discussed in \Sec{sec:parity}).}}
% The time-dependent solution for the current in the spinless noninteracting Anderson model was also used as benchmark for a 
% real-time RG study~\cite{Andergassen11a} of the interacting resonant level model, in the limit where the nonlocal interaction coupling between 
% the quantum dot and the reservoirs vanishes.
% Our benchmark result provides the complete time-dependent propagator, density operator as well as the current
% and also includes the spin.
Interaction effects bring additional complications, and also here the causal superfermions bring about
simplifications, some of which were not yet noted in \Cite{Saptsov12a}
\HL{and are pointed out here.}
\\
%
% Benchmark idea
%
Second, the \HL{analysis} of the noninteracting limit provides an important benchmark
for studies employing the real-time transport theory.
% This approach is typically applied to deal with strongly interacting problems
% and the exact solution of the noninteracting limit has not been discussed much on a general level,
% in particular, when including the effect of multiple spin-orbitals.
\HL{This approach is tailored to deal with strongly interacting problems,
but it has proven difficult to see on a general level how the exact solution of the noninteracting limit is recovered, in particular, when including the effect of multiple spin-orbitals.}
Without spin, the exact solutions were checked to be reproduced explicitly by nonperturbative diagrammatic \HL{summation}~\cite{Schoeller97hab,Schoeller99tut}
but the simplifications due to the vanishing of interactions arise only after a detailed analysis of cancellations.
In the presence of spin- and orbital degeneracies, this is even less obvious when working with explicit, model-dependent matrix representations of superoperators in large Liouville spaces \HL{required for interacting systems.}
Our application of the causal superfermion approach to the real-time formalism
shows immediately how the noninteracting limit is \me{correctly} reached on the general \emph{superoperator} level.
\me{This illustrates that it may actually pay off to understand the noninteracting limit in the best possible way,
when one is interested in addressing interacting problems and
even when one is not expanding around the noninteracting limit.
So far this aspect has not been given much attention within the real-time framework.}
\\
%
% Comparison with other approaches
% 
\HL{Third}, the causal superfermion formulation also facilitates comparison of the real-time transport theory with other approaches, which usually take the noninteracting limit as a reference
(using, e.g., path-integrals or Green's functions) and therefore always contain its solution explicitly on a general level.

% RG motivates U=0 study
Our previous study, \Cite{Saptsov12a}, provides an example for which the stationary, noninteracting limit of the Anderson model \me{-- exhaustively analyzed here -- functions as} a benchmark.
The one- plus two-loop real-time renormalization group approach worked out there
has the important property that it includes the exact solution for the noninteracting limit $U=0$,
while for large $U$ it still provides a good approximation that is nonperturbative in $\Gamma$.
% Question
It was found that even for the noninteracting limit one still needs a full one- plus \emph{two}-loop RG to obtain the exact density operator.
Reformulated, the exact effective Liouvillian of the quantum dot turns out to be \emph{quartic} (instead of quadratic) in the  field
superoperators.
This may seem to be surprising at first in view of the absence of interactions. 
% Answer
\HL{\me{As mentioned above, in the present paper} we show that although the renormalized perturbation theory terminates,
it does so only at the \emph{second} loop order (i.e., terms quartic in the fields),
finding explicit agreement with the stationary RG results~\cite{Saptsov12a} for the full density matrix, self-energy and charge current.}
% Moreover, the lack of interactions on the quantum dot results in an additional simplification,
% namely that the two-loop part of the full time-evolution superoperator factorizes.
% As we show, this is less clear when considering the two-loop self-energy in Laplace space.
% %
% Most importantly, we show that this truncation and factorization directly follow from the fundamental anticommutation relations of the causal fermionic superoperators.
% The truncation is in fact a consequence of the \emph{super Pauli principle}, the Liouville-space analog of the corresponding principle in the Hilbert-space of the quantum mechanics of closed systems.
The time-dependent solution for the current in the spinless noninteracting Anderson model was also used as benchmark for \HL{another} real-time RG study~\cite{Andergassen11a}, \HL{dealing with} the interacting resonant level model, in the limit where the nonlocal interaction coupling between 
the quantum dot and the reservoirs vanishes.
\HL{The benchmark result of the present paper} provides the complete time-dependent propagator, density operator as well as the current
and also includes the spin.

% Real-time approach good for noninteracting
Combined with the second quantization tools of causal superfermions,
the real-time Liouville-space approach becomes a more accessible tool
for dealing with noninteracting problems.
Such problems continue to attract attention,~\cite{Maciejko06,Jin10} especially regarding non-Markovian effects that arise beyond the wideband limit,
for which no general analytic solution seems to be known.
In the wideband limit, noninteracting problems can
 be solved by means of various other techniques,
 for both the stationary limit~\cite{Caroli71} and for the transient approach.~\cite{Langreth91}
However,  we demonstrate how in the real-time approach the full time-evolution can be calculated quite straightforwardly in this limit, i.e.,
avoiding a self-energy calculation and without transforming to Laplace space and back.
In the {description of the noninteracting decay} we include the effects of spin and pairing coherence and of two-particle 
correlations in the initial quantum dot state \refAB{that have been ignored so far}.
Solving such problems with the real-time approach has the additional advantage
of directly allowing one to gauge the effect of interactions,
and to include them, in either a perturbative or nonperturbative way.
\me{A case in point is the key result for fermion-parity protected decay mode,
\Eq{keyresult}.}
Finally, our formulation also provides a framework for calculating corrections beyond the wideband limit.

%%%%%%%%%%%%%%%%%%%%%%%%%%%%%%%%%%%%%%%%%%%%%%%%%%%%%%%%%%%%%%%%%%%%%%%%%%%%%%%%%%%%%%%%%%%%%%%%%%%%%%%%%%%
\emph{Outline}
The paper is organized as follows.
In \Sec{sec:model}, we directly formulate the model in Liouville-space notation,
introduce the causal superfermion fields, construct the Liouville-Fock space
and formulate the super-Pauli principle.
In \Sec{sec:timeevolution}, we formulate the time-dependent Liouville-space perturbation theory
and derive a renormalized series that explicitly incorporates the wideband limit on the superoperator level.
We give general rules for the simplifications that arise in the noninteracting ($U=0$) limit.
This critically relies on the causal structure which is made explicit by the causal superfermions.
\HL{Importantly, in this limit the renormalized series naturally terminates at the second loop, and higher-order corrections are identically 
equal to zero.}
This analysis also reveals the special importance of the physical infinite temperature limit $T \rightarrow \infty$,
serving as a reference point for both the construction of Liouville-Fock space and the renormalized perturbation theory.
In \Sec{sec:results}, we perform the explicit one- and two-loop calculations
giving the exact, full time-evolution propagator, the density operator and the charge current
in the noninteracting limit ($U=0$).
To better understand the stationary limits of these results and to directly compare with the real-time RG-results  of \Cite{Saptsov12a},
we additionally perform the calculation directly in Laplace space.
We summarize our results in \Sec{sec:conclusion}
and discuss their generalization to multiple orbitals,
the implications for real-time renormalization-group schemes,
and possible further application of the developed ideas.

\section{Anderson model in Liouville-Fock space\label{sec:model}}
\subsection{Fermion-parity superselection rule\label{sec:parity}}

In this paper, we  make explicit  use  of the postulate of fermion-parity (or univalence) superselection  in quantum mechanics and quantum field theory.~\cite{Wick52,Aharonov67}
It is a part of quantum kinematics and can therefore be discussed before any model of the dynamics is formulated.
Here, we briefly illustrate the main substance of this postulate and discuss one of its aspects, which is crucial for starting up the formulation
of our approach in the following.
For example, for a single level quantum dot with field operators $d_\sigma$ and $d_\sigma^\dag$, where $\sigma=\pm$ corresponds to spin up ($\uparrow$) 
and down ($\downarrow$) along the $z$-axis,
the fermion-parity operator, recurring at many crucial steps in the paper, is defined as
\begin{align}
  (-1)^n := e^{i \pi n} = \prod_\sigma (1-2n_\sigma),
\end{align}
where $n=\sum_\sigma n_\sigma$
is the fermion number operator.
For this simple case, the fermion-parity superselection rule can be phrased as follows:~\cite{Bogolubov89} the density operator and the operator of any physical observable $A$ must commute with the total fermion parity operator of the system:
\begin{align}
  [(-1)^n,A]_{\HL{-}} =   [(-1)^n,\rho]_{\HL{-}} = 0.
  \label{supersel}
\end{align}
This excludes the possibility of interference (superpositions of) states with even and odd number of fermions.
The operator $(-1)^n$ has been applied in \Cite{Gardiner04} to make field operators for different fermion species commute (rather than anticommute), and it plays a key role in setting up the second quantization in Liouville space.

Here and in the following, it is convenient to introduce an additional particle-hole index $\eta$:
\begin{align}
  {d}_{\eta \sigma}
  &= \left
    \{ \begin{array}{rl}
      {d}^{\dagger}_{\sigma}, ~ \eta=+ \\
      {d}_{\sigma}, ~ \eta=-
    \end{array} \right.
  .
  \label{etadef}
\end{align}
Throughout the paper we will denote the inverse value of a two-valued index with
a bar, e.g.,
\begin{align}
  \bar{\eta} = -\eta
\end{align}
We combine all indices into a multi-index variable written as a number:
\begin{align}
  1 = \eta,\sigma; \quad \bar{1}= \bar{\eta},\sigma
\label{bar1},
\end{align}
where, by way of exception, the bar denotes inversion of the particle-hole index
only.
 If we have more than one multi-index, we distinguish their components by
 using the multi-index number as a subscript:
 $1=\eta_1,\sigma_1,r_1,\omega_1,~2=\eta_2,\sigma_2,r_2,\omega_2$
and use a multi-index Kronecker symbol
\begin{align}
  \delta_{12} = \delta_{\eta_1, \eta_2} \delta_{\sigma_1, \sigma_2}
  .
  \label{delta12}
\end{align}
For clarity, we usually omit these subscripts if there is only one multi-index as in \Eq{bar1}.  Then ${d}_{1}= {d}_{\eta\sigma}$ and
 $ {d}_{\bar{1}}= {d}_{-\eta \sigma}$,
and we can summarize all fundamental relations simply by $(d_{1})^\dag = d_{\bar{1}}$ and $[d_1,d_2]_{+} = \delta_{1\bar{2}}$.
\HL{Throughout the paper, we denoted the (anti)commutators by $[A,B]_-=AB-BA$ and $[A,B]_+=AB+BA$.}

A first application of the fermion-parity arises when we connect the quantum dot to reservoirs with field operators $a_{\sigma r k}$, where $\sigma$ is the spin index, $k$ the orbital index, and 
the reservoir index $r= \pm$ corresponds to left/right.
We have to make a \emph{choice} for commutation relations of $a_{\eta \sigma r k}$ relative to $d_{\eta'\sigma'}$:
{in setting up the second quantization}, one is free to choose either commutation or anticommutation relations for fermions of different states / particles, whereas one \emph{must} have anticommutation relations for fermions in the same state.~\cite{LandauLifshitz3}
Both choices produce identical, correctly antisymmetrized multi-particle states.
Usually, the most elegant choice, indicated here by a prime on the field operators, is to let them all anticommute:
\begin{subequations}
  \label{anticomm}
  \begin{align}
    [ a'_1 , d'_2 ]_{+} & = 0
    ,
    \label{fcanticomm}
    \\
    [d'_1 , d'_2 ]_{+} & =  \delta_{1\bar{2}}
    ,
    \label{canticomm}
    \\
    [ a'_1 , a'_2 ]_{+} & = \delta_{1\bar{2}}
    .
    \label{fanticomm}
  \end{align}
\end{subequations}
The fields are pairwise Hermitian adjoints, $(d_1')^{\dag}=d_{\bar{1}}'$ and $({a'_1})^{\dag}=a'_{\bar{1}}$.
The fermion number operator of the dot is expressed as $n=\sum_\sigma  {d'_{\sigma}}^\dag d'_{\sigma}$ and the corresponding fermion-parity operator anticommutes with the dot fields $d'_1$ (using $[(-1)^n]^2=e^{2i\pi n}=1$),
\begin{align}
  (-1)^n d'_1(-1)^n=e^{i\pi n} d'_1 e^{-i\pi n}=-d'_1
  ,
  \label{dprimeanticommute}
\end{align}
but commutes with the reservoir operators $a'_1$ (like any operator local to the dot),
\begin{align}
  \label{f-n-com}
  a'_1 (-1)^n = (-1)^n a'_1.
\end{align}

In approaches where the reservoir degrees of freedom are eliminated by a partial trace operation, it is much more convenient to let reservoir and dot fields \emph{commute} by definition, allowing operators of different subsystems to be separated easily.
By doing this from the start many unnecessary canceling sign factors can be avoided.
Such fields are used throughout this paper and are indicated by leaving out the prime.
The fields on the different (the same) systems mutually (anti)commute:
\begin{subequations}
  \label{comm}
  \begin{align}
    [a_1,d_2]_- & =0,
    \\
    [a_1,a_2]_+ & =\delta_{1\bar{2}},
    \\
    [d_1,d_2]_+ & =\delta_{1,\bar{2}},
  \end{align}
\end{subequations}
with $d_1^{\dag}=d_{\bar{1}}$ and $a_1^{\dag}=a_{\bar{1}}$.
This choice of commutation relations was used in \Cites{Schoeller09a} and \onlinecite{Saptsov12a}, and will be used here as well,
unless stated otherwise.

The fermion-parity operator now appears as the formal device relating the above two choices, which is convenient to have at hand for a direct comparison with other approaches, 
e.g., the Green's {function} approach,~\cite{Saptsov13b} starting from the choice \Eq{anticomm}.
One way of obtaining the choice in \Eq{comm} from the fields satisfying \Eq{anticomm} is the following change of variables:
\begin{subequations}
  \label{primetrans}
  \begin{alignat}{5}
    a_1        &= (-1)^n a'_1      & &=         a'_1     (-1)^n,
    \label{ad-oper}
    \\
    d_1        &= -\eta_1 (-1)^n d'_1 & & =  \eta_1 d'_1 (-1)^n.
    \label{dprime}
  \end{alignat}
\end{subequations}
Here,  $(a_1)^{\dag}=a_{\bar{1}}$ and the $\eta$-sign ensures that the adjoint relation $(d'_1)^{\dag}=d'_{\bar{1}}$ is also preserved:
using \Eq{dprimeanticommute} $
(d_1)^{\dag}
 = \eta_1 (-1)^n (d'_1)^\dag
 = - \eta_1  (d'_1)^\dag (-1)^n
 = \eta_{\bar{1}}  d'_{\bar{1}} (-1)^n
 = d_{\bar{1}}
$.
A key point, needed later, is that when tracing out the reservoirs only averages of products of an even number of reservoir fermions can appear, and the quantum dot operator $(-1)^n$ in \Eq{ad-oper} cancels out in
$\text{Tr}_R a_1 \ldots a_{2k} = \text{Tr}_R a'_1 \ldots a'_{2k}$ since $(-1)^{2kn}=1$.
The transformation \eq{primetrans} is only canonical \emph{locally} on the quantum dot and on the reservoirs. Since it is not globally canonical we must check how observable operators are transformed. This is done in \Sec{sec:anderson} once we have specified the dynamics and the physical operators of interest.

\subsection{Anderson model\label{sec:anderson}}
The model that we consider was already sketched in \Fig{fig:sketch}.
The usual formulation of the single-level Anderson model
specifies the Hamiltonian
\begin{align}
  \label{dot_ham}
  H  =   \epsilon n + B S_z  + U  n_\uparrow  n_\downarrow 
\end{align}
where $\epsilon$ denotes the energy of the orbital and
\begin{align}
  n=\sum_\sigma  n_\sigma , ~~~~  n_\sigma=  {d}^{\dagger}_\sigma  {d}_\sigma
  ,
  \label{eq:n}
\end{align}
is the fermion number operator.\footnote{The operators $n_\sigma$ and $\vec{S}$ have the same form in terms of $d_1$ operators as in terms of $d_1'$, see
discussion after \Eq{banticomm}.}
Furthermore, $S_z=\tfrac{1}{2}\sum_\sigma \sigma n_\sigma$ is the $z$ component of the
spin vector operator
$\vec{S}=
\sum_{\sigma \sigma'}  \tfrac{1}{2} \vecg{\sigma}_{\sigma,\sigma'} 
{d}^{\dagger}_{\sigma}{d}_{\sigma'}
$ along the external magnetic field $\vec{B}=B\vec{e}_z$ (in units $g
\mu_B =1$)
and
$\vecg{\sigma}$ is the vector of Pauli matrices.
The dot is attached to electrodes which are treated as free electron reservoirs:
\begin{align}
  H^{R}=\sum_{\sigma,r,k} \epsilon_{\sigma r k}   {a}^{\dagger}_{\sigma r k} 
  {a}_{\sigma r k}
  .
\end{align}
The reservoir electron number and spin 
\begin{align}
  {n}^{R}  = \sum_r {n}^r, \quad  \vec{s}^{R} &= \sum_r \vec{s}^r,
  \label{eq:nR}
\end{align}
respectively, can be decomposed into their contributions
$
{n}^r =
\sum_{\sigma,k}  {a}^{\dagger}_{\sigma r k}  {a}_{\sigma r k}
$
and $
\vec{s}^r =
\sum_{\sigma,k}  \tfrac{1}{2} \vecg{\sigma}_{\sigma,\sigma'}{a}^{\dagger}_{\sigma r k}  {a}_{\sigma' r k}$.
Before we introduce the coupling, we introduce the notation of \Cite{Saptsov12a} to conveniently deal with the continuum limit.
The reservoirs are described by a density of states
$
\nu_{r\sigma} (\omega)= \sum_k \delta (\omega - \epsilon_{\sigma r k} +\mu_{r} )
$ and we go to the energy representation of the fermionic operators,
\begin{align}
  {a}_{\sigma r} (\omega)
  = \frac{1}{\sqrt{\nu_{r} (\omega) } } \sum_k  {a}_{\sigma r k}\delta (\omega -
  \epsilon_{\sigma r k} +\mu_{r} )
  ,
\end{align}
with the anticommutation relations:
\begin{align}
  [ {a}_{\sigma r} (\omega),  {a}_{\sigma' r'}^{\dagger} (\omega')]_{+}
  &=
   \delta_{\sigma,\sigma'} \delta_{r,r'} \delta (\omega - \omega'),
  \\
  [ {a}_{\sigma r} (\omega),  {a}_{\sigma' r'} (\omega')]_{+} &= 0
 .
\end{align}
Here, we denote (anti)commutators by $[A,B]_{\mp} = A B \mp B A$.
The continuous reservoir Hamiltonian is thus
\begin{align}
  H^{R} = \sum_{\sigma,r} \int d\omega (\omega + \mu_r ) 
  {a}^{\dagger}_{\sigma r} (\omega)  {a}_{\sigma r} (\omega)
  ,
  \label{resHam}
\end{align}
with the electron energy $\omega$ taken relative to $\mu_r$ for reservoir $r$.
The junctions connecting the dot and reservoirs are modeled by the tunneling
Hamiltonian
\begin{align}
   {V} &=  \sum_{r} V^r,
   \label{Vdef}
  \\
   {V^r} &=  \sum_{\sigma} \int d\omega \sqrt{\nu_{r\sigma} (\omega)}
   \left( t_{r\sigma} (\omega) {a}^{\dagger}_{\sigma r} (\omega)  {d}_{\sigma}+ h.c.
\right),
   \label{Vrdef}
\end{align}
with real spin-dependent amplitudes $t_{r\sigma} (\omega)$. Using the spectral density
\begin{align}
  \Gamma_{r\sigma} (\omega) =2\pi \nu_{r\sigma} (\omega) |t_{r\sigma}(\omega)|^2,
  \label{spectdens}
\end{align}
it is convenient to rescale the field operators:
\begin{align}
\label{rescaling}
   {b}_{\sigma r} (\omega) =\sqrt{\frac{\Gamma_{r\sigma} (\omega)}{2\pi}} 
{a}_{\sigma r} (\omega).
\end{align}
We thus incorporate two sources of spin-polarization of the tunneling rates:
either the attached electrodes are ferromagnetic [$\nu_{r\sigma} (\omega)$]
or the tunnel junctions are magnetic [$t_{r\sigma}$], or both.
We made the simplifying assumption that the magnetizations of the electrodes are collinear
(either parallel or antiparallel), and also collinear with the spin-polarization axes of the tunnel junctions.
Moreover, the external magnetic field $\vec{B}$ is assumed to be collinear with this axis.

As before, we introduce an additional particle-hole 
index:
\begin{align}
  {b}_{\eta\sigma r} (\omega)
  & = \left
    \{ \begin{array}{rl}
      {b}^{\dagger}_{\sigma r} (\omega), ~ \eta=+ \\
      {b}_{\sigma r} (\omega), ~ \eta=-
    \end{array} \right.
  .
\end{align}
and combine all indices into a multi-index variable written as a number, which now includes an additional continuous index $\omega$:
\footnote{
  Although the multi-indices of the reservoirs
  contain additional variables,
  we use the same notation for them. 
  No confusion should arise
  since it is always clear from the context to
  which type of operator (dot, reservoir) the given multi-index belongs.
}
\begin{align}
  1 = \eta,\sigma,r,\omega; \quad \bar{1}= \bar{\eta},\sigma,r, \omega
  .
\end{align}
Then ${b}_{1}= {b}_{\eta\sigma r} (\omega)$ and $ {b}_{\bar{1}}= {b}_{-\eta,
\sigma r} (\omega)$
and the (anti)commutation relations are
\begin{align}
  [ d_1, b_2 ]_{-} & =  0,
  \\
  [ d_1, d_2 ]_{+} & =  \delta_{1\bar{2}},
  \label{danticomm}
  \\
  [ b_1, b_2 ]_{+} & = \frac{\Gamma_1}{2\pi} \delta_{1\bar{2}},
  \label{banticomm}
\end{align}
where $\Gamma_1=\Gamma_{r\sigma}(\omega)$.
In \Eq{banticomm}, it is left implicit that the multi-index $\delta$-function $\delta_{1\bar{2}}$ contains an additional delta-function $\delta(\omega_1-\omega_2)$ relative to the Kronecker-delta \eq{delta12} in \Eq{danticomm}.

Since we formulated our model in terms of fields obtained by a noncanonical transformation, we should now check the form of the model in terms of the (primed) anticommuting fields [\Eq{anticomm}].
First, any local reservoir observable has the same form in terms of $b_1$ operators as in terms of $b'_1$. This immediately follows from the fermion-parity superselection rule:
by \Eq{supersel} local reservoir observables always 
contain products of even numbers of the primed reservoir field operators.
Second, any  operator local to the quantum dot also has the same form due to fermion-parity superselection rule \emph{if} it conserves the fermion number $n$.
\footnote{
  For models with particle nonconserving terms
  additional signs may however appear
  which change the \emph{form} of the model Hamiltonian expressed in terms of the new variables.
  This is only a change of variables and the change of the Hamiltonian form does not alter the physics.
}
Locally, we can thus express everything in terms of $b_1$ and $d_1$ by simply omitting the primes.
However, the interaction operator
$V
=
\sum_{\sigma,r} \int d\omega
({b'}^\dag_{\sigma r \omega} d'_{\sigma}+ {d'_{\sigma}}^\dag b'_{\sigma r \omega})
=
\sum_1 \eta b'_{\bar{1}} d'_1$ now changes its form. In fact, it simplifies by loosing its $\eta$-sign:
\begin{align}
  V &=    b_{\bar{1}} d_{{1}},
  \label{Vshort}
\end{align}
Here, we implicitly sum over all discrete parts of the multi-index
$1$ (i.e., $\eta,\sigma,r$) and integrate over its continuous 
part ($\omega$).
An alternative discussion of the above not explicitly referring to the fermion parity can be found in~\Cite{Schoeller09a}

Finally, the reservoirs are assumed to be in thermal equilibrium with temperature $T$,
each described by its own grand-canonical density operator,
\begin{align}
  \label{grand_canon}
  \rho^{R}=\prod_{r}\rho^{r}, \quad \rho^r = \frac{1}{Z^r} e^{-\frac{1}{T} (
H^{r}-\mu^r {n}^r)},
\end{align}
where
$Z^r = \Tr{r} e^{-\frac{1}{T} ( H^{r}-\mu^r {n}^r)}$.
For the example setup shown in \Fig{fig:sketch}
one can give the electrochemical potentials by, e.g., assuming 
a symmetrically applied bias voltage, i.e., 
$\mu_{L,R}=\pm V_b/2$.
However, most of our results apply to any number of electrodes and
do not depend on this.

Together with the the Hamiltonian of the total system,
\begin{align}
  \label{tot_Ham}
  H^{{\tot}}= H  +  H^{R} +  {V},
\end{align}
this specifies the model.
The noninteracting resonant-level model (NRLM) is obtained by setting
$U=0$ in the dot Hamiltonian \Eq{dot_ham} in \Eq{tot_Ham}
and discarding the spin.

\subsection{Reduced time-evolution propagator}
In order to calculate the dynamics of the reduced density operator of the quantum dot,
 we first need to consider the evolution of the total system density operator.
It is generated by the Liouville--von Neumann equation:
\begin{align}
  \partial_t \rho^{\mathrm{tot}} (t)
  = -i \left[  H^{\mathrm{tot}} ,  \rho^{\mathrm{tot}} (t) \right]_{-}
  = -i L^{\mathrm{tot}} \rho^{\mathrm{tot}} (t)
  ,
  \label{von-Neumann}
\end{align}
with the Liouvillian superoperator
$L^{\mathrm{tot}}\bullet=[H^{\mathrm{tot}} , \bullet]_{-}$.
Superoperators are linear 
transformations of operators
and throughout the paper (if needed) we let the solid bullet $\bullet$ indicate the operator on which a
superoperator acts.
In the following, we will make the common assumption that
 the initial state of the total system at time $t_0$ is a direct product
\begin{align}
  \rho^{\mathrm{tot}}(t_0) =  \rho^{R} \rho (t_0).
  \label{initialstate}
\end{align}
However, some of the developments reported in the following do not depend on this assumption.
In a forthcoming work~\cite{Saptsov13b} we will show that the causal superfermion approach
is useful also when initial reservoir-dot correlations are present.
The formal solution of \Eq{von-Neumann} is:
\begin{align}
  \label{vonNeu}
  \rho^{\mathrm{tot}} (t)
  & = e^{-i L^\mathrm{tot} (t-t_0)} \rho^{\mathrm{tot}} (t_0)
  .
\end{align}
The reduced dot density operator is obtained by integrating out of
reservoirs degrees of freedom:
\begin{align}
  \rho (t) = \Tr{R} \rho^{\mathrm{tot}} (t) =\Tr{R}\left( e^{-i L^\mathrm{tot} (t-t_0)}
    \rho^{R} \right) \rho (t_0)
  .
  \label{rhot}
\end{align}
Equation \eq{rhot} is the starting point for a perturbation theory
for the propagator superoperator
\begin{align}
  \Pi (t,t_0) = \Tr{R}\left( e^{-i L^\mathrm{tot} (t-t_0)}\rho^{R} \right) \bullet
  .
  \label{pit}
\end{align}
Decomposing $L^\mathrm{tot}= L + L^{R} +L^V$, with
$L    =[H  ,\bullet]_{-}$,
$L^{R}=[H^{R},\bullet]_{-}$
we expand in the tunnel coupling $L^V=[V,\bullet]_{-}
\sim \sqrt{\Gamma}$.
Usually two additional steps are taken, in either order.
First, one derives a Dyson equation for the exact propagator and introduces a self-energy $\Sigma(t,t')$.
\begin{align}
  \Pi(t,t_0) &=  e^{-iL(t-t_0)}
  \nonumber
  \\
               & -i \timeint{t}{t_2}{t_1}{t_0} e^{-iL(t-t_2)} \Sigma(t_2,t_1) \Pi(t_1,t_0)
               .
   \label{dyson}
\end{align}
The reduced density operator is then found to satisfy Nakajima-Zwanzig / generalized master / kinetic equation
\begin{align}
  \partial_t \rho(t)=-i \int \limits_{t_0}^{t}d t'  L(t,t') \rho (t')
  ,
  \label{effective-l}
\end{align}
where
\begin{align}
  L(t,t^{'})=L \bar{\delta}(t-t') + \Sigma(t,t')
  \label{effective}
\end{align}
is the so called \emph{effective Liouvillian} for the quantum dot.
We introduced $\bar{\delta}(t-t') := 2 \delta(t-t')$ such that \begin{align}
  \int_{t_0}^t dt' \bar{\delta}(t-t') = 1
  .
  \label{deltabar}
\end{align}
to absorb the factor 2 that is required to recover the Liouville equation 
for   $\partial_t \rho(t)= -iL \rho (t')$ from \Eq{effective-l} for $\Sigma(t,t')=0$ (since $\int_{t_0}^{t}\delta(t-t') dt'=1/2$).
See Refs. \onlinecite{Timm08} and \onlinecite{Koller10} for a discussion of the equivalence of the Nakajima-Zwanzig and real-time approaches, 
and \Cite{Jin10} for a derivation using Feynman path-integrals in Keldysh space in the context of relaxation dynamics of the NRLM.
The problem is then reduced to the calculation of the self-energy and the subsequent solution of the kinetic equation \eq{effective-l}.
In many cases, this step is indeed advantageous.
Second, if one is mostly interested in the stationary state of the dot, it is more
convenient to change to a Laplace representation.~\cite{Schoeller09a}
Equation \eq{effective-l} in the Laplace representation is then the starting point for the calculation of stationary quantities using different 
approximate calculation schemes, e.g., perturbative~\cite{Koenig96prl,Leijnse08a,Emary09} and renormalization group approaches.~\cite{Schoeller09a,Pletyukhov12a,Saptsov12a}
The time evolution can be obtained by calculating the full Laplace image of the reduced density operator by means of perturbation theory or 
renormalization group approaches,~\cite{Pletyukhov10,Andergassen10,Andergassen11a,Hoerig12} and then performing the inverse Laplace transformation.

However, a direct approach in the time representation is of interest.
For instance,  even if one is interested in stationary properties in the end, some manipulations may be easier or clearer in the time representation, for instances, in problems of noise 
and counting statistics,~\cite{Thielmann04a,Braggio05, Flindt08, Flindt10}
Markovian approximations~\cite{Breuer} and adiabatic driving corrections~\cite{Splettstoesser05,Splettstoesser06,Kashuba12} or 
simplifications for higher-order tunnel rates in the stationary limit.~\cite{Koller10}
Whereas the above cited works mostly deal with strongly interacting (Anderson) quantum dots,
here, the noninteracting limit of the Anderson model ($U=0$) has our interest.
We show that for this case it is convenient to work directly with the propagator $\Pi(t,t_0)$ in the time-representation. The self-energy is only used in an intermediate renormalization step of the perturbation series for $\Pi(t,t_0)$  to deal with the wideband limit.
For now, however, we make no assumption on $U$ unless stated otherwise.

Although we only calculate Schr\"odinger picture quantities, it is useful to extend the standard interaction representation~\cite{LandauLifshitz9} to the Liouville space.~\cite{Fickbook}
The solution of the von Neumann equation \eq{vonNeu} for the total
density operator has a form familiar from the Hamiltonian time-evolution of the state vector in Hilbert-space quantum mechanics:
\begin{align}
  \label{rho-inter}
  \rho^{\mathrm{tot}}(t) = e^{-i\left( L+L^R\right)(t-t_0)}\hat{T}
  e^{-i\int\limits_{t_0}^t  L^V (\tau) d\tau} \rho^{\mathrm{tot}}(t_0),
\end{align}
where $\hat{T}$ denotes the time-ordering of superoperators
and $ L^V(\tau)$ are the tunnel Liouvillians in the interaction picture:
\begin{align}
\label{general_interaction}
  L^V (\tau)=e^{i\left(L+L^R\right)(\tau-t_0)} L^V
e^{-i\left(L+L^R\right)(\tau-t_0)}.
\end{align} 
Expanding \Eq{rho-inter} in $L^V(t)$, one obtains the time-dependent perturbation expansion
\begin{align}
\label{total-expansion}
 &\rho^{\mathrm{tot}}(t)= e^{-i\left(L+L^R\right)(t-t_0)}\left[1-i\int\limits_{t_0}^t
dt_1   L^V (t_1) \right.\\ \nonumber
&\left.+(-i)^2\int\limits_{t_0}^t dt_2\int \limits_{t_0}^{t_2} dt_1   L^V
(t_2)  L^V (t_1)+ ...\right]\rho^{\mathrm{tot}}(t_0)
.
\end{align}
Making use of $\mathrm{Tr}_{R} L^R =0$,
the perturbation expansion for the time-dependent reduced density operator in powers of $L^V$
 reads as
\begin{widetext}
\begin{subequations}
\label{pt_lv}
\begin{align}
  \rho(t)
  &
  =  e^{-i{L}(t-t_0)}\Tr R [ 1 \, + 
    \sum_{m=0}^{\infty} (-i)^{m}
    \underset{t \geq t_m \ldots \geq t_1 \geq t_0}{\int dt_1 \ldots dt_m}
    L^V (t_m) \ldots L^V (t_1)
  ]\rho^{\mathrm{tot}}(t_0) 
  \label{reduced-expansion}
  \\
  &= e^{-i{L}(t-t_0)} \rho(t_0)
  + \sum_{m=0}^{\infty} (-i)^{m}
  \underset{t \geq t_m \ldots \geq t_1 \geq t_0}{\int dt_1 \ldots dt_m}
   e^{-iL(t-t_m)} \Tr R
  \Big[L^V e^{-i(L+L^R)(t_m- t_{m-1})} \ldots L^V e^{-i(L+L^R)(t_1-t_0)} \rho^{\mathrm{tot}}(t_0) \Big]
.
  \label{reduced-expansion-usual}
\end{align}
\end{subequations}
\end{widetext}
A direct analysis of the series \Eq{reduced-expansion} is cumbersome, even for the noninteracting case. 
To derive a manageable series, we now introduce the \emph{causal} fermionic-superoperators in the next section.

\subsection{Second quantization in Liouville Fock space}

The above Liouville space formulation of quantum mechanics is well-known~\cite{Fano1957}
and has found many applications.~\cite{Mukamelbook}
However, when applied to quantum many particle systems, it becomes more powerful
if analogs of field-theoretical techniques are introduced, in particular,
field \emph{super}operators and Liouville-Fock space.~\cite{Schmutz,Prosen08,Kosov,Kosov09,Mukamel08,Esposito09rev,Saptsov12a}

\subsubsection{Causal superfermions
\label{sec:superfermion}}

Special \emph{causal} field superoperators may be constructed:
\footnote{
 The definitions $G^{\pm}$ in \Eq{superG} and $J^{\pm}$ in \eq{superJ} differ from those in \Cite{Saptsov12a} [Eqs.(51) and (52) respectively]
 by using the opposite convention for the sign of $q$.
 This is effected by placing the field operators $d_1$ and $b_1$ to the right of the operators $(-1)^n$ and $(-1)^{n^R}$ in \Eqs{superG} and \eq{superJ}, respectively.
 In contrast to Eqs.(51) and (52) in \Cite{Saptsov12a},
   we here explicitly write the superoperators $p^{L^n}\bullet=p^{n}\bullet p^{n}$, ($p=\pm$, $p^{2n}=1$).
 However, the definitions of $\bar{G}$ and $\tilde{G}$ \Eq{GJ-def} coincide with those of \Cite{Saptsov12a}.
 Here, we favor $q=+$ for ``creation'' ($G^+=\bar{G}$) and $q=-$ for ``destruction'' superoperators ($G^-=\tilde{G}$) [but opposite in the reservoirs, cf. \Eq{superJ}],
 tying in better with the discussion of second quantization to Liouville space.
 The opposite convention used in \Cite{Saptsov12a} agrees better with that
 used for ``quantum'' and ``classical'' fields obtained by the Keldysh rotation in the path integral approach of Kamenev et. al.~\cite{Kamenev09}.}
for the quantum dot they read as
\begin{align}
  \label{superG}
  G^q_1\bullet =
  \frac 1{\sqrt{2}} \Big\{  d_1\bullet + q(-1)^n\bullet (-1)^n d_1 \Big\}
  ,
\end{align}
and for the reservoir
\begin{align}
  \label{superJ}
  J^q_1\bullet=
   \frac 1{\sqrt{2}} \Big\{ b_1\bullet - q(-1)^{n^{R}}\bullet   (-1)^{n^{R}} b_1 \Big\}
  .
\end{align}
Here, $q=\pm$ labels the components obtained after a ``Keldysh rotation'' of simpler superoperators defined by left and right multiplication with a Hilbert-Fock space field.
Moreover, $(-1)^n = e^{i\pi n}$ and $(-1)^{n^R} = e^{i\pi n^R}$ are the fermion-parity operators of the dot and reservoirs, respectively [cf. \Eqs{eq:n} and \eq{eq:nR}].
The additional minus sign in the second term of the definition of $J^q_1$, relative to the definition of $G^q_1$ is purely conventional but is advantageous later [cf. \Eqs{ret-contr}-\eq{keld-contr}].
The tunneling Liouvillian $L^V$ can be written compactly as
\begin{align}
  L^V
  =\sum_{q=\pm} G^{q}_1 J^{q}_{\bar{1}}
  =\bar{G}_1 \bar{J}_{\bar{1}} + \tilde{G}_1\tilde{J}_{\bar{1}}
  ,
  \label{q-li}
\end{align}
where in the second equality we used ``bar-tilde'' notation of \Cites{Schoeller09a} and \onlinecite{Saptsov12a} for the $q=\pm$ components, respectively:
\begin{equation}
  \begin{alignedat}{2}
    \bar{G}_1 &= G_1^{+}, &\quad  \bar{J}_1 & =J_1^{+} ,
    \\
    \tilde{G}_1 &= G_1^{-}, & \tilde{J}_1 & =J_1^{-} ,
  \end{alignedat}
  \label{GJ-def}
\end{equation}
which is sometimes more convenient.
We note that in the rewriting of $L^V$ as \Eq{q-li} the fermion-parity superselection rule is already used, see~\Cite{Saptsov12a} for a discussion.
The superoperators $G_1^q$ and $J_1^q$ obey fermionic anticommutation relations:
\begin{alignat}{3}
\label{anticommut_d}
  \Big[ \bar{G}_1,\tilde{G}_2\Big]_+ &= \delta_{1,\bar{2}}
  ,
  &~~
  \Big[ \bar{G}_1,\bar{G}_2\Big]_+ &= \Big[ \tilde{G}_1,\tilde{G}_2\Big]_+ & =0
  ,
  \\
  \label{anticommut_r}
  \Big[ \bar{J}_1,\tilde{J}_2\Big]_+ &= \HL{\frac{\Gamma_1}{2\pi}} \delta_{1,\bar{2}}
  ,
  &
  \Big[ \bar{J}_1,\bar{J}_2\Big]_+ &= \Big[ \tilde{J}_1,\tilde{J}_2\Big]_+ &=0
  ,
\end{alignat}
where $[\ ,\ ]_+$ now denotes the anticommutator of superoperators.
The {superoperators} of the dot and the reservoirs commute with each other,
\begin{align}
  [\tilde{J}_1,\tilde{G}_2]_-
  = [\bar{J}_1,\bar{G}_2]_-
  = [\bar{J}_1,\tilde{G}_2]_-
  = [\tilde{J}_1,\bar{G}_2]_- = 0
  .
  \label{anticommut_dr}
\end{align}
This follows from our assumption that the dot and the reservoir \emph{operators} commute [see \Sec{sec:parity}].
Furthermore, the field superoperators are pairwise related by the super-Hermitian conjugation (which will be defined shortly hereafter):
\footnote{
  Although one should in general carefully distinguish between Hermitian and super-Hermitian conjugation,
  we use the same notation $\dagger$ for both
  since it is always clear from the context  whether we deal with a super- or  usual operator.
}
\begin{alignat}{2}
  \bar{G}_1^\dagger  & = \tilde{G}_{\bar{1}},
  & \quad
  \bar{J}_1^\dagger = \tilde{J}_{\bar{1}}
  \label{dagger}
  .
\end{alignat}
In this relation, note the reversal of the causal index ($\bar{q}$) as well as the  multi-index ($\bar{1}$):
  $(G_1^q)^\dagger = G_{\bar{1}}^{\bar{q}}$ and  $(J_1^q)^\dagger = J_{\bar{1}}^{\bar{q}}$.

The crucial properties of the \emph{causal} fermionic field superoperators
\eq{superG}-\eq{superJ}, distinguishing them from previously introduced field superoperators (see \Cite{Saptsov12a} for a detailed comparison) are
\begin{alignat}{2}
  \Tr{D} \bar{G}_1\bullet & =0,
  &\quad
  \Tr{R}\tilde{J}_1\bullet & = 0,
  \label{trace}
  \\
  \bar{G}_1 (-1)^n & =0,
  &\quad
  \tilde{J}_1 (-1)^{n^R} & = 0,
  \label{parity}
\end{alignat}
for \emph{all} values of the multi-index $1$.
In the following, we will see that \Eq{trace} relates to the fundamental causal structure of the correlation functions expressed in these superfields and plays a key role in maximally simplifying them [cf. \Eq{ret-contr}-\eq{keld-contr}].
Equation \eq{parity} arises since the fermion-parity operator is used to ensure that the field superoperators anticommute.
This property leads to an interesting exact result for the interacting Anderson model discussed in \Sec{sec:generalWBL}.

It is of interest to also give the superoperators $G_1^q$ in terms of the anticommuting dot fields $d'_1$ [\Eq{dprime}]:
\begin{subequations}
  \begin{align}
    G_1^q \bullet
    & =
    \eta \frac{1}{\sqrt{2}}
    [
    d'_1 , (-1)^{n} \bullet
          ]_{-q}
    \label{G-crep}
    \\
    & =
          \eta (-1)^{n+1} \frac{1}{\sqrt{2}} \left( d'_1 \bullet + q\bullet d'_1 \right)
    .
 \end{align}
\end{subequations}
The causal field superoperators in Liouville-Fock space are thus simply the commutator and anticommutator of these Hilbert-Fock space fields operators (cf. \Cite{Mukamel08}), but only after the fermion-parity has been applied to its argument.
In \Eq{G-crep}, taking the commutator and anticommutator of an operator is the superoperator-equivalent of performing the Keldysh rotation.~\cite{Saptsov12a}
One useful aspect of the form \eq{G-crep} is that for $q=+$ the two fundamental properties \eq{trace}-\eq{parity} are immediately clear:
the property \eq{trace} follows from the vanishing of the trace of a commutator
and \eq{parity} from the vanishing of any commutator with the unit operator.
\footnote{Here, we insert $(-1)^n$ for the argument of \Eq{G-crep} (the bullet $\bullet$) and make use of $(-1)^n(-1)^n=1$.}
These are dual properties with respect to the scalar product in Liouville-Fock space, see \Sec{sec:L-F-s}.
The form \eq{G-crep} is also convenient for a direct comparison with expressions which occur in formalisms using the local dot Green's functions.~\cite{Saptsov13c}

\subsubsection{Causal basis for Liouville-Fock space: Super-Pauli principle
\label{sec:L-F-s}}
As was shown in \Cite{Saptsov12a}, the causal field superoperators $G^\pm$ [\Eq{superG}]
 generate a basis for the Liouville space of the quantum dot,
 in close analogy with the construction of the usual fermion Fock-basis in the many-particle Hilbert space.
The 4-dimensional Hilbert-Fock space of the quantum dot is spanned by the orthogonal state vectors
\begin{align}
  \ket{0},
  \quad
  \ket{\uparrow}=d_\uparrow^\dagger\ket{0},
    \quad
  \ket{\downarrow}=d_\downarrow^\dagger\ket{0}, 
    \quad
  \ket{2}=d_\uparrow^\dagger d_\downarrow^\dagger\ket{0}.
\end{align}
Operators $A=\sum_{k,l=0,\uparrow,\downarrow,2} A_{k,l}|k\rangle\langle l|$
acting on this space themselves form a 16 dimensional linear space $\mathfrak{L}$
with the inner product $(A|B)=\mathrm{Tr}_D A^\dagger B$, which we refer to as the \emph{Liouville space} of the quantum dot.
By $|A)$ we denote an operator $A$ considered as a supervector in $\mathfrak{L}$, and use
the rounded bracket notation {to avoid} possible confusion with Hilbert space state vectors $\ket{\psi}$.
A set of mutually orthogonal supervectors, i.e., operators $A_i, i=1,...,16$ satisfying
\begin{align}
  \label{scalar}
  (A_i|A_j)=\Tr D A_i^\dagger A_j=\delta_{i,j}
\end{align}
form an orthonormal basis in the quantum-dot Liouville space.
Superoperators are linear maps $S: \mathfrak{L}\rightarrow\mathfrak{L}$
and can be expressed in this basis as:
\begin{align}
  S=\sum\limits_{i,j} S_{i,j}|A_i)(A_j|, ~~i,j=0,...,16
  .
\end{align}
where $(A| \bullet =\mathrm{Tr}_D A^\dag \bullet$ denotes the linear function on operators $\bullet$ built from the operator $A$.
The super-Hermitian conjugation \Eq{dagger} is defined with respect to the Liouville-space inner
product, i.e., $(A|S^\dag |B)= (B|(SA))^{*}$.

For a Liouville space of a many-particle system, a \emph{Liouville-Fock space}, a super Fock-basis can be constructed starting from some operator defining a vacuum superstate.
(In \Sec{sec:examples}, we indicate that some care must taken when using the term ``superstate''.)
For the causal field superoperators \eq{superG} the vacuum superstate of the quantum dot is given by
\begin{subequations}
  \label{basis-boson}
  \begin{alignat}{3}
    |Z_L)&=\tfrac{1}{2} \unit,
    \label{ZL}
  \end{alignat}
which is indeed destroyed by the annihilation superoperators $G^{-}_1=\tilde{G}_1$:
  $\tilde{G}_{\eta\sigma}|Z_L)=0$
  for all $\eta$, $\sigma$ by the superhermitian conjugate of the identity \eq{trace}.
  From this vacuum another seven
  \emph{bosonic} operators are created by application of 
  all possible products of an \emph{even} number of fermionic creation superoperators $G^{+}_1=\bar{G}_{\eta\sigma}$.
  The doubly occupied superstates are
  \begin{alignat}{3}
    |\chi_\sigma) & = \bar{G}_{+\sigma}\bar{G}_{-\sigma}|Z_{L})
                 &&=-\tfrac 1 2 e^{i\pi n_\sigma} 
    ,\label{chichi}\\
    |T_+) & = \bar{G}_{+\uparrow}\bar{G}_{+\downarrow}|Z_{L})
          && =d^{\dagger}_{\uparrow} d^{\dagger}_{\downarrow},
          &&
    \label{T+}\\
    |T_-) & = -\bar{G}_{-\uparrow}\bar{G}_{-\downarrow}|Z_{L})
         && =d_{\downarrow}  d_{\uparrow},
         &&
    \label{T-}
    \\ 
    |S_\sigma)   & = \bar{G}_{+\sigma}\bar{G}_{-{\bar{\sigma}}}|Z_{L})
               && =d_\sigma^\dagger d_{\bar{\sigma}}, 
    \label{S}
\end{alignat}
where $\sigma=\pm=\uparrow,\downarrow$, cf. \Eq{eq:n}.
The operators $\chi_\sigma$ are proportional to the fermion-parity operator for a single spin-$\sigma$ orbital of the dot,
$e^{i\pi n_\sigma}  = 1-2n_\sigma$.
The ``most filled'' superstate $|Z_R)$ (quadruply occupied) equals the total fermion-parity operator of the dot $e^{i\pi n} =\prod_\sigma e^{i\pi n_\sigma} =(1-2n_\uparrow)(1-2n_\downarrow)$,
normalized to the Liouville-space scalar product:
\begin{alignat}{3}
 \label{ZR}
  |Z_{R})   & = \bar{G}_{{+}\uparrow}\bar{G}_{{-}\uparrow}\bar{G}_{{+}\downarrow}\bar{G}_{{-}\downarrow}|Z_{L})
           && = \tfrac 1 2 e^{i\pi n}
  .
\end{alignat}
\end{subequations}
In addition, another eight \emph{fermionic} superoperators, are created by the action of products of \emph{odd} numbers of fermionic creation superoperators $\bar{G}_{\eta\sigma}$,
either with one excitation
\begin{subequations}
\label{basis-fermion}
\begin{align}
  |\alpha^+_{\eta\sigma})=\sigma^{\frac{1-\eta}{2}} \bar{G}_{\eta,(\sigma\eta)}|Z_L)
  = \tfrac{1}{\sqrt{2}}\sigma^{\frac{1-\eta}{2}} d_{\eta,(\sigma\eta)},
  \label{alpha+}
\end{align}
or with three ($\bar{\sigma} \eta = \sigma \bar{\eta} = -\sigma \eta$)
\begin{align}
  |\alpha^-_{\eta\sigma})
  & =\sigma^{\frac{1-\eta}{2}}
  \bar{G}_{\eta,(\sigma\eta)}
  \bar{G}_{\eta,(\bar{\sigma}\eta)}
  \bar{G}_{\bar{\eta},(\sigma\bar{\eta})} |Z_L)
  \nonumber \\
  &= e^{i\pi n} \tfrac{1}{\sqrt{2}}\sigma^{\frac{1-\eta}{2}} d_{\eta,(\sigma\eta)}
  .
  \label{alpha-}
\end{align}
\end{subequations}
Using the anticommutation relations \eq{anticommut_d}, one shows that the above 16 supervectors (operators) \eq{basis-boson}-\eq{basis-fermion} form a complete, orthonormal basis in the sense of \Eq{scalar} for the Liouville-Fock space.
For the central applications of this paper, we do not need the fermionic part of this basis except for the next \Sec{sec:examples}.
However, since second-quantized expressions for superoperators act on the entire Liouville space, one must be aware of this linear subspace, as we will illustrate in several cases
in the following.

In the above construction of Liouville-Fock space, the fermion-parity operator $(-1)^n=2Z_R$ plays a fundamental role.
% 1
First, it was included into the definition of the field superoperators to ensure fermionic anticommutation relation. This, in particular, results in
\begin{align}
  (\bar{G}_1)^2 = 0
  ,
  \label{superpauli}
\end{align}
which expresses that one cannot doubly occupy a superstate (labeled by $1=\eta,\sigma$).
We will refer to \Eq{superpauli} as the \emph{super-Pauli exclusion principle} by analogy with the Pauli principle for fermionic fields in Hilbert-Fock space, for which $(d_\sigma^\dag)^2=0$.
A consequence of central importance to the paper is that any product of more than four creation (or destruction) superoperators necessarily contains at least one duplicate and therefore vanishes:
\begin{align}
  \bar{G}_m \ldots \bar{G}_1 = 0 \quad \text{for $m>4$}
  .
    \label{superpauli2}
\end{align}
This is to be compared with the vanishing of products of more than two field creation operators
in Hilbert-Fock space, i.e., $d_{\sigma_m}^\dag..d_{\sigma_1}^\dag=0$ for $m>2$.
Equation \eq{superpauli2} can be generalized to an Anderson model with $N$ spin-orbitals by replacing the condition with $m>2N$.
% 2
Second, the singly and triply occupied superstates are constructed from the same set of four Hilbert-Fock field operators $d_{\eta\sigma}$, but they differ by the application of the parity operator, 
$|\alpha^-_{\eta\sigma}) = (-1)^n |\alpha^+_{\eta\sigma})$.
It is this factor that ensures their orthogonality.\footnote{
  The fermion-parity guarantees orthogonality,
  $
  ( \alpha^{+}_1 | \alpha^{-}_2 )
  \propto  {\protect \tr_{D}}  d_{1}^{\dag} (-1)^n d_2 
  =       {\protect \tr_{D}} (-1)^n d_{2} d_{1}^{ \dag}   = 0
  $,
  since
  $
  {\protect \tr_{D}} ((-1)^n \bullet)  =
  \protect \sum_{n_\uparrow n_\downarrow}
  (-1)^{ n_\uparrow + n_\downarrow}
  \protect \bra{ n_\uparrow n_\downarrow}
  \bullet
  \protect \ket{  n_\uparrow n_\downarrow}
  =
  0
  $
  for any one-particle operator since the matrix element
  is independent of ${\protect n_\uparrow}$  or ${\protect n_\downarrow}$ (or both).
}
% 3
Finally, the left multiplication by the fermion-parity operator implements a superoperator analog of a particle-hole transformation, mapping basis supervectors with $\mathcal{N}$ superparticles onto those with $4-\mathcal{N}$ superparticles.\footnote{
  This connects to the other possible construction of the Liouville-Fock space,
  starting from the fermion-parity operator $|Z_R)=\tfrac{1}{2}(-1)^n$ as the vacuum state,
  which is annihilated by the creation operator ${ G^{+}_1 = \bar{G}_1 }$ by the fundamental relation \eq{parity}:
  $\bar{G}_{\eta\sigma}|Z_R)=0$.
  See \Cite{Saptsov12a}, Appendix E, for a systematic discussion.
}

In analogy to the usual second quantization, one can introduce a super occupation operator [cf. \Eq{dagger}]:\footnote{
  Note that the total superoperator $\sum_{\eta,\sigma} \mathcal{N}_{\eta\sigma}$
  counting the occupation of the basis superkets \Eqs{basis-boson}-\eq{basis-fermion}
  does \emph{not} equal the superoperator defined by the commutator with the particle number operator, $[n,\bullet]$.
}
\begin{align}
  \mathcal{N}_{\eta \sigma}  =  \bar{G}_{\eta \sigma} \tilde{G}_{\bar{\eta} \sigma} 
  =
\bar{G}_{\eta \sigma} \bar{G}_{\eta \sigma}^\dag
.
\label{Netasigma}
\end{align}
By construction, due to the anticommutation relations it satisfies
\begin{align}
  [ \mathcal{N}_{\eta \sigma},  \bar{G}_{\eta \sigma} ] =  \bar{G}_{\eta \sigma}
  .
\end{align}
It therefore simply counts the number of times that the creation superoperators $\bar{G}_{\eta \sigma}$ appears in a basis superket, which is restricted to 0 or 1 by \Eq{superpauli}: with $\mathcal{N}_1 := \mathcal{N}_{\eta \sigma}$
\begin{align}
  \mathcal{N}_i \bar{G}_{m} \ldots \bar{G}_{1} |Z_L)
  = (\delta_{i,m} + \ldots + \delta_{i,1}) \bar{G}_{m} \ldots \bar{G}_{1} |Z_L)
  .
\end{align}

Finally, we note that operators \Eqs{chichi}-\eq{S} are closely related
to the group generators of the spin- ($S$) and charge-rotation ($T$) symmetry of the Anderson model
(they transform as irreducibly under the symmetry group).
By working in the causal Liouville-Fock space basis \eq{basis-boson}-\eq{basis-fermion} one thus not only profits from the fundamental causal properties of interest here,
but one also maximally exploits these model-specific symmetries, see the study \Cite{Saptsov12a}, where this was of crucial importance.

\subsubsection{Examples of second quantization in Liouville space\label{sec:examples}}
Before we move on,
we first illustrate the above introduced second quantization in Liouville-Fock space using causal superfermions.
We discuss the expansion of a supervector, using the density operator as an example, and {the expansion} of a superoperator, the Liouvillian $L$.

\paragraph{Mixed state supervector $\rho$.}
We can construct the most general form of the reduced density operator for the quantum dot 
by accounting for the restrictions on a physical mixed state:
the operator $\rho$ must (i) be positive, (ii) be self-adjoint, (iii) have unit trace, and (iv) satisfy the fermion-parity superselection rule (univalence).~\cite{Wick52,Aharonov67,Saptsov12a}
The latter requires that any
density operator $\rho$ has no off-diagonal matrix elements with respect to the fermion-parity quantum number [cf. \Eq{supersel}]:
\begin{align}
  \label{superselection-simple}
  [\rho,(-1)^n]_{-} = 0.
\end{align}
The linear space containing such operators satisfying (ii)-(iv) is spanned by the bosonic operators
in \Eqs{basis-boson}.
The reduced density operator is a supervector in this space, and
 is thus generated by application of products of an \emph{even} number of creation superfields from the vacuum superket
with, in general, seven coefficients $\Omega_\pm(t), \Phi_\pm(t), \Upsilon_\pm(t),$ and $\Xi(t)$:
\begin{align}
  \rho(t)
  = & 
  \Big\{
    \tfrac{1}{2}  + \sum\limits_\sigma\Phi_\sigma(t) \bar{G}_{+\sigma}\bar{G}_{-\sigma} 
    +\sum\limits_\sigma\Omega_\sigma(t)\bar{G}_{+\sigma}\bar{G}_{-\bar{\sigma}}
    +
  \nonumber
  \\
  &
    \sum\limits_\eta \eta\Upsilon_\eta(t) \bar{G}_{\eta \uparrow}\bar{G}_{\eta \downarrow}
    +\Xi(t) \bar{G}_{+\uparrow}\bar{G}_{-\uparrow}\bar{G}_{+\downarrow}\bar{G}_{-\downarrow}
  \Big\}|Z_L)
 \nonumber
  \\
  = &
  \tfrac 1 2 |Z_L) + \sum\limits_\sigma\Phi_\sigma(t) |\chi_\sigma)+\Xi(t) |Z_R)
  +
  \nonumber
  \\
  &\sum\limits_\eta \Upsilon_\eta(t)
  |T_\eta)+\sum\limits_\sigma\Omega_\sigma(t)|S_\sigma)
  .
  \label{general-initial-dot}
\end{align}
With appropriate restrictions imposed by the positivity condition (i),
these coefficients thus parametrize an arbitrary dot state,
e.g., the complicated time-dependent density operator $\rho(t)$ of the $U\neq0$ Anderson model [\Eq{reduced-expansion-usual}].
The coefficients are the non-equilibrium averages $ \langle \bullet \rangle (t) = \mathrm{Tr}_{D} [ \bullet \rho(t)]$ of the complete set of local observable operators \eq{basis-boson}.
The coefficient
\begin{align}
  \Phi_\sigma (t)  = (\chi_\sigma|\rho(t) )  =  -\tfrac{1}{2} \braket{ e^{i\pi n_\sigma} }(t)
  \label{phi-zr}
\end{align}
gives the average occupation: $\langle n_\sigma \rangle(t)=1/2 + \Phi_\sigma(t)$
by using $e^{i\pi  n_\sigma }= (1-2n_\sigma)$.
The   coefficient 
\begin{align}
  \Xi (t)         = (Z_R|\rho(t))          = \tfrac{1}{2} \braket{ e^{i\pi n} }(t),
  \label{zr-phi}
\end{align}
the average of the fermion-parity operator,
$\Xi(t)
= \tfrac 1 2 \braket{ \prod_\sigma e^{i\pi n_\sigma} }(t)
= 2 \braket{ n_\uparrow n_\downarrow}(t) -\braket{n}(t) +1/2
$,
takes into account the correlations of the occupancies:
$\langle n_\uparrow n_\downarrow\rangle(t)
\neq
\langle n_\uparrow\rangle(t) \langle n_\downarrow\rangle(t)$
is equivalent to
$
\Xi(t)\neq 2\prod_\sigma \Phi_\sigma (t)
$.
Furthermore, the average of an ``anomalous'' and a spin-flip combination of Hilbert-Fock space field operators
\begin{alignat}{3}
 \Upsilon_{\bar{\eta}}(t) &=
 (T_{\bar{\eta}}|\rho(t)) && =
 \eta \langle d_{\uparrow}^\eta d_\downarrow^{{\eta}}\rangle (t)
 ,
  \label{Upsilon}
  \\
  \Omega_{\bar{\sigma}}(t) &=
  (S_{\bar{\sigma}}|\rho(t)) &&=
  \langle d_{\sigma}^\dagger d_{\bar{\sigma}}\rangle (t)
  ,
  \label{Omega}
\end{alignat}
describe the transverse spin ($\uparrow$-$\downarrow$) coherence and the electron-pair ($0$-$2$) coherence of the state at time $t$.
At the initial time $t_0$ such coherences can be prepared:
the transverse spin coherence by contact with a ferromagnet with a polarization transverse to the magnetic field $B$
and the electron-pair coherence by contact with a superconductor.
At finite times $t$ such coherences will persist, but in the stationary limit $t \rightarrow \infty$ they must vanish since the Anderson model has spin-rotation symmetry (with respect to the magnetic field axis) and charge-rotation symmetry.~\cite{Saptsov12a}
Likewise, two-particle correlations can be initially present on the quantum dot if it has been in contact with an interacting system.
These will decay, in the sense that
$\lim_{t\rightarrow \infty} \Xi(t) = 2\prod_\sigma \lim_{t\rightarrow \infty} \Phi_\sigma (t)$,
if the quantum dot is noninteracting ($U=0$).

Of the eight bosonic operators \eq{basis-boson}, only $Z_L$ has a nonzero trace, and the physical requirement $\mathrm{Tr} \rho=1$ completely fixes its coefficient in \Eq{general-initial-dot}.
By itself, the operator
\begin{align}
  \rho_\infty := \frac{1}{2} |Z_L)=\frac{1}{4} {\unit}
  \label{rhomax}
\end{align}
represents the physical stationary state of the quantum dot coupled to reservoirs at infinite temperature (i.e., $T$ much larger than any other energy scale, i.e., $U$, $\epsilon-\mu_r$, $B$).
In any finite-temperature mixed state \eq{general-initial-dot}, there are in general two- and four-superfermion excitations. In our formalism, such super excitations correspond to a ``cooling'' relative to the infinite-temperature supervacuum \eq{rhomax}.
Although this point of view is opposite to that in the Hilbert-Fock space (where excitations rather describe a ``heating'' of the zero-temperature vacuum $\ket{0}$), the causal superfermion approach is thus entirely physical and brings definite 
insights and advantages in the study of open quantum systems.
However, care must be taken to import physical concepts from second quantization in Hilbert-Fock space.
For instance, it should be noted that of the basis supervectors only $|Z_L)$ can represent a physical state \emph{on its own}:
the other 15 basis supervectors, such as $\bar{G}_1|Z_L)$, are traceless by construction [\Eq{trace}] and cannot fulfill the probability normalization condition $\text{Tr}_D \rho = 1$ by themselves.
Moreover, the fermionic basis vectors \eq{basis-fermion}, such as $\bar{G}_1|Z_L)$,  do not have the right fermion-parity.
It is only in superpositions with $|Z_L)$ of the form \eq{general-initial-dot} that the bosonic basis supervectors \eq{basis-boson} take part in real mixed states described by a density operator, whereas the fermionic basis vectors \eq{basis-fermion} only play a role in \emph{virtual} intermediate mixed states, 
see discussion of \Eq{virtual} in the following.
This should be kept in mind when speaking formally about ``superstates'', ``superparticles'' or ``superexcitations'',
a terminology which we do consider to be useful.

\paragraph{Liouvillian superoperator $L$.\label{sec:Lil}}
As a next illustration, we discuss the second quantized form of the Liouvillian superoperator of the isolated Anderson model in terms of the field superoperators:
\begin{subequations}
 \label{L2ndquant}
\begin{align}
  &L=\sum\limits_{\eta,\sigma}
  \eta\Big( \left[\epsilon+ U/ 2\right]+\sigma B/2\Big)
  \bar{G}_{\eta\sigma} \tilde{G}_{\bar{\eta}\sigma}+
   \label{quadratic}
  \\
  &+\frac U 2 \sum\limits_{\eta,\sigma}
  \left(
    \bar{G}_{\eta\sigma}
    \tilde{G}_{\bar{\eta}\sigma}
    \tilde{G}_{\eta\bar{\sigma}}
    \tilde{G}_{\bar{\eta}\bar{\sigma}}
    +
    \bar{G}_{\eta\sigma}
    \bar{G}_{\bar{\eta}\sigma}
    \bar{G}_{\eta\bar{\sigma}}
    \tilde{G}_{\bar{\eta}\bar{\sigma}}
  \right),
   \label{quartic}
 \end{align}
\end{subequations}
which is verified by substitution of \Eq{superG} to give $L=[H,\bullet]_{-}$ with $H$ given by \Eq{dot_ham}.
Similar to the usual second quantization technique, this expression directly reveals a number of general properties.
For instance, particle number conservation is expressed by the fact that in the field superoperators only conjugate pairs of $\eta$, $\bar{\eta}$ appear.
Furthermore, since only products of an \emph{even} number of field superoperators appear, the superoperator $L$ preserves the fermion-parity superselection rule of the density operator \Eq{superselection-simple}:
The off-diagonal supermatrix elements between a bosonic [$|B)$, \Eq{basis-boson}] and a fermionic [$|F)$, \Eq{basis-fermion}] basis operator vanish, $(B|L|F)=(F|L|B)=0$,
simply because these are created from $|Z_L)$ by the action of an even and an odd number of field superoperators, respectively.
As a result, if initially $\rho(t_0)$ satisfies \Eq{superselection-simple} then so does $\rho(t) = e^{-iL(t-t_0)} \rho(t_0)$ at later times $t>t_0$ for a closed system.
A property more specific to the use of \emph{causal} superfermions in the second quantized form \Eq{L2ndquant}, is that the conservation of probability is immediately obvious \emph{term-by-term}: each term ends with a destruction superoperator $\bar{G}$ on the left, ensuring by \Eq{trace} that the trace is preserved $\text{Tr}_{D} e^{-iL(t-t_0)} \rho(t_0) = \text{Tr}_{D}  \rho(t_0)$.

More specific to the Anderson model is that \Eq{L2ndquant} automatically achieves an interesting decomposition of the interaction term.
In particular, the \emph{quartic} term \eq{quartic} $\propto U$  commutes with the simple quadratic term \eq{quadratic} which also contains $U$. Importantly, the quartic term acts \emph{only} in the fermionic sector of the Liouville-Fock space [spanned by the superkets \eq{basis-fermion}]:
for any two bosonic operators $B$ and $B'$ we have $(B|L|_\text{quartic}|B') = 0$.
This follows from
\begin{align}
  \label{BBmatel}
  (B|\sum_{\eta,\sigma}\bar{G}_{1}\tilde{G}_{\bar{1}}\tilde{G}_{2}\tilde{G}_{\bar{2}}|B') = 0,
\end{align}
where $1=\eta,\sigma$ and $2=\eta,\bar{\sigma}$ and the same for the second term in \Eq{quartic} (which is the Hermitian superadjoint of this).
Equation \eq{BBmatel} immediately follows from the structure of the quartic term
using reasoning very similar to that used in the usual second quantization in Hilbert-Fock space.
The result \eq{BBmatel}, together with the fermion-parity superselection rule \eq{superselection-simple}, now implies that in the time-evolution expansion \Eq{reduced-expansion-usual} 
the quartic interaction term \Eq{quartic} \emph{plays no role} in the free quantum-dot propagator $e^{-iL(t_{k+1}-t_{k})}$
when it occurs after an \emph{even} number ($k$) of tunneling Liouvillians $L^V$.
For example, inserting in the fourth-order expression of \Eq{reduced-expansion-usual}
a complete set of superstates (expansion of the unit superoperator) for the quantum dot between $L^V(t_3)$ and $L^V(t_2)$
and substituting \Eq{q-li} we obtain the structure\footnote{In \Eq{virtual} all free dot propagators $e^{-iL(t_{k+1}-t_k)}$ between the vertices $G_{k+1}^{q_{k+1}}$
and $G_k^{q_k}$ for $k\neq 2$ are denoted as ellipsis ``...'' for compactness.}
\begin{widetext}
\begin{align}
  \cdots G_4^{q_4}\cdots G_3^{q_3} e^{-iL(t_{3}-t_{2})} G_2^{q_2} \cdots G_1^{q_1}\cdots \rho(t_0) 
  =   \sum_{B,B'} \cdots G_4^{q_4}\cdots G_3^{q_3} |B)(B| e^{-iL(t_{3}-t_{2})}|B')(B'| G_2^{q_2}\cdots G_1^{q_1}\cdots \rho(t_0)
  ,
  \label{virtual}
\end{align}
\end{widetext}
where  only the quantum-dot part of the expressions are shown.
The sums over  $|B)$, $|B')$ are restricted to the bosonic superkets \Eq{basis-boson}:
since $\rho(t_0)$ is a bosonic operator,
application of an even number of superfields brings us back to the bosonic sector.
Now the quartic $U$-term drops out in the matrix elements
$(B| e^{-iL(t_{3}-t_{2})}|B')$ due to \Eq{BBmatel}.
The propagation of the bosonic \emph{virtual intermediate} states in \Eq{reduced-expansion-usual} is thus defined entirely by the \emph{quadratic} part
of the dot Liouvillian,  \Eq{quadratic}, and is thus \emph{effectively noninteracting}, with a renormalized single-particle energy level: $\epsilon\rightarrow \epsilon+U/2$.
This general rule leads to very useful simplifications in perturbative~\cite{Saptsov13b} and nonperturbative~\cite{Gergs13thesis} studies of the interacting Anderson model that will not be explored further here.

Another point revealed by the second quantization of $L$, \Eq{L2ndquant},
is that the
 the essential two-particle operator to which the interaction couples in the \emph{Hamiltonian} $H$, \Eq{dot_ham}, is the fermion-parity operator $(-1)^n= 2Z_R$:
\begin{align}
   L|_\text{quartic} = \big[ \tfrac{1}{4}U(-1)^n, \, \bullet \, \big]_{-}
   =
  \frac U 2 \sum_{\nu,\eta,\sigma}|\alpha_{\eta\sigma}^{\nu})(\alpha_{\eta\sigma}^{\bar{\nu}}|
  .
  \label{quartic-term}
\end{align}
The term \eq{quartic-term} captures the essential many-particle effect of the interaction $U$ since in the quadratic term \Eq{quadratic} the \emph{effect} of $U$ can be compensated by tuning the level position to the particle-hole symmetry point $\epsilon=-U/2$.
The first rewriting in \Eq{quartic-term} shows that the quartic term corresponds to the operator $\frac{U}{4}(-1)^n$ contained in $H$.
This also shows that it acts only in the Liouville-Fock space spanned by fermionic operators \Eq{basis-fermion}, again leading to \Eq{BBmatel}, since the fermion-parity operator by construction (anti)commutes with all bosonic (fermionic) operators by \Eq{basis-boson} [\Eq{basis-fermion}]. The second rewriting in terms the fermionic superbras and superkets [\Eq{basis-fermion}] explicitly confirms this.

Finally, we emphasize that a particularly, useful aspect of the above
reasoning, based \emph{directly} on the Liouville-Fock representation \Eq{L2ndquant},
is that it also allows one to  infer general physical properties of a superoperator describing an open fermionic system,
even when it does not have the commutator form which $L$ has.

\subsubsection{Interaction picture of causal superfermions\label{sec:intpic}}
Using the second quantization, we can also easily work out the explicit form of interaction Liouvillian $L^V$ [\Eq{q-li}] in the interaction picture,
$L^V(t)= \sum_q e^{i(L+L^R)(t-t_0)}G^q_{1} J^q_{\bar{1}} e^{-i(L+L^R)(t-t_0)}$,
which is required in the next section.
For a noninteracting quantum dot ($U=0$) we have the following simplifying property:
for $1=\eta,\sigma$
\begin{align}
  \label{L-commut-G}
  \Big[ L, G_1^q\Big]_- & = \eta\epsilon_\sigma  G_1^q,
  \quad \text{for $U=0$}
  \\
  \epsilon_\sigma & =\epsilon+B\sigma/2
  ,
  \label{es}
\end{align}
which follows from the quadratic part \eq{quadratic} of $L$ using the superfermion commutation relations \eq{anticommut_d}.
Note that the right-hand side is independent of $q$, i.e., the creation and annihilation superoperators have the same frequency
$\eta\epsilon_\sigma$, in contrast to Hilbert-Fock space fields.
The interaction-picture field superoperators
\begin{subequations}
\begin{align}
  G_1^{q}(t) & :=e^{iL(t-t_0)}G_1^{q}e^{-iL(t-t_0)}       
  \label{dortex}
  \\
             & = e^{i\eta\epsilon_\sigma(t-t_0)} {G}^q_1 \quad \quad \text{for $U=0$}
\label{simple-td}
\end{align}
\end{subequations}
in the noninteracting case $(U=0)$
are then simply proportional to those in the Schr\"odinger picture, ${G}^q_1$,
since we can commute $e^{-iL(t-t_0)}$ through ${G}^q_1$ using \Eq{L-commut-G},
resulting only in a phase factor.
This is the crucial simplification, which allows the exact solution of the noninteracting problem to be obtained quite simply
once we have taken the wideband limit, as discussed in the next section.
The field superoperators of the noninteracting reservoirs, have the same simple time dependence as those of the dot for $U=0$.
Since by our definitions in \Eq{superG} and \eq{superJ} \cut{of} $J^{\bar{q}}_1$ and $G^q_1$ with opposite $q$ index play the same role, one can 
write analogous to \HL{\Eq{L2ndquant}, accounting for a factor due to the rescaling \eq{rescaling}},
\begin{align}
  L^R & = [H^R,\bullet ]_{-}
  =
  %\int d\omega \sum_{\eta,\sigma}
\HL{{\frac{2\pi}{\Gamma_1}}}
  \eta(\omega+ \mu_r) %\tilde{J}_{\eta\sigma \HL{r}} \HL{(\omega)}\bar{J}_{\bar{\eta}\sigma\HL{r}}\HL{(\omega)}
\tilde{J}_{\HL{1}} \bar{J}_{\HL{\bar{1}}}
  ,
  \label{LR}
\end{align}
\HL{(with the usual implicit integration over the index $\omega$ and summation over indices $\eta,\sigma,r$ of the multi-index $1=\eta,\sigma,\omega,r$)}
 from which it follows that
\begin{align}
  [L^R,J_1^q]_{-} &=\eta (\omega+\mu_r) J_1^q.
  \label{LR-commute}
\end{align}
% for $1=\eta,\sigma,\omega,r$. 
As a result, for $\bar{1}=-\eta,\sigma,\omega,r$
\begin{subequations}
  \begin{align}
    J_{\bar{1}}^{q}(t)
    & :=e^{iL^R(t-t_0)}J_{\bar{1}}^{q}e^{-iL^R(t-t_0)}       
    \\
    & = e^{-i\eta\left( \omega + \mu_r \right)(t-t_0)} {J}^q_{\bar{1}}
    .
    \label{Jt}
  \end{align}
\end{subequations}
Therefore, implicitly summing (integrating) over discrete (continuous) indices,
\begin{subequations}
  \begin{align}
    L^V(t) &=
    e^{-i\eta\left( \omega + \mu_r \right)(t-t_0)} J_{\bar{1}}^q G_1^{q}(t)
    \label{interaction-time-g}
    \\
    &= e^{-i\eta\left( \omega + \mu_r-\eta\epsilon_\sigma \right)(t-t_0)} J_{\bar{1}}^q  G_1^{q}, \quad \text{for $U=0$}
    .
  \end{align}
\end{subequations}

\section{Time-evolution and causal
superfermions\label{sec:timeevolution}}
We now first set up the time-dependent perturbation theory for the general, \emph{interacting}
case ($U \neq 0$), explicitly incorporating the wideband limit
from the start on the level of \emph{superoperators}.  This leads to a
renormalized version of the perturbation series~\cite{Schoeller09a}
for which the crucial result \Eq{simple-td} can be directly exploited
to solve the noninteracting problem ($U=0$) exactly by a
next-to-leading-order perturbative
calculation.

\begin{figure}[t]
  \includegraphics[width=1.0\columnwidth]{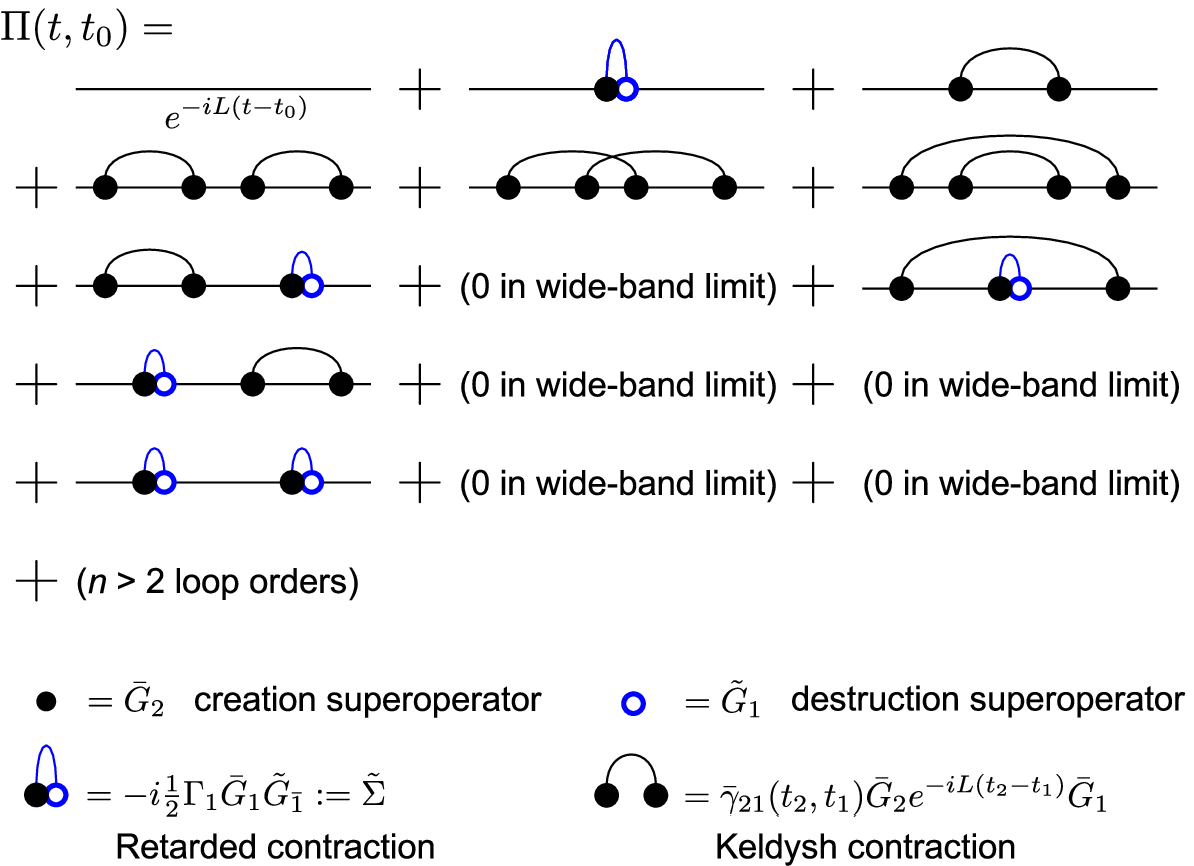}
  \caption{
    Real-time perturbation expansion in the wideband limit:
    Diagrams contributing to the full time-evolution {propagator} $\Pi(t,t_0)$ 
    in the first three loop orders of perturbation theory as given by \Eq{mth-expansion},
    $\Pi_{0}=e^{-iL(t-t_0)}$, $\Pi_{2}$ and $\Pi_{4}$.
    Within each column, the two-loop diagrams have the same contraction configuration but differ by the contraction functions involved ($\bar{\gamma}$ or $\tilde{\gamma}$).
    Diagrams and expressions for the corresponding terms for the {self-energy} $\Sigma(t,t_0)=\Sigma_2+\Sigma_4+\ldots$ are obtained by
    (i) retaining only the diagrams in the second and third column
    and
    (ii) discarding from these diagrams the free time-evolution parts before the first and after the last vertex.
    The full evolution $\Pi(t,t_0)$ is then generated by $\Sigma(t_2,t_1)$ through the Dyson equation \eq{dyson}.
    As indicated, the retarded contractions are ``Markovian'', i.e.,
    in the wideband limit they act \emph{instantaneously} and do not allow for internal time-integrations ($\delta$-function constraint), cf. \Eq{time-ret}.
    We indicate the number of two-loop diagrams that give zero due to this constraint, but do not draw them.
    These are the only contractions that survive in the $T\rightarrow \infty$ limit
    and are considered further in \Fig{fig:resum}a.
  }
  \label{fig:diagrams}
\end{figure}

\subsection{Real-time perturbation expansion for the reduced propagator
\label{sec:pt}}
We are now in a position to exploit the causal superoperator second quantization technique 
to the expansion for the propagator $\Pi = \sum_{m=0}^\infty \Pi_m$ defined by \Eq{reduced-expansion}.
For each term in the expansion of order $m$ in $L_V$, denoted by $\Pi_m$, 
one can collect all reservoir superoperators by commuting them to the left through the $G^q$'s:
\begin{align}
  &\Pi_m=(-i)^m e^{-iL(t-t_0)}
  \Tr R \left[ L^V(t_m) ...  L^V(t_1) \rho^{\mathrm{tot}}(t_0)\right]
  =
  \nonumber
  \\
  &(-i)^m
  \braket{
    J^{q_m}_{\bar{m}}(t_m) ... J^{q_{1}}_{\bar{1}}(t_1)
   }_R
   \, 
   e^{-iL(t-t_0)}  G_m^{q_m}(t_m)... G_1^{q_1}(t_1)
  ,
  \nonumber
  \\
  \label{mth-expansion}
\end{align}
where we implicitly perform a time-ordered integration such that
$t\geq t_m \geq ... \geq t_1 \geq t_0$,
as well as a summation over all dummy indices $m,..,1$.
Here, $\langle S \rangle_R$ denotes the reservoir average of a \emph{super}operator $S$:
\begin{align}
 \langle S \rangle_R = \Tr{R} ( S \rho^{R} ).
\end{align}
To eliminate the reservoirs, we need the multi-particle correlation functions of the reservoirs.
Their time-dependence amounts to a simple phase factor by \Eq{Jt},
$  \braket{
    J^{q_m}_{\bar{m}}(t_m) ... J^{q_{1}}_{\bar{1}}(t_1)
   }_R
=$
$\prod_{i=1}^m e^{-i\eta_i(\omega_i+\mu_{r_i})(t_i-t_0)}
\braket{
    J^{q_m}_{\bar{m}} ... J^{q_{1}}_{\bar{1}}
   }_R
$, and the remaining equal time correlation functions follow from the Wick theorem~\cite{Schoeller09a} for the $J_1^q$ superoperators:~\cite{Saptsov12a} for even $m$
\begin{align}
  \langle J^{q_m}_m ... J^{q_{1}}_{1} \rangle_R
  =
  \sum_\text{contr} (-1)^{P} \prod_{ \langle i,j \rangle}
  \langle J_i^{q_i} J_j^{q_j} \rangle_{R}
  .
  \label{wick}
\end{align}
Here, $(-1)^P$ denotes the usual fermionic sign of the permutation $P$ that disentangles the contractions over which we sum,
$\langle i,j \rangle$ denoting the product over contracted pairs.
For odd $m$ the average vanished by the fermion-parity superselection rule.
The Wick theorem \eq{wick} was shown in \Cite{Saptsov12a} to follow from the standard derivation of Gaudin~\cite{Gaudin60} when using the  superoperator expression for the equilibrium fluctuation-dissipation theorem for the reservoirs:
\begin{align}
  \label{FDT}
  \bar{J}_1 |\rho^{R}) = \tanh (\eta_1 \omega_1/2T) \tilde{J}_1 |\rho^{R}).
\end{align}
The field superoperators \eq{superJ} are called ``causal'' since
they make the constraints imposed by causality~\cite{Kamenev09,Jakobs10a} explicit:
there are only two possible types of contraction functions $\langle J_2^{q_2} J_1^{q_1}\rangle_R$ in the expansion \Eq{wick} that are nonzero.
These are~\cite{Saptsov12a} the
retarded function
\begin{align}
  \label{ret-contr}
  \tilde{\gamma}_{2,1} (\eta_1\omega_1)
  :=
  \langle \bar{J}_2 \tilde{J}_1 \rangle_{R}
  &
  = \frac{\Gamma_1}{2\pi} \delta_{2,\bar{1}}
  ,
\end{align}
and the Keldysh function
\begin{align}
  \label{keld-contr}
  \bar{\gamma}_{2,1} (\eta_1\omega_1)
  :=
  \langle \bar{J}_2\bar{J}_1 \rangle_{R}
  &
  =\frac{\Gamma_1}{2\pi}
  \tanh(\eta_1\omega_1/2T)\delta_{2,\bar{1}}
  ,
\end{align}
while all other possible pair contractions are equal to zero.
These properties of the contractions give a corresponding \emph{causal structure} to the perturbation theory which is revealed only when using the causal field superoperators~\eq{superG},
as we will see in the following.
We have thus explicitly integrated out the reservoir degrees of freedom,
and obtained the real-time perturbation theory~\cite{Schoeller09a,Leijnse08a} for the reduced-time evolution superoperator:
\begin{align}
  & \Pi_m= 
  (-i)^m \left(
    \sum_\text{contr} (-1)^{P} \prod_{ \langle i,j \rangle} 
    \gamma_{i,j}^{q_i}(t_i-t_j)
  \right) \times
  \label{pim}
  \\
  &
  e^{-iL(t-t_1)}G_m^{q_m}e^{-iL(t_m-t_{m-1})}G_{m-1}^{q_{m-1}} \ldots G_1^{q_1}e^{-iL(t_1-t_0)}
  .
  \nonumber
\end{align}
An individual term consists of a sequence of free dot evolutions, generated by $L$ [\Eq{L2ndquant}],
interrupted by the pair-wise action of quantum dot field superoperators $G^q$ [\Eq{dortex}],
which is weighted by the time-dependent reservoir correlation function (Fourier transform)
\begin{align}
  \gamma_{i',j'}^{q_i}(t_i-t_j)
  & :=  \int d \omega_i e^{-i\eta_i(\omega_i+\mu_i)(t_i-t_j)} \gamma_{i,j}^{q_i}(\eta_i \omega_i)
  .
  \label{gammaijprimed}
\end{align}
On the right hand side, we also make use of both the $q$-index as well as the ``bar-tilde'' notation, as in \Eq{GJ-def}:
\begin{align}
  \gamma_{i,j}^{q_j}
  := 
  \langle {J}_i^{+} {J}_j^{q_j} \rangle
  =
  &
      \tilde{\gamma}_{i,j}\delta^{q_j, -} + \bar{\gamma}_{i,j}\delta^{q_j,+}
  .
  \label{gammaij}
\end{align}
We note that the initial time $t_0$ cancels out in the reservoir dynamical phase factor since $\gamma_{i,j}^{q_j} \propto \delta_{1\bar{2}} \propto \delta(\omega_i-\omega_j)\delta_{\bar{\eta}_i,\eta_j}$ [by \Eqs{ret-contr}-\eq{keld-contr}].
The primed multi-indices $i'$, $j'$ on the left hand side of \Eq{gammaijprimed} indicate that we leave out the reservoir frequencies $\omega_i$ and $\omega_j$  from the multi-indices $i$, $j$, respectively.
At this stage these frequencies have been integrated out of the theory,
and from hereon we omit the primes, i.e.,
the multi-indices ($1$, $\bar{1}$, etc. in \Eq{pim}) do not contain $\omega_i$ anymore, unless stated otherwise.

In \Fig{fig:diagrams}, we represent individual terms in \Eq{pim}
diagrammatically~\cite{Schoeller09a, Saptsov12a}
and the total evolution is the sum of such terms over all possible Wick pairings of an even number of discrete indices $m,\ldots,1$, integrated over ordered times, i.e., $t \geq t_m \ldots \geq t_1 \geq t_0$.
We will refer to an $m$-th order diagram contributing to $\Pi_m$ with $m/2$ contractions ($\gamma$) as a $m/2$-loop diagram.
Importantly, due to the structure of the reservoir correlation function \Eq{gammaij}, each term in \eq{pim} has a \emph{causal structure}:
the destruction superoperator
$G^{-}_{1}=\tilde{G}_1$ can never appear on the left of the field superoperator (either $G^{\pm}=\bar{G}$ or $\tilde{G}$)
 with which it is contracted.
One implication of this structure is that $\bar{G}$ always stands on the far left, at the latest time $t_m$. 
This ensures by \Eq{trace} that \emph{term-by-term} $\tr \Pi(t,t_0) = \tr $ and therefore probability is conserved, $\tr \rho(t) = \tr \rho(t_0) = 1$, since $\tr \Pi_0 = \tr$ and $\tr \Pi_m = 0$ for  $m \geq 1$.
We now turn to further implications of this causal structure in the wideband limit.

\subsection{Wide-band limit\label{sec:widebandlimit}}
The perturbation theory \Eq{pim} applies generally without further assumptions to the interacting Anderson model ($U \neq 0$).
However, even when considering the noninteracting limit ($U=0$) in combination with the wideband limit (WBL)
for the stationary state ($t \rightarrow \infty$)
it is not directly obvious how to explicitly evaluate \Eq{pim}.
To obtain the exact solution in that case one needs to identify
 which contributions vanish
 in each loop order of the time-evolution superoperator, and then sum up the remaining ones from all orders.~\cite{Schoeller97hab,Schoeller99tut}
We now show how in the wideband limit the time-dependent perturbation series \eq{pim} for the interacting case ($U \neq 0$) 
can be transformed with the help of our causal superoperators.
In this new formulation, the solution of the noninteracting limit ($U=0$) also becomes obvious,
even allowing for the direct calculation of the full time-evolution $\Pi(t,t_0)$.

\subsubsection{Retarded reservoir correlations,
- elimination of annihilation superfields
\label{sec:eliminate}}

\begin{figure}[t]
  \includegraphics[width=1.0\columnwidth]{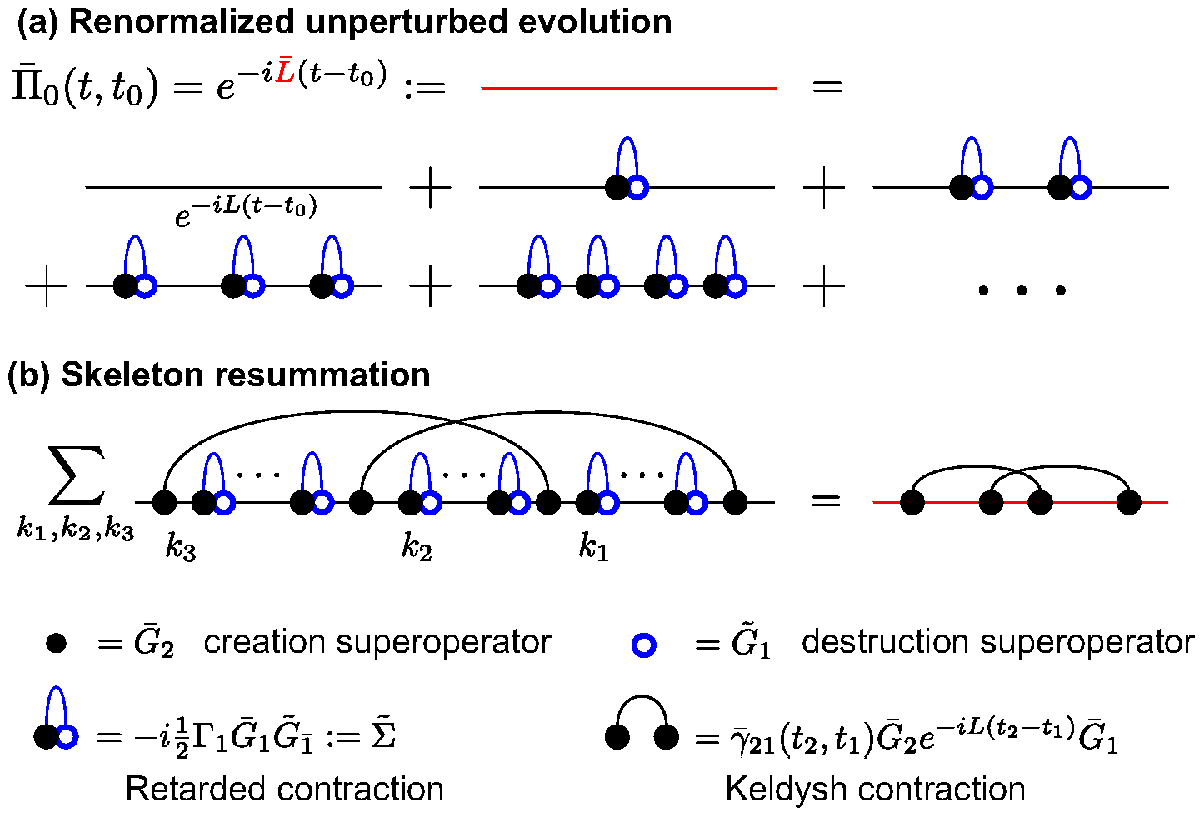}
  \caption{
    Wide-band limit:
    (a)
    Free dot evolutions (the black horizontal lines) interrupted by a sequence of $k$ retarded contractions $\tilde{\gamma}$ (the blue curved lines)
    are resummed to define a renormalized Liouvillian $\bar{L} := L +\tilde{\Sigma}$ [\Eq{barl}].
   The retarded ``Markovian'' contractions give rise to {instantaneous} $\tilde{\Sigma}$ blocks [\Eq{full-tildesig}],
    cf. \Fig{fig:diagrams}.
    The sum defines a \emph{renormalized} unperturbed evolution $\bar{\Pi}_0$ (the red horizontal line)
    which is \emph{dissipative}, see \Eq{InfTrelax},
    and provides a starting point for a new perturbation theory. 
    (b)
    Next, diagrams with a fixed configuration of Keldysh contractions $\bar{\gamma}$ (the black curved lines) are summed over all possible
    insertions of retarded contractions, here illustrated for two Keldysh loops.
    What remains is a renormalized perturbation theory in which only $\bar{L}$ and creation superoperators $\bar{G}$ appear explicitly with Keldysh contractions $\bar{\gamma}$, see \Fig{fig:diagramsren}.
  }
  \label{fig:resum}
\end{figure}

The key simplification in the wideband limit, in which the rates $\Gamma_{r\sigma}$ are constant, is that the retarded contraction function becomes 
energy independent, corresponding to a $\delta$-function in time [cf. \Eq{deltabar}]
(see also \Cites{Langreth91, Jauho94, Jin10}):
\begin{align}
  \tilde{\gamma}_{2,1}(t_2,t_1)
  &= \frac{ \Gamma_1 }{ 2\pi } \delta_{1,\bar{2}} \int d \omega_{1} e^{-i\eta_1 (\omega_{1}+\mu_{r_1}) (t_2-t_1)}
  \nonumber
  \\
  & = \Gamma_1 \delta(t_2-t_1)\delta_{2,\bar{1}}= \frac{\Gamma_1}{2} \bar{\delta}(t_2-t_1)\delta_{2,\bar{1}}
  ,
\label{time-ret}
 \end{align}
where $\Gamma_1=\Gamma_{r\sigma}$ does not depend on the frequency $\omega_1$ or time $t_1$.
By working with causal field superoperators we thus automatically collect a Markovian part of the dynamics:
the ``Markovian contraction'' $\tilde{\gamma}$ appears only when a \emph{destruction} superoperator $\tilde{G}$ is contracted with a $\bar{G}$ (necessarily so by the causal structure).
This allows one to easily eliminate the $\tilde{G}$ from the perturbation series \eq{mth-expansion}, thereby isolating the remaining, nontrivial part of the time-evolution.
To do this, we note that ``processes'' described by $\tilde{\gamma}$ occur instantly in time.
\footnote{
  Also here one must be careful with physically interpreting the expressions: 
the action of a field superoperator $G^{q}_1$ results in a Liouville-space superposition of terms with a definite Keldysh contour index. Only the latter can be identified with a ``process'' generated by the total Hamiltonian $H^{\mathrm{tot}}$ [\Eq{tot_Ham}].
}
Therefore, all $\Pi_m$-diagrams vanish in which one or more vertices appear between any two vertices connected by a $\tilde{\gamma}$-contraction: 
there is no phase space left for the integration of the time variable of such vertices due to the $\delta$-function 
constraint \eq{time-ret}.\footnote{See Secs. II 3b and II 3c in \Cite{Saptsov12a} for a corresponding argument in frequency space.}
This means that in the surviving diagrams the $\tilde{\gamma}$ contractions form a ladder series, see \Fig{fig:resum}a, which can be summed up.
The skeleton diagram for this resummation is shown in \Fig{fig:resum}b and consists of a single term:
with \Eq{time-ret} 
\begin{subequations}
\label{full-tildesig}
\label{tildesigma}
\begin{align}
  \tilde{\Sigma}(t_1-t_2) &=
  -i \sum\limits_r\bar{G}_1 \Gamma_1 \delta(t_1-t_2)e^{-iL(t_1-t_2)} \tilde{G}_{\bar{1}}
  \label{tildeSig_r}
  \\
  &= 2\tilde{\Sigma} \, \delta(t_1-t_2)
   = \tilde{\Sigma} \, \bar{\delta}(t_1-t_2)
   ,
\end{align}
\end{subequations}
with the time-independent superoperator
\begin{align}
  \tilde{\Sigma}
  =-i \sum_1 \frac{1}{2} \Gamma_1 \bar{G}_1\tilde{G}_{\bar{1}}
  =-i \sum_{\sigma} \frac{1}{2} \Gamma_{\sigma} \sum_{\eta} \bar{G}_{\eta \sigma} \tilde{G}_{\bar{\eta} \sigma}
  .
  \label{Sigmatilde}
\end{align}
Note that in the sum over $1=\eta,\sigma,r$, $\Gamma_1=\Gamma_{r\sigma}$ does not depend on $\eta$
and we again introduced the function $\bar{\delta}$ of \Eq{deltabar}.
The sum of the spin-resolved tunnel rates over the reservoirs is denoted by
\begin{align}
  \Gamma_\sigma  =  \sum_{r} \Gamma_{r\sigma}
  .
  \label{gammasigmadef}
\end{align}
The superoperator \eq{Sigmatilde} is just the (constant) Laplace transform of $\tilde{\Sigma}(t_1-t_2)$
and is skew-adjoint, $\tilde{\Sigma}^\dag =-\tilde{\Sigma}$,
since $\tilde{G}_{\bar{1}} = \bar{G}_1^\dagger$.
By resumming diagrams as illustrated in \Fig{fig:resum}b,
we can now simply leave out all terms with retarded contractions $\tilde{\gamma}$ from the series
and we can incorporate their effect into a
simple renormalization of the bare dot Liouvillian by the skeleton term \eq{tildesigma}:
\begin{align}
  L\rightarrow \bar{L}=L+\tilde{\Sigma}.
  \label{barl}
\end{align}
In this way, we have eliminated the \emph{annihilation} field superoperators $\tilde{G}$ of the quantum dot which enter only through the retarded reservoir correlation function $\tilde{\gamma}$ [cf. \Eq{ret-contr}].
This elimination was first pointed out in the more general framework of the real-time renormalization-group as formulated in \Cite{Schoeller09a}
(where it is referred to a discrete RG step), which is not limited to the wideband limit and which explicitly constructs the corrections due to the frequency dependence (e.g., vertex renormalization).
However, the above simpler derivation~\footnote{See \Cite{Saptsov12a} for a Laplace space argument.} may be of broad practical interest since in most studies the wideband limit is assumed from the start anyway.
Also, the use of $\delta$-restrictions on time integrations reveals a mathematical analogy to the theory of disordered metals where spatial $\delta$-correlations of the disorder suppress crossing impurity contractions.~\cite{Levitov-Shytov}

\subsubsection{Renormalized perturbation theory for finite temperature
\label{sec:ftpt}}

Having eliminated the destruction superoperators $\tilde{G}$ and the retarded reservoir correlation functions $\tilde{\gamma}$ by the replacement \Eq{barl},
we obtain a new time-ordered expansion for the propagator,
$\Pi(t,t_0) = \sum_{m=0}^\infty \bar{\Pi}_m(t,t_0)$,
for which the $m$th-order term is analogous to \Eq{mth-expansion}
\begin{subequations}
\label{mth-expansion-bar-all}
\begin{align}
  \bar{\Pi}_m = &
  (-i)^m
  \langle
  \bar{J}_{\bar{m}}(t_m)  ...  \bar{J}_{\bar{1}}(t_1)
  \rangle_R
  e^{-i\bar{L}(t-t_0)}
  \bar{G}^{'}_m(t_m)... \bar{G}^{'}_1(t_1)
  \nonumber
  \\
  = &
  (-i)^m
  \sum_\text{contr} (-1)^{P} \prod_{ \langle i,j \rangle}^{\phantom{\,}}   \bar{\gamma}_{i,j}(t_i-t_j)
  \label{mth-expansion-bar2}
  \\
  &
  \times
  e^{-i\bar{L}(t-t_0)}
  \bar{G}^{'}_m(t_m)... \bar{G}^{'}_1(t_1)
  \nonumber
  \\
  = &
  (-i)^m
  \sum_\text{contr} (-1)^{P} \prod_{ \langle i,j \rangle}^{\phantom{\,}}
  \bar{\gamma}_{i,j}(t_i-t_j)
  \label{mth-expansion-bar}
  \\
  &
  \times
  e^{-i\bar{L}(t-t_m)}\bar{G}_m e^{-i\bar{L}(t_m-t_{m-1})}  \ldots \bar{G}_1 e^{-i\bar{L}(t_1-t_0)}
  .
  \nonumber
\end{align}
\end{subequations}
with the same conventions as in \Eq{pim}, but with a \emph{renormalized} interaction picture
for the causal {creation} superoperators
\begin{align}
  \bar{G}_j^{'}(t)=e^{i\bar{L}(t-t_0)}\bar{G}_j e^{-i\bar{L}(t-t_0)}
  \label{WBL-intervertex}
\end{align}
whose difference from \Eq{dortex} is indicated by the prime.

\begin{figure}[t]
  \includegraphics[width=1.0\columnwidth]{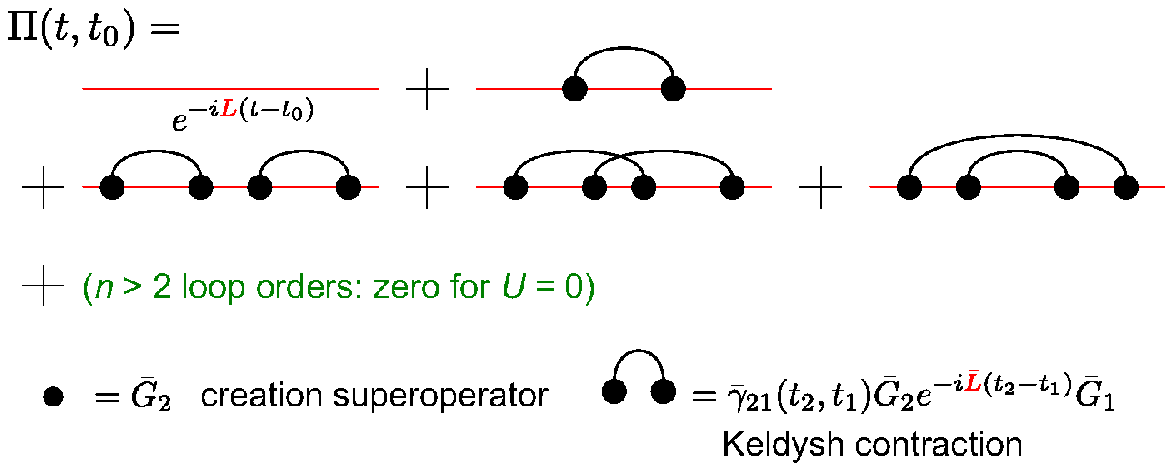}
  \caption{
    \emph{Renormalized} real-time perturbation expansion in the wideband limit:
    contributing to the full time-evolution {propagator} $\Pi(t,t_0)$
    in the first few loop orders of \emph{renormalized} expansion \eq{mth-expansion-bar},
    $\bar{\Pi}_{0}=e^{-i\bar{L}(t-t_0)}$, $\bar{\Pi}_{2}$ and $\bar{\Pi}_{4}$.
    Diagrams and expressions for the \emph{renormalized} self-energy $\bar{\Sigma}(t,t_0)=\bar{\Sigma}_2+\bar{\Sigma}_4+\ldots$ are obtained by the same steps indicated in \Fig{fig:diagrams}.
    The full evolution $\Pi(t,t_0)$ is now obtained by  $\bar{\Sigma}(t,t_0)$ from the alternative Dyson equation \eq{dysonren} in which the unperturbed evolution is generated by the {renormalized} Liouvillian $\bar{L}=L + \bar{\Sigma}$, see \Fig{fig:resum}a.
    In contrast to the series in \Fig{fig:diagrams}, the renormalized series for both $\Pi(t,t_0)$ and $\bar{\Sigma}(t,t_0)$, and therefore also for $\Sigma(t,t_0)= \tilde{\Sigma}(t,t_0)+\bar{\Sigma}(t,t_0)$, \emph{terminates} at loop order \emph{two} for the noninteracting Anderson model ($U=0$)
    due to the super-Pauli principle \eq{superpauli}\HL{-\eq{superpauli2}}, see \Eq{pitrunc}-\eq{sigmatrunc}.
  }
  \label{fig:diagramsren}
\end{figure}

The renormalized perturbation theory for $\Pi(t,t_0)$ is expressed entirely in terms of the  wideband limit form of Keldysh contraction function \eq{keld-contr}
(see \App{sec:keldysh}),
\begin{align}
  &
  \bar{\gamma}_{2,{1}}(t_2-t_1)
  \nonumber
  \\
   &=
   \delta_{2,\bar{1}}
   \frac {
     \Gamma_{\HL{2}}
   } {2\pi}
   \int d\omega_2 e^{-i\eta_{\HL{2}} (\omega_2+\mu_{r_2}) (t_2-t_1)} \tanh(\eta_2\omega_2/2T)
  \nonumber
  \\
  & =
  \delta_{2,\bar{1}}
  \frac {-i
    \Gamma_{\HL{2}}
    T }{\sinh\left(\pi T(t_2-t_1)\right)} e^{-i\eta_2 \mu_{r_2}  (t_2-t_1)} 
  ,
  \label{time-keldysh}
\end{align}
the \emph{creation} field superoperators $\bar{G}$,
and the renormalized Liouvillian $\bar{L}$
generating the \emph{renormalized}  free evolution 
$\bar{\Pi}_0(t,t_0) = e^{-i \bar{L}(t-t_0)}$.
The diagrammatic expansion, shown in \Fig{fig:diagramsren},
is much simpler than the original one in \Fig{fig:diagrams}.
Since all appearing creation superoperators $\bar{G}$ anticommute,
the key difficulty in the superoperator structure of \Eq{mth-expansion-bar}
lies in the failure of the $\bar{G}$ to commute with the renormalized Liouvillian, more precisely,
$[\bar{G},L]_{-} \not\propto \bar{G}$ due to the quartic interaction term \eq{quartic} in $L$ for $U\neq0$.

The expansion \eq{mth-expansion-bar} captures the time-dependent, finite-temperature effects.
For $T\rightarrow  \infty$ the renormalized perturbation theory is exact already in zeroth order:
in this limit, all higher order $m \geq 1$ corrections \eq{mth-expansion-bar} vanish
since the Keldysh contraction \eq{time-keldysh} goes to zero for $T \rightarrow  \infty$
[even without taking the wideband limit, cf. \Eq{keld-contr}].
Thus, $\bar{L}$ generates the exact, dissipative, Markovian effective Liouvillian [cf. \Eq{effective}] in the infinite temperature limit:
\begin{align}
  \lim_{T \rightarrow \infty} L(t,t')  &=\bar{L} \bar{\delta}(t-t')
  ,
  \\
 \lim_{T \rightarrow \infty} \Pi(t,t_0) & = e^{-i \bar{L}(t-t_0)}
 ,
 \label{piinf}
\end{align}
with $\bar{L}=L+\tilde{\Sigma}$.
Since this renormalized time-evolution serves as a reference for the renormalized perturbation theory \eq{mth-expansion-bar}, 
it will be considered in more detail in the next section.

Although the causal superfermion approach is crucial in setting up the renormalized series \eq{mth-expansion-bar},
one can calculate $\tilde{\Sigma}$, and therefore $\bar{L}$, using any equivalent density operator
technique~\cite{Leijnse08a,Timm08,Koller10,Timm11}
(Nakajima-Zwanzig, etc.) simply by taking the leading order in the tunnel coupling in the wideband limit and then letting $T \rightarrow \infty$.

\subsection{Infinite temperature limit and fermion-parity\label{sec:infiniteT}}

Before we continue our analysis of the renormalized perturbation theory \eq{mth-expansion-bar} for the noninteracting limit ($U=0$) in \Sec{sec:Pauli},
we point out an interesting immediate consequence of the above general structure 
of \eq{mth-expansion-bar} which applies to the interacting Anderson model ($U \neq 0$).
In fact, it applies to a broad class of quantum-dot models, i.e., for other model Hamiltonians instead of $H$, \Eq{dot_ham}, coupled to the reservoirs by a bilinear, particle-conserving $H_T$.

We have taken the $T \rightarrow \infty$ limit to define the starting point for both the construction of Liouville-Fock space 
[namely, the vacuum superket $|Z_L) = \tfrac{1}{2} \unit$] and for the renormalization of the perturbation theory \eq{mth-expansion-bar}
[$\tilde{\Sigma}\bar{\delta}(t-t') = \lim_{T \rightarrow \infty} \Sigma(t,t')$].
The result for $\tilde{\Sigma}$ holds nonperturbatively in all parameters in the limit $T \rightarrow \infty$
even though it results from the leading term in the perturbation theory in $\Gamma_{r\sigma}$.
It is all the more surprising that it has observable implications in a \emph{finite}-temperature experiment,~\cite{Contreras12}
for arbitrary values of $\Gamma_{r\sigma}$, the interaction $U$, applied voltages and magnetic field (only restricted  by the wideband limit).
This result was first noted in the perturbative study \Cite{Contreras12} and subsequently related to the $T \rightarrow \infty$ limit and the fermion parity, generalizing it nonperturbatively in $\Gamma$ and arbitrary Anderson-type models.~\cite{Saptsov12a}
We now analyze this fermion-parity protected decay mode within the time-dependent perturbation theory
in order to directly compare with the analysis in \Sec{sec:results},
which avoids the Laplace space analysis of \Cite{Saptsov12a} altogether.
Moreover,
we now also include the spin-dependent tunneling which \Cite{Contreras12} also considered.
For this discussion and that following in \Sec{sec:generalWBL}, it is useful to elaborate more on the self-energy,
although most parts of this work emphasize the possibility of calculating the time-evolution propagator $\Pi(t,t_0)$ directly from \Eq{mth-expansion-bar} by using field superoperators.
The self-energy also facilitates comparison with results from real-time RG and other density operator approaches.

\subsubsection{Renormalized self-energy\label{sec:self-energy}}

The self-energy superoperator is defined either by the kinetic equation \eq{effective-l} for $\rho(t)$ or the equivalent Dyson equation for $\Pi(t,t_0)$, \Eq{dyson}.
Diagrammatically it is defined by collecting those parts of diagrams of $\Pi(t,t_0)$ that are \emph{irreducibly} contracted (i.e., diagrams pieces obtained by \emph{only} cutting through free time-evolutions, without cutting contractions) and summing these.
The perturbation theory for the self-energy superoperator is then simply obtained from the perturbation theory for $\Pi(t,t_0)$ by 
(i) restricting the sum to irreducible contractions and
(ii) omitting the initial ($t_0 \rightarrow t_1$)
and final ($t_m \rightarrow t$)
free time-evolutions.
The perturbation theory can then be resummed in terms of these self-energy diagram blocks.
If this is done for the original perturbation theory \Eq{mth-expansion},
taking $e^{-iL(t-t_0)}$ as the free time-evolution, we obtain the Dyson \Eq{dyson}
with self-energy $\Sigma(t,t_0)$ (equal to the Nakajima-Zwanzig kernel).
However, the renormalized perturbation theory \Eq{mth-expansion-bar} takes $e^{-i\bar{L}(t-t_0)}$ as a reference.
This series can be resummed as well in terms of different self-energy diagram blocks, now denoted by $\bar{\Sigma}(t,t')$.
This gives an equivalent Dyson equation for the same superoperator $\Pi(t,t_0)$,
\begin{align}
  \Pi(t,t_0) &=  e^{-i\bar{L}(t-t_0)}
  \nonumber
  \\
            & -i \timeint{t}{t_2}{t_1}{t_0} e^{-i\bar{L}(t-t_2)} \bar{\Sigma}(t_2,t_1) \Pi(t_1,t_0)
  \label{dysonren}
  ,
\end{align}
The renormalized self-energy superoperator $\bar{\Sigma}(t,t_0)$~\cite{Schoeller09a,Saptsov12a}
is obtained from \Eq{mth-expansion-bar} by keeping \emph{irreducible} Keldysh contractions.
This corresponds to a decomposition of the effective dot Liouvillian,
\begin{align}
  \label{effective-bar}
  {L}(t,t^{'}) = \bar{L}\bar{\delta}(t-t')+\bar{\Sigma}(t,t')
  ,
\end{align}
alternative to \Eq{effective}.
Since the kinetic equation \eq{effective-l} only depends on the \emph{sum} of the reference Liouvillian
$L\bar{\delta}(t-t')$ [$\bar{L} \bar{\delta}(t-t')$]
and the self-energy
$\Sigma(t,t')$ [$\bar{\Sigma}(t,t')$]
appearing in the Dyson equation \Eq{dyson} [\Eq{dysonren}], this results in the same time-evolution $\Pi(t,t_0)$ in the wideband limit.

\subsubsection{Fermion-parity protected decay mode\label{sec:generalWBL}}

We now discuss how the infinite temperature self-energy $\tilde{\Sigma}$ affects the  \emph{finite-temperature} time-evolution of the density operator $\rho(t)$.
The key observation is that by the causal structure [cf. \Sec{sec:pt}] of \Eq{mth-expansion-bar},
also the renormalized self-energy $\bar{\Sigma}$ always has a creation superoperator $\bar{G}_1$ standing on the \emph{far right}, i.e., at the time $t_1$ of the initial tunnel ``process''.
An immediate consequence of our Liouville-Fock space construction using \emph{causal} superfermions is that the maximally filled superket, i.e., the fermion-parity operator $|Z_R)=\tfrac{1}{2}(-1)^n$, 
is an exact right zero eigenvector of the nontrivial self-energy since $\bar{G}_m|Z_R)=0$ [\Eq{parity}],
\begin{align}
  \bar{\Sigma}(t,t_0)|Z_R)=0 \quad \text{for any $t,t_0$}
  .
  \label{barSigmaZR}
\end{align}
This a consequence of the super-Pauli principle in Liouville space, \Eq{superpauli}.
The time-evolution of the excitation mode $|Z_R)$ is then completely determined by $\tilde{\Sigma}$ which we obtained exactly in the wideband limit. 
We emphasize that it is determined \emph{completely} by the \emph{leading order} term in $\Gamma$ in the limit $T \rightarrow \infty$.
The action of $\tilde{\Sigma}$ on this mode follows directly from the superfermion anticommutation relation and $\bar{G}_1|Z_R)=0$ [\Eq{parity}],
\begin{align}
  \tilde{\Sigma}(t,t')|Z_R)
  &=
  -i \bar{\delta}(t-t') \sum_1 \tfrac{1}{2} \Gamma_1 \bar{G}_1\tilde{G}_{\bar{1}}|Z_R)
  \label{SigmatildeZR}
  \\
  & =
  -i \bar{\delta}(t-t') \sum_1 \tfrac{1}{2}\Gamma_1\HL{(1-\tilde{G}_{\bar{1}}\bar{G}_1)}  |Z_R)
  \nonumber
  \\
  &=
  -i \bar{\delta}(t-t') \Gamma |Z_R) \quad \text{for all $t \geq t' \geq t_0$}
  .
  \nonumber
\end{align}
The \emph{fermion-parity eigenvalue} is simply the sum of all tunnel rates over both reservoirs and spins
  \begin{align}
    \Gamma & =  \sum_{\sigma} \Gamma_{\sigma r}  = \sum_{\sigma} \Gamma_{\sigma}
    \label{gammadef}
  \end{align}
times $-i$.
The renormalized time-dependent perturbation theory directly shows that for the fermion-parity mode $|Z_R)$ the $T \rightarrow \infty$ evolution remains exact at \emph{all finite temperatures}:
\begin{align}
  \Pi(t,t_0)|Z_R)=e^{-\Gamma(t-t_0)} |Z_R)  \quad \text{for all $t \geq t_0$}
  .
  \label{piZR}
\end{align}
All higher order corrections given by \Eq{mth-expansion-bar},
responsible for dependence on $U$, $\epsilon$, $B$, $\mu_r$, and $T$,
vanish:
$\bar{\Pi}_m(t,t_0)|Z_R)=0$ for $m \geq 1$.
This follows since  $|Z_R)$ is a super eigenvector of $\bar{L}$ 
and $\bar{G}_m|Z_R)=0$ by the super-Pauli principle.
The former follows from $L|Z_R) \propto [H,(-1)^n]_{-} = 0$ by the superselection rule \eq{supersel}
and \Eq{SigmatildeZR}.
The exact result \eq{piZR}
can be also formulated for the original self-energy:
$\Sigma$ has $|Z_R)$ as an exact eigenmode with eigenvalue $-i \Gamma$,
\begin{align}
  \Sigma(t,t') |Z_R)=-i \Gamma |Z_R)  \quad \text{for all $t \geq t' \geq t_0$}
  .
  \label{SigmaZR}
\end{align}
For multi-orbital models this generalizes to $\Gamma =\sum_{r\sigma l}\Gamma_{r\sigma l}$ and $|Z_R)=\prod_{\sigma l} e^{i \pi n_{\sigma l}} / N$ for $N$ spin-orbitals, where $l$ is the orbital quantum number.~\cite{Saptsov12a}

It should be noted that \Eqs{piZR}-\eq{SigmaZR} hold nonperturbatively both in the tunneling $\Gamma$ as well as in Coulomb interaction $U$ and down to zero temperature $T=0$: only the wideband limit is used here.
Due to the fundamental fermion-parity superselection principle
the eigenvalue is thus prevented from picking up any dependence on energies other than $\Gamma=\sum_{r\sigma} \Gamma_{r\sigma}$ [\Eq{gammadef}] and the decay remains strictly exponential.
Since only the sum of spin-dependent rates enters,
the spin-polarization of the tunneling also has no influence.
Because we explicitly used this fermion-parity superselection principle in the construction of the causal field superoperators [cf. \Eq{parity} and following discussion], the property \eq{piZR} becomes directly clear on the superoperator level once the wideband limit has been taken [\Eq{mth-expansion-bar}].

We note that $\bar{G}_1$ has only one nontrivial zero right eigenvector,  $|Z_R)$, for all values of the multi-index $1$:
in analogy to usual the second quantization,
only the maximally filled state is a common zero eigenvector of \emph{all} creation superoperators.
Therefore the above argument applies only to the special fermion-parity superket $|Z_R)$.

The exact result \eq{piZR} implies for the time-evolution of $\rho(t) = \Pi(t,t_0) \rho(t_0)$,
starting from an initial state $\rho(t_0)$ with the general form \Eq{general-initial-dot} at $t=t_0$,
that
\HL{
\begin{align}
  \rho(t) = \left[ \Xi(t_0) e^{-\Gamma(t-t_0)} + \ldots \right] |Z_R) + \ldots
  ,
  \label{rhotxi}
\end{align}
see the introductory discussion of \Eq{keyresult} and \Fig{fig:sketch}.}
The decay of the initial two-particle correlation $\Xi(t_0)$ on the quantum dot thus happens on a time scale $t-t_0 \lesssim \Gamma^{-1}$
which is independent of all further energy scales mentioned above.
 As pointed out in \Cite{Contreras12},
the appearance of the sum of all rates \eq{gammadef} in the decay rate of the the two-particle correlation seems to have a simple 
origin in the noninteracting limit 
($U=0$):
using a Markovian approximation,
$\braket{ n_\uparrow n_\downarrow} = \braket{ n_\uparrow} \braket{ n_\downarrow} \propto
\prod_\sigma e^{-\Gamma_\sigma(t-t_0)}=e^{-\Gamma(t-t_0)}$.
However, as emphasized there, it is all the more surprising that the decay maintains this exponential form and the value of the decay rate in the interacting limit, 
even when attaching a ferromagnet or superconductor \me{or when the dot is initially in a correlated state,} where this factorization breaks down.
Also, note that the Markovian approximation remains exactly valid for this decay mode.

The real-time renormalization group approach,~\cite{Saptsov12a} was found to be consistent with
the eigenvalue equation \eq{SigmatildeZR},
even without any truncations of the exact hierarchy of RG equations or any approximations other than the wideband limit (as it should).
In the continuous RG-flow of the effective Liouvillian towards low energies, the coefficient of $|Z_R)(Z_R|$ is given by the eigenvalue $-i\Gamma$ and this coefficient does not flow.

Finally, we recall that the superket $|Z_R)$ in \Eq{piZR} on its own does not represent a physical state, only in linear combination with the vacuum superket $|Z_L)$ of the form
\Eq{general-initial-dot} it does.
Measurement setups that can target specifically the fermion-parity protected decay mode contained in this mixed state were analyzed in \Cites{Contreras12} and \onlinecite{Schulenborg13thesis}.
In the next section, we will calculate the full time dependence of such physical states
for $T=\infty$, and finite interaction $U$ and in \Sec{sec:results} for finite $T$ but $U=0$.

\subsubsection{Infinite temperature limit and Markovian relaxation\label{sec:infinite}}

As mentioned in \Sec{sec:ftpt},
the decay of all modes is Markovian and exponential in the infinite-temperature limit,
which surprisingly continues to hold for the special fermion-parity mode $|Z_R)$ at any finite $T$
as we have just seen.
To calculate the $T \rightarrow \infty$ time evolution, $\bar{\Pi}(t,t_0)=e^{-i \bar{L}(t-t_0)}$ [\Eq{piinf}],
which serves as a reference for the renormalized perturbation theory \eq{mth-expansion-bar},
we need (some of) the super eigenvectors of $\bar{L} = L + \tilde{\Sigma}$.
These follow easily from the second quantized forms \eq{Sigmatilde} and \eq{L2ndquant} of the superoperators,
as we now show.

First, the self-energy
$
\tilde{\Sigma}
=
-i \tfrac{1}{2} \sum_{\eta \sigma} \Gamma_{\sigma} \mathcal{N}_{\eta \sigma}
$
[\Eq{Sigmatilde}] simply counts the mode occupation through the superoperator $\mathcal{N}_{\eta \sigma}$ [\Eq{Netasigma}],
and multiplies it with the half of the spin-resolved decay rate \eq{gammasigmadef}, $\Gamma_\sigma = \sum_r \Gamma_{r\sigma}$.
The bosonic basis superkets \Eqs{basis-boson} are thus super eigenvectors of $\tilde{\Sigma}$,
and the latter can be expressed in projectors onto the bosonic part of the basis \eq{basis-boson}
\begin{subequations}
  \label{sigmatilde_basis}
  \begin{align}
    \tilde{\Sigma} 
    =
    &
    -i \sum_\sigma  \Gamma_{\sigma}
    |\chi_\sigma)(\chi_\sigma|
    \label{chitildeSigma}
    \\
    &
    -i \Gamma |Z_R)(Z_R|
    \\
    &
    - i\tfrac{1}{2} \Gamma
   \Big[ \sum_\sigma |S_\sigma)(S_\sigma|+\sum_\eta |T_\eta)(T_\eta| \Big]
    +F
    ,
\end{align}
\end{subequations}
where again $\Gamma = \sum_\sigma \Gamma_{\sigma}$ [\Eq{gammadef}] and
$F$ denotes the irrelevant fermionic part [cf. \Sec{sec:L-F-s}].
Since $\tilde{\Sigma}^\dag =-\tilde{\Sigma}$ [cf. \Eq{tildesigma}],
the eigenvalues are necessarily imaginary or zero [bosonic eigenvector $|Z_L)$],
and the left super eigenvectors are the super adjoints of the right ones
with the same eigenvalues.

Similarly, we can rewrite the Liouvillian:
\begin{subequations}
  \begin{align}
    L =
    & \sum_{\eta,\sigma}
    \left(
      \eta \left[ \epsilon+ U/ 2+\sigma B/2 \right]
    \right)
    \mathcal{N}_{\eta\sigma}
    + L|_\text{quartic}
    \label{L}
    \\
    =
    &
    \sum_\sigma  \sigma B
    |S_\sigma)(S_\sigma|
    +
    \sum_\eta  \eta ( 2\epsilon+U )
    |T_\eta)(T_\eta|
    +F
    ,
    \label{L_basis}
  \end{align}
\end{subequations}
where $F$ again denotes the irrelevant fermionic part.
\footnote{For the full expression for \Eq{L_basis} see Eq. (135) in \Cite{Saptsov12a}.}
In writing \Eq{L_basis} we used that the quartic part of $L$ [\Eq{quartic}] acts \emph{only} on the fermionic part of the Liouville-Fock space [cf. \Eq{quartic-term}],
whereas the quadratic part has a component in both the bosonic and the fermionic part.

For the particular case of the Anderson model, the bosonic blocks of $L$ and $\tilde{\Sigma}$ commute,
as \Eqs{L_basis} and \eq{sigmatilde_basis} explicitly show.
\footnote{
  The $T\rightarrow \infty$ self-energy $\tilde{\Sigma}$ is quadratic in general due to the wideband limit.
  Therefore, the Liouville-Fock basis \Eqs{basis-boson}-\eq{basis-fermion} is an eigenbasis,
  also when generalized to multi-orbital Anderson models.
  However, when including more general interaction terms in $L$ for such models,
  it may be that the bosonic blocks of $L$ and $\tilde{\Sigma}$ do not commute and have no common eigenbasis.
  In this case, the diagonalization of $\bar{L}$ may be less simple.
}
Therefore, when working in the basis naturally provided by the causal superfermions,
we obtain the diagonal form for (bosonic part of) $\bar{L}=L +\tilde{\Sigma}$
when simply adding\footnote{
  Note that \emph{fermionic} basis superkets are \emph{not} super eigenkets of $L$ for $U \neq 0$,
  due to \Eq{quartic-term}], see \Cite{Saptsov12a}.
} \Eqs{sigmatilde_basis} and \eq{L_basis},
which we quote here for future reference:
\begin{subequations}
  \label{barL_basis}
  \begin{align}
    \bar{L}
    =
    &
    - i \sum_\sigma \Gamma_{\sigma}
    |\chi_\sigma)(\chi_\sigma|
    \label{barL_chi}
    \\
    &
    - i \Gamma
    |Z_R)(Z_R|
    \label{barL_ZR}
    \\
    &
    +
    \sum_\sigma  \left( \sigma B - i \tfrac{1}{2} \Gamma \right)
    |S_\sigma)(S_\sigma|
    \\
    &
    +
    \sum_\eta \left( \eta ( 2\epsilon+U ) - i\tfrac{1}{2} \Gamma \right)
    |T_\eta)(T_\eta|
    +F
    ,
\end{align}
\end{subequations}
Notably, this implies that for the time-evolution in the limit $T \rightarrow \infty$
-- fully determined by the bosonic part of $\bar{L}=L+\tilde{\Sigma}$ --
the nontrivial quartic part \eq{quartic-term} has no influence:
the interaction parameter $U$ only enters via the quadratic part of the interaction Liouvillian.
\footnote{
  For finite temperature \emph{and} $U \neq 0$ the fermionic part of $\bar{L}$, not given in \Eq{barL_basis},
  is needed
  to describe virtual intermediate renormalized time evolution, see the discussion of \Eq{virtual}.
  In the $U=0$ illustrations in this paper, it is not needed.
}

From the diagonal form \eq{barL_basis}
we immediately obtain the $T \rightarrow \infty$ time evolution of the reduced density operator
expressed in terms of the coefficients of the initial dot density operator $\rho(t_0)$ [cf. \Eq{general-initial-dot}]:
\begin{align}
  \lim_{T \rightarrow \infty} \rho(t)
  =
  &
  e^{-i\bar{L}(t-t_0)} \rho(t_0)
  \\
  =
  &
  \frac 1 2 |Z_L)
  + \sum_\sigma e^{- \Gamma_\sigma(t-t_0)} \Phi_\sigma (t_0) |\chi_\sigma)
  \nonumber
  \\
  &+e^{-\Gamma(t-t_0)}\Xi(t_0)|Z_R)
  \nonumber
  \\
  &
  +\sum_\eta e^{ -\left[ i\eta (2\epsilon+U) + \tfrac{1}{2}\Gamma \right] (t-t_0)}\Upsilon_\eta (t_0)|T_\eta)
  \nonumber
  \\
  &
   +\sum_\sigma e^{ -\left[ i\sigma B + \tfrac{1}{2}\Gamma \right](t-t_0)}\Omega_\sigma(t_0)|S_\sigma) 
  .
  \label{InfTrelax}
\end{align}
The result \eq{InfTrelax} explicitly illustrates that renormalized free evolution dissipative,
involving exponential decay with rates $\Gamma_\sigma$, $\Gamma/2$ and $\Gamma$.
The oscillations during the decay \eq{InfTrelax} indicate the presence of coherent spin $\uparrow$-$\downarrow$ excitations (frequency $B$) or coherent electron-pair 0-2 excitations (frequency $2\epsilon+U$) in the initial state $\rho(t_0)$.

At infinite temperature, we obtain from \Eq{InfTrelax} in the stationary limit the maximum von Neumann entropy state \eq{rhomax},
$  \lim_{T \rightarrow \infty} \rho(\infty)=\rho_\infty =\tfrac{1}{4} \unit =\tfrac{1}{2}|Z_L)$, the vacuum superket \eq{ZL} in our Liouville-Fock space construction.
Indeed, $|Z_L)$ is the right zero eigenvector of $\bar{L}$ since $|Z_L)(Z_L|$ is missing in \Eq{barL_basis},
a point that will be important later on [cf. discussion of \Eq{barSigma}].
Generally,
the left zero super eigenvector $(Z_L|$ guaranteeing probability conservation, $(Z_L|\bar{L} =0$,
only implies the \emph{existence} of at least one right zero supervector, the stationary state $|\rho_{\mathrm{st}})$ (assuming it is unique).
However, in general this stationary state is not equal to the (properly normalized) super adjoint of $(Z_L|$.
The above discussion now shows that in general the deviation of $|\rho_{\mathrm{st}})$ from $\tfrac{1}{2}|Z_L)$ is generated by the finite temperature corrections described by $\bar{\Pi}_m$ for $m \geq 1$ (and $\bar{\Sigma}$),
which makes good physical sense.

\subsection{Noninteracting limit\label{sec:Pauli}}
Nearly all considerations of the perturbation series \Eq{mth-expansion} 
and \eq{mth-expansion-bar} so far apply to the interacting Anderson model ($U \neq 0$) and, where mentioned, its generalizations.
We now identify the simplifications that the noninteracting limit $U=0$ brings.

In \Sec{sec:intpic}, we found that the interaction picture field superoperators $G_1^q(t)= e^{iL(t-t_0)}G_1^qe^{-iL(t-t_0)}$ [\Eq{simple-td}] simplify for $U=0$ since in this case $L$ is quadratic in these fields.
However, even in this simple limit, the original perturbation theory \eq{mth-expansion} still contains an infinite series of terms.
Although this series can be resummed, this can be avoided if one instead starts from the physically motivated renormalized perturbation series 
\eq{mth-expansion-bar-all} as we now show.
For $U=0$ the renormalized interaction picture of the creation superfields $\bar{G}'_1(t) = e^{i\bar{L}(t-t_0)}\bar{G}_1 e^{-i\bar{L}(t-t_0)}$ [\Eq{WBL-intervertex}] with respect to the quadratic renormalized Liouvillian
\begin{align}
   \bar{L} =
   &
   \sum_{\eta,\sigma}
   \left(
     \eta \epsilon_\sigma - i \tfrac{1}{2} \Gamma_\sigma
   \right)
   \bar{G}_{\eta\sigma} \tilde{G}_{\bar{\eta}\sigma}
   ,
   \label{barL}
\end{align}
simplifies, since now
\begin{align}
  \Big[\bar{L},\bar{G}_1\Big]_{-} &=
  \left( \eta \epsilon_\sigma -i\tfrac{1}{2} \Gamma_\sigma  \right)
  \bar{G}_1
  ,
  \label{barLbarGcom}
\end{align}
which follows from the anticommutation relations \eq{anticommut_d}.
Analogous to \Eq{simple-td} we obtain:
\begin{align}
  \label{barG-evol}
  \bar{G}'_1(t)=
     e^{\big( i\eta\epsilon_\sigma + \tfrac{1}{2}\Gamma_\sigma \big) (t-t_0)}
  \bar{G}_1
  ,
\end{align}
and therefore these field superoperators also anticommute:
\begin{align}
  [\bar{G}_2'(t_2), \bar{G}_1'(t_1) ]_{+} = 0
  .
  \label{Gprimeanticom}
\end{align}

Inserting \Eq{barG-evol} into \eq{mth-expansion-bar} we obtain the key result
\begin{subequations}
  \label{mth-expansion-bar-non}
  \begin{align}
    \bar{\Pi}_m 
    = 
    (-i)^m
    & \sum_\text{contr} (-1)^P
    \nonumber
    \\
    &
     \times 
    \prod_{\langle i,j \rangle}  \bar{\gamma}_{i,j}(t_i-t_j)
    e^{i\eta_i\epsilon_{\sigma_i}(t_i-t_j)}
    \label{cohfactor}
    \\    
    & \times
    \prod_{k=1}^{m} e^{\tfrac{1}{2}\Gamma_{\sigma_k}(t_k-t_0)}
    \label{dissfactor}
    \\
    & \times
    e^{-i\bar{L}(t-t_0)} {\bar{G}}_m...{\bar{G}}_1
    .
    \label{superoperatorfactor}
\end{align}
\end{subequations}
Note that the dissipative factors \eq{dissfactor} depend explicitly on $t_0$, in contrast to the coherent phase factors \eq{cohfactor}, in which $t_0$ cancels out [cf. \Eq{gammaijprimed}].
Their exponential increase leads to no problems as it is always dominated by the exponentially decaying term $e^{-i\bar{L}(t-t_0)}$ in \Eq{superoperatorfactor}.

The result \eq{mth-expansion-bar-non} shows that
 all terms  of order $m>4$ are \emph{identically zero} due to their \emph{superoperator structure}.
Like \Eq{barSigmaZR}, this is another manifestation of the super-Pauli principle~\eq{superpauli}\HL{-\eq{superpauli2}}.
The exact result for the case $U=0$ is thus obtained in just the first two nonvanishing loop orders
of the renormalized perturbation theory,
represented by precisely those diagrams shown in \Fig{fig:diagramsren}:
\begin{align}
  \Pi(t,t_0) &
  =   \bar{\Pi}_0(t,t_0) + \bar{\Pi}_2(t,t_0) +\bar{\Pi}_4(t,t_0)
  ,
  \label{pitrunc}
  \\
  \bar{\Sigma}(t,t_0) &
  =  \bar{\Sigma}_2(t,t_0) +\bar{\Sigma}_4(t,t_0)
  ,
  \label{sigmatrunc}
\end{align}
and the exact total self-energy follows from $\Sigma(t,t_0)=\tilde{\Sigma}(t,t_0)+\bar{\Sigma}(t,t_0)$.
The super-Pauli principle also shows that the renormalized Dyson equation can be solved exactly,
relating  \Eq{pitrunc} to \Eq{sigmatrunc}:
\footnote{
  To obtain \Eqs{barpi2}-\eq{barpi4},
  (i) insert the finite expansions \Eq{pitrunc} and \eq{sigmatrunc} into the renormalized Dyson equation \eq{dysonren},
  (ii) use that $\bar{\Sigma}_2$ and $\bar{\Sigma}_4$ commute with $\bar{\Pi}_0$ up to unimportant c-number factors
  and
  (iii) use that -- ignoring $\bar{\Pi}_0$ -- by \Eq{superpauli}: $(\bar{\Sigma}_2)^\alpha (\bar{\Sigma}_4)^\beta =0$ when
  it contains $2\alpha+4\beta \geq 4$ superfields, leaving only four combinations $(\alpha,\beta)=(0,0),(1,0),(2,0),(1,1)$.
}
\begin{align}
  \bar{\Pi}_2(t,t_0) &
  =   \bar{\Pi}_0 \ast \bar{\Sigma}_2 \ast \bar{\Pi}_0
  (t,t_0)
  ,
  \label{barpi2}
  \\
  \bar{\Pi}_4(t,t_0) &
  =
  \bar{\Pi}_0 \ast
  \left[
    \bar{\Sigma}_4 + \bar{\Sigma}_2 \ast \bar{\Pi}_0 \ast \bar{\Sigma}_2
  \right] \ast \bar{\Pi}_0
  (t,t_0)
  ,
  \label{barpi4}
\end{align}
where by $A \ast B(t,t_0)=\int_{t_0}^t dt_1 A(t,t_1)B(t_1,t_0)$ we denote the time convolution.
This illustrates the computational advantage of working with field superoperators directly in Liouville space,
in particular when incorporating the causal structure into  these fields.
Equations~\eq{pitrunc}-\eq{sigmatrunc} show that it becomes practical to avoid
 the calculation of the self-energy in the limit $U=0$,
since the structure of the time propagator $\Pi$ is no less complicated.
In fact, as we will see explicitly in the next section [\Eq{formal-2loop2}, \Fig{fig:factorize} and \Eq{2loopPT-gamma}],
the inclusion of the second reducible term in $\bar{\Pi}_4$ on the right-hand side in \Eq{barpi4} makes it a simpler object than $\bar{\Sigma}_4$ in the $U=0$ limit.
We emphasize that the simple structure of the noninteracting problem appears only
in the \emph{renormalized} series, i.e., after explicitly exploiting the wideband limit.
Similar, but less drastic, simplifications on the \emph{superoperator level} can be exploited for interacting problems as well.~\cite{Saptsov13b}

We emphasize that the truncation in \Eqs{pitrunc}-\eq{sigmatrunc} does not rely on the spin- and charge-rotation~\cite{Saptsov12a} symmetry of the Anderson model:
only the number of spin orbitals of the model (=2) 
and the absence of quartic terms and higher in $L$ matter ($U=0$).
The crucial observation for this was that in \Eq{mth-expansion-bar-non} we were able to commute $\bar{L}$ through all the $\bar{G}$ fields to the left side, thereby only generating $c$-number factors, without changing the structure of
 the appearing field superoperators.
For the interacting Anderson model, however, the expansion \eq{pim} must be used:
here $\bar{L}$ cannot be commuted through the fields $\bar{G}$ without generating additional, higher order terms.
Then even the renormalized perturbation theory has nonzero terms beyond the two-loop order.
(Note however, that the self-energy $\tilde{\Sigma}$ {remains quadratic} even for finite $U$ due to the wideband limit.)
Yet even in this case the causal superfermions bring simplifications, cf. \Eq{virtual}.
\footnote{
  For $U\neq 0$ \Eq{mth-expansion-bar-non} is no longer valid.
  In this case,  nonzero terms arise beyond the second order in the general expansion \Eq{mth-expansion-bar}.
  The reason is that if one tries to
  commute $\bar{L}=L+\tilde{\Sigma}$ to the left -- containing a quartic term [\Eq{L2ndquant}] --
  this generates terms containing destruction superoperators $\tilde{G}_1$
  which are not required to vanish by the super-Pauli principle.
  We note that by our discussion of \Eq{virtual} (taking $q_4=q_3=q_2=q_1= +$) this effect of the quartic terms in $L$
  is restricted to only half of the propagators in virtual intermediate states.
}

The result \Eqs{pitrunc}-\eq{sigmatrunc} thus explicitly shows that the higher order terms in the renormalized perturbation theory are
\emph{generated} by the nontrivial part of the Coulomb interaction, \Eq{quartic},
which is the part that couples $U$ to the fermion-parity operator, \Eq{quartic-term}.
The real-time renormalization group in the one- plus two-loop approximation~\cite{Saptsov12a}
encodes these higher order terms into a renormalization of $\bar{L}$ and $\bar{G}$.
An important property of this approach is that while it provides a good approximation for large $U$ as well, it exactly includes the solution for the noninteracting limit $U=0$.
The above simple analysis explains why \emph{at least} the one- and two-loop orders must be included to exactly recover the noninteracting limit (i.e., for the full self-energy and the full density operator, not just selected quantities~\footnote{As pointed out in \Cite{Saptsov12a}, for selected quantities such as the charge current a one-loop RT-RG approximation already includes the noninteracting limit. A recent perturbative study indicates that for thermoelectric transport such fortuitous simplifications are absent~\cite{Gergs13thesis}.}).

Finally, we note that \Eq{sigmatrunc} shows that the quantum-dot self-energy $\Sigma$ and therefore the effective Liouvillian \eq{effective-bar} is \emph{quartic}\footnote{
  Here ``quartic'' refers to the total power of field operators $\bar{G}'_1(t)$ in the (renormalized) interaction picture [\Eq{WBL-intervertex}].
  For $U=0$ this is proportional to the original creation superoperator $\bar{G}_1$ [\Eq{barG-evol}].
 However, for $U \neq 0$ explicit calculation of $\bar{G}'_1(t)$ shows it additionally contains destruction superoperators $\tilde{G}$.
  In this case, the super-Pauli principle cannot truncate the perturbation series after the second order.
}
 in the field superoperators, even in the noninteracting limit ($U=0$).
This is to be contrasted with, e.g., Green's {function} and path-integral approaches where only quadratic expressions appear for 
noninteracting problems.
To understand better how this difference arises we turn to the explicit evaluation of all contributions 
in \Eq{pitrunc} for $U=0$, which will reveal a further simplification.
\section{Exact noninteracting time-evolution\label{sec:results}}

In this section, we specialize to the noninteracting limit $U=0$, unless stated otherwise.
\subsection{Time-evolution\label{sec:time-evolution}}
\subsubsection{Time-evolution propagator}
We now calculate the one- and two-loop corrections to the renormalized free propagator \eq{piinf},
which is defined by the time-local renormalized Liouvillian \eq{barL_basis}
\begin{align}
  \label{Pibar0}
  \bar{\Pi}_0(t,t_0) 
   =
   &
   \lim_{T\rightarrow \infty} \Pi(t,t_0)= e^{-i \bar{L}(t-t_0)}
   \\ \nonumber
   =
   &
   \, e^{-\Gamma(t-t_0)}|Z_R)(Z_R| +  \sum_\sigma  e^{-\Gamma_{\sigma}(t-t_0)}
  |\chi_\sigma)(\chi_\sigma|
  \\
  &+
  \sum_\sigma e^{-\left(i\sigma B+\tfrac{1}{2} \Gamma\right)(t-t_0)} |S_\sigma)(S_\sigma|
  \nonumber
  \\
  &+\sum_\eta e^{-\left(i\eta 2\epsilon+\tfrac{1}{2} \Gamma\right)(t-t_0)}|T_\eta)(T_\eta|
  ,
  \nonumber
\end{align}
(the fermionic part is omitted) and then discuss the time-dependent density operator $\rho(t)$.

\paragraph{One-loop propagator.}
The one-loop ($m=2$) contribution to \Eq{mth-expansion-bar}
can be written as:
  \begin{align}
    &\bar{\Pi}_2(t,t_0)
    \label{1loopdef}
    \\
    &=
    -
    e^{-i\bar{L}(t-t_0)}
    \sum_{21}
    \timeint{t}{t_2}{t_1}{t_0} 
    \bar{G}'_2(t_2) \bar{G}'_1(t_1) \bar{\gamma}_{21}(t_2-t_1)
    \nonumber
    \\
    &
    =
    -
    e^{-i\bar{L}(t-t_0)}
    \sum_{r,\sigma,\eta}
    \bar{G}_2\bar{G}_{\bar{2}} 
    \label{1loopftpt}
    \\
    &
    \quad
    \times
    \timeint{t}{t_2}{t_1}{t_0}
    \bar{\gamma}_{2,\bar{2}} (t_2-t_1)
    e^{i\eta\epsilon_\sigma (t_2-t_1)}
    e^{\tfrac{1}{2}\Gamma_\sigma (t_2+t_1-2t_0)}
    ,
    \nonumber
  \end{align}
making use of the factor $\delta_{2,\bar{1}}$ in the contraction $\bar{\gamma}_{2,{1}}$
and writing $2=\eta,\sigma,r$.
The one-loop contributions only generate transitions which increase the superfermion number by two.
This is expected since in \Eq{1loopftpt} only \emph{creation} superoperators appear.
In the bosonic sector \eq{basis-boson} these transitions proceed from the supervacuum state $|Z_L)$ to the doubly excited superkets $|\chi_\sigma)$
and from there to the most occupied superket $|Z_R)$.
The corresponding matrix elements of the superoperator $\bar{G}_2\bar{G}_{\bar{2}}$ are again easily determined using the algebra of superfermions:
first, we note that
$\bar{G}_{\eta \sigma}\bar{G}_{\bar{\eta}\sigma}
=
 \eta \bar{G}_{+ \sigma}\bar{G}_{- \sigma}$ since the fields anticommute.
Next, we see that for transitions between 0 and 2 superparticles,
by definition $\bar{G}_{+ \sigma}\bar{G}_{- \sigma}|Z_L) = |\chi_\sigma)$,
and between 2 and 4 superparticles,
$(Z_R|\bar{G}_{+ \sigma}\bar{G}_{- \sigma}
= [ \tilde{G}_{+ \sigma}\tilde{G}_{- \sigma}|Z_R) ]^\dag
= [ \bar{G}_{+ \bar{\sigma}}\bar{G}_{- \bar{\sigma}}|Z_L) ]^\dag
= [ |\chi_{\bar{\sigma}}) ]^\dag
= (\chi_{\bar{\sigma}}|
$, using \Eq{dagger}.
\footnote{
  Alternatively, one can make use of the explicit basis expansion of the creation superoperator $\bar{G}$ defined by Eq. (118) and Eq. (120) in \Cite{Saptsov12a}.
  }
As a result,
\begin{align}
  \bar{G}_{\eta \sigma}\bar{G}_{\bar{\eta}\sigma}
  = \eta \Big(
  |\chi_\sigma)(Z_L|+|Z_R)(\chi_{\bar{\sigma}}|
  \Big) + F
  ,
  \label{GG}
\end{align}
where $F$ is again an irrelevant fermionic part.
Inserting this in \Eq{1loopftpt},
we obtain after  summing over $\eta$:
\begin{align}
  &\bar{\Pi}_2(t,t_0)
  =
  -   e^{-i\bar{L}(t-t_0)}
  \sum_{\sigma}
  \Big[ |\chi_\sigma)(Z_L|+ |Z_R)(\chi_{\bar{\sigma}}|\Big]
  \label{1loopbetter}
  \\
  &
  \times
  \sum_r 2 \Gamma_{r \sigma}
  \timeint{t}{t_2}{t_1}{t_0}
  \frac{ T \sin\left(\epsilon_{r\sigma}(t_2-t_1)\right) }{ \sinh(\pi T (t_2-t_1)) }
  e^{\tfrac{1}{2}\Gamma_\sigma(t_2+t_1-2t_0)}
  ,
  \nonumber
\end{align}
where we denote the quantum-dot energies $\epsilon_{\sigma}=\epsilon + \sigma B/2$ relative to the electrochemical potential $\mu_r$ by
\begin{align}
  \label{ers}
  \epsilon_{r\sigma}=\epsilon_{\sigma}- \mu_{r}.
\end{align}
Next, the diagonal form \eq{barL_basis} of the renormalized free propagator is used
\begin{alignat}{3}
  \bar{L}|\chi_\sigma) & =-i\Gamma_\sigma |\chi_\sigma)
  ,
  & \quad
  \bar{L}|Z_R)        & =-i\Gamma |Z_R)
  ,
  \label{Lbaraction}
\end{alignat}
and we change variables,
$\Theta  = t_2+t_1-2t_0$ and $\tau  =t_2-t_1$
in the integration,
$ \int_{t_0}^{t} \int_{t_0}^{t_2} dt_2 dt_1
= 
\tfrac{1}{2}
\int_0^{\Delta} d \tau
\int_{\tau}^{2\Delta -\tau} d\Theta
$,
denoting $\Delta  = t-t_0$,
and then perform the $\Theta$-integral,
\begin{align}
  \bar{\Pi}_2(t,t_0)
  =
  &
  \sum_{\sigma}
  \left\{
    |\chi_\sigma)(Z_L| +   e^{-\Gamma_{\bar{\sigma}} \Delta}  |Z_R)(\chi_{\bar{\sigma}}| 
  \right\}
  \nonumber
  \\
  &
  \times
  \sum_{r}
  \frac{2 \Gamma_{r\sigma} }{\Gamma_\sigma }
  \left( F^{+}_{r\sigma}(\Delta)  + F^{-}_{r\sigma}(\Delta)  \right)  
  .
   \label{oneloop-integral}
\end{align}
The explicit expressions for the time-dependent coefficients (see \App{sec:Ffunc_ap}),
\begin{subequations}
  \label{Fresult}
  \begin{align}
  & F^{+}_{r\sigma}(\Delta)
  :=
  -\int_0^{\Delta} d \tau \frac {T \sin(\epsilon_{r\sigma}\tau)}{\sinh\left(\pi T\tau\right)} e^{-\tfrac{1}{2}\Gamma_\sigma \tau}
  \nonumber
  \\
  &=\frac 1 {\pi} \Im \Psi\left(\frac 1 2 - \frac {i\epsilon_{r\sigma}-\tfrac{1}{2}\Gamma_\sigma}{2\pi T}\right)
  \label{digamma-lerch-phi}
  \\
  &
  +
  \frac{1}{\pi} \Im
  e^{(-\pi T +i\epsilon_{r\sigma}-\tfrac{1}{2}\Gamma_\sigma) \Delta}
  \Phi \Big(
    e^{-2\pi T\Delta};1;\frac{1}{2}+\tfrac {i\epsilon_{r\sigma} - \tfrac{1}{2}\Gamma_\sigma}{2\pi T}
  \Big)
  ,
 \nonumber
  \\
  &F^{-}_{r\sigma}(\Delta)
  := e^{-\Gamma_\sigma\Delta}
  \int_0^{\Delta} d\tau \frac{T\sin(\epsilon_{r\sigma}\tau)} {\sinh(\pi T \tau)}
  e^{\tfrac{1}{2}\Gamma_\sigma \tau}
  \nonumber
  \\
  & =-
    e^{-\Gamma_\sigma\Delta}   \cdot
  \left.
    F^{+}_{r\sigma}(\Delta)
  \right|_{ \Gamma_\sigma  \rightarrow
             -\Gamma_\sigma }
  ,
  \label{Frelation}
  \end{align}
\end{subequations}
can be expressed through the digamma-function $\Psi(z)$,
with $\Im \Psi(z) = - \Im \sum_{n=0}^{\infty} 1/(n+z)$
and the Lerch transcendent function~\cite{Erdely}
$
\Phi(z;s;\nu) = \sum_{n=0}^{+\infty} \tfrac{z^n}{(n+\nu)^s}
$.
The asymptotic values are
\begin{subequations}
  \begin{align}
\label{inf_F}
&F^{+}_{r\sigma}(\infty)=\frac 1 {\pi} \Im \Psi\left(\frac 1 2 - \frac {i\epsilon_{r\sigma}-\tfrac{1}{2}\Gamma_\sigma}{2\pi T}\right),\\
&F^{-}_{r\sigma}(\infty)=0.
\end{align}
\end{subequations}
The decay to these stationary values, given by the second term on the right-hand side of \Eq{digamma-lerch-phi}, has the asymptotic form:
\begin{align}
  & F^{+}_{r\sigma}(\Delta) -   F^{+}_{r\sigma}(\infty) =
  \sin \left(
    \epsilon_{r\sigma}\Delta + \arctan \Big( \frac{\epsilon_{r\sigma}}{\varsigma_\sigma} \Big)
  \right)
  \label{asymptotic}
  \\
  &
  \times
  \begin{dcases}
     \frac{  T  }{ \sqrt{(\pi T)^2+(\epsilon_{r\sigma})^2} } 2e^{-\pi T \Delta}   
    & \Delta \gg \Gamma_\sigma^{-1} \gg T^{-1}
    \\
    \frac { T  }{ \sqrt{ (\Gamma_\sigma/2)^2+(\epsilon_{r\sigma})^2} }
    \dfrac{e^{-\tfrac{1}{2}\Gamma_\sigma \Delta}}{\sinh(\pi T \Delta)}
    &  \Delta \gtrsim T^{-1} \gg \Gamma_\sigma^{-1},
  \end{dcases}
  \nonumber
\end{align}
where $\varsigma_\sigma=\mathrm{max}\{\tfrac 1 2 \Gamma_\sigma,\pi T\}$.
Thus for low temperature $T \ll \Gamma_{r\sigma}$,
the oscillatory decay of both $F^{+}_{r\sigma}(\Delta)$ and $F^{-}_{r\sigma}(\Delta)$ [following from \Eq{Frelation}]
is at least as fast as
$e^{-(\Gamma_\sigma/2) \Delta}$.

\begin{figure}[t]
  \includegraphics[width=0.94\columnwidth]{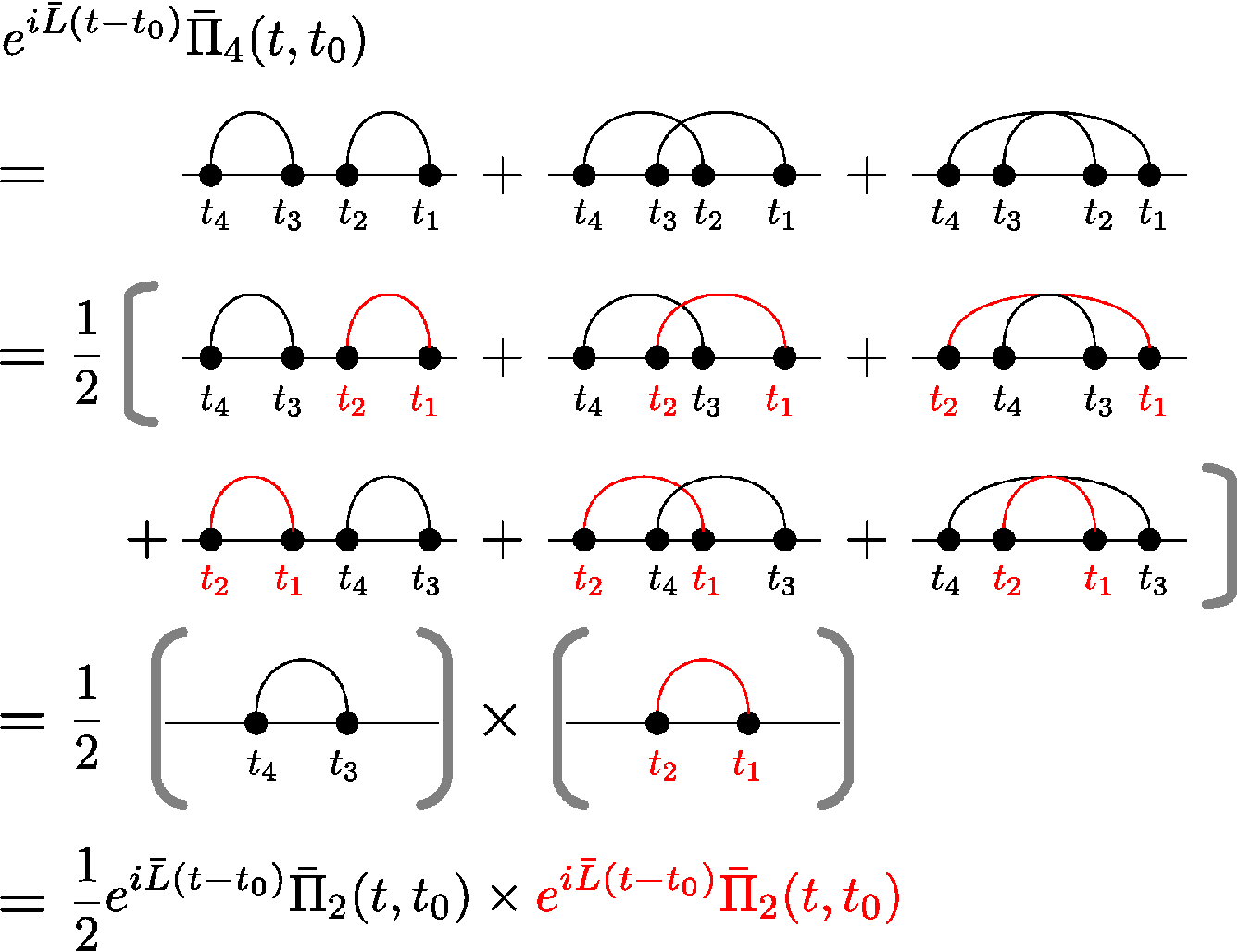}
  \caption{
      Factorization \eq{formal-2loop2} of the renormalized two-loop propagator $\bar{\Pi}_4(t,t_0)$:
    diagrammatic proof, see \App{sec:twoloop_ap} for explicit expressions.
    We use the modified convention that diagrams represent the same expressions \Eq{mth-expansion-bar-non} as in  \Fig{fig:diagramsren}
    but now without the overall renormalized unperturbed propagator $e^{-i\bar{L}(t-t_0)}$
    in \Eq{superoperatorfactor}.
    First equality: Two loop terms in \Fig{fig:diagramsren}
    with the time-ordered integrations $t \geq t_4 \geq t_3 \geq t_2 \geq t_1 \geq t_0$.
    Second equality: We
    first duplicated the terms while compensating with a factor $1/2$
    and then relabeled the dummy time variables as indicated.
    The dummy multi-indices, are relabeled correspondingly:
    a vertex with time $t_i$ thus stands for $\bar{G}_i(t_i)$ with multi-index $i=\eta_i,\sigma_i$.
    In each term the integrations are such that vertices maintain their order,
    e.g., in the second term we integrate over $t \geq t_4 \geq t_2 \geq t_3 \geq t_1 \geq t_0$.
    Since the vertex superoperators anticommute [\Eq{Gprimeanticom}],
    we can factorize the \emph{integrand} superoperators and bring together the black and the red parts.
    For the second and fifth diagram (crossed contraction),
    this introduces a quantum-dot fermion sign $(-1)$ that exactly cancels the Wick sign of the reservoir  [$(-1)^P$ in \Eq{formal-2loop}].
    As a result all integrands are given by the same superoperator.
    Third equality: by summing all the integrals with all possible time-orderings, the resulting \emph{integral} also factorizes.
    The resulting expression in the factors is precisely the one-loop propagator
    (i.e., the one-loop terms in \Fig{fig:diagramsren}, again using the mentioned convention).
    Multiplying the equation by $e^{-i\bar{L}(t-t_0)}$ gives \Eq{formal-2loop2}.
      }
  \label{fig:factorize}
\end{figure}

\paragraph{Two-loop propagator.}
The two-loop ($m=4$) contribution to \Eq{mth-expansion-bar2} reads as
 \begin{align}
  &\bar{\Pi}_4(t,t_0)
  =
    e^{-i\bar{L}(t-t_0)}
  \nonumber
  \\
  &  \timeintfour{t}{t_4}{t_3}{t_2}{t_1}{t_0}
  \bar{G}'_4(t_4) \bar{G}'_3(t_3)\bar{G}'_2(t_2)\bar{G}'_1(t_1)
  \times
  \nonumber
  \\
  &
  \sum_{\langle i,j,k,l\rangle} (-1)^P \bar{\gamma}_{i j}(t_i-t_j) \bar{\gamma}_{k l}(t_k - t_j)
  ,
  \label{formal-2loop}
\end{align}
where $\langle i,j,k,l\rangle$ denotes the sum over the following possible contractions:
$i,j,k,l=4,3,2,1$ (reducible), $4,2,3,1$, and $4,1,3,2$ (both irreducible).
The second major simplification occurring in the noninteracting limit ($U=0$),
besides the truncation \eq{pitrunc} is that this contribution can be factorized as \HL{(see \App{sec:twoloop_ap} and \Fig{fig:factorize}):}
\begin{align}
  \bar{\Pi}_4(t,t_0)
  = \tfrac 1 2 \bar{\Pi}_2(t,t_0) e^{i\bar{L}(t-t_0)} \bar{\Pi}_2 (t,t_0).
  \label{formal-2loop2}
\end{align}
This general insight is again readily obtained by making use of the causal superfermions: since without interaction, the time-dependent interaction-picture fields anticommute 
[\Eq{Gprimeanticom}],
relabeling of dummy indices and integration variables in \Eq{formal-2loop} directly leads to the factorization \eq{formal-2loop2} on the superoperator level.
This is shown diagrammatically in \Fig{fig:factorize} and written out in  \App{sec:twoloop_ap}.
Our approach thus clarifies how in the \emph{many-body} Liouville-space simplification arises for $U=0$, which is important since interacting theories are necessarily formulated in this large space.
Clearly, when making use of the collinearity of the spin-dependence of the tunneling and due to the magnetic field, the Liouville spaces of superfermions with different spin can be considered independently and one arrives also at \Eq{formal-2loop2}.
However, this consideration, formulated in \App{sec:factorize}, does not make clear how in the formalism applicable to interacting cases
this factorization comes about.
Moreover, it is unnecessary when one makes use of the causal superfermions.
%%% WARNING
%\vspace{-0.5cm}
%%% WARNING
Independent of the result \Eq{formal-2loop2},
 the structure of the superoperator \eq{formal-2loop}
can also be determined easily
using the Liouville-space second quantization.
Inserting \Eq{barG-evol}, the superoperator part of the expression \eq{formal-2loop} is
$\propto \, e^{-i\bar{L}(t-t_0)} \bar{G}_4 \bar{G}_3\bar{G}_2\bar{G}_1$.
The product of four creation superoperators can only be nonzero
if all multi-indices $1,2,3,4$ are different
and it is therefore \emph{always} proportional to the transition superoperator
taking the vacuum superket $|Z_L)$
into the most filled fermion-parity superket $|Z_R)$ [\Eq{ZR}]:
\begin{align}
  \bar{\Pi}_4(t,t_0) =\vartheta(t-t_0) |Z_R)(Z_L|.
  \label{4term}
\end{align}
%%% WARNING
%\vspace{-0.5cm}
%%% WARNING
All that remains is to calculate the coefficient $\vartheta$ as a function of $\Delta=t-t_0$
by substituting \Eq{oneloop-integral} into \Eq{formal-2loop2}:
\begin{subequations}
\label{varthetaresult}
\begin{align}
\nonumber
 \vartheta(\Delta)=
 &
 2
 \sum_{\sigma}
 \sum_{r,r'}
 \frac{ \Gamma_{r\sigma} }{\Gamma_\sigma }
 \Big( F^{+}_{r\sigma}(\Delta)  + F^{-}_{r\sigma}(\Delta)  \Big) \\
 &\times\frac{\Gamma_{r'{\bar{\sigma}}}}{\Gamma_{\bar{\sigma}} }
 \Big( F^{+}_{r'\bar{\sigma}}(\Delta)  + F^{-}_{r'\bar{\sigma}}(\Delta)  \Big) \\ \nonumber
 =
 &
 4
 \sum_{r} 
 \frac{ \Gamma_{r\uparrow} }{\Gamma_\uparrow }
 \Big( F^{+}_{r\uparrow}(\Delta)  + F^{-}_{r\uparrow}(\Delta)  \Big) \\ 
 &\times
\sum_{r'} 
 \frac{\Gamma_{r'{\downarrow}}}{\Gamma_{\downarrow} }
 \Big( F^{+}_{r'\downarrow}(\Delta)  + F^{-}_{r'\downarrow}(\Delta) \Big).
\end{align}
\end{subequations}
That the fourfold, time-ordered integral \Eq{formal-2loop} reduces to the simple product of the two spin-resolved functions depending only on the difference $\Delta=t-t_0$
is expected from the considerations in \App{sec:factorize}.
However, without the renormalized formulation of the perturbation theory in the causal superfermion framework the origin of such simplifications in real-time calculations remain  unclear.

Finally, we note that the superoperator form \eq{4term}
 as well as the truncation of the perturbation series \eq{pitrunc}
 both remain valid for the case of the \emph{noncollinear} magnetizations of the (ferromagnetic) reservoirs and / or of tunnel junctions.
They are based on the very general causal structure of the perturbation series, which is independent of the spin-rotation and other symmetries of the problem.
Summarizing, the full time-evolution propagator reads as
\begin{widetext}
\begin{align}
  \Pi(t,t_0) 
  & =
  [ \bar{\Pi}_0(t,t_0)+\bar{\Pi}_2(t,t_0)+\bar{\Pi}_4(t,t_0) ]
  \label{integral-solution-wbl}
  \\
  & =
  \lim_{T\rightarrow \infty} \Pi(t,t_0)
  +
  \sum_{r,\sigma}
  \frac{2 \Gamma_{r\sigma} }{\Gamma_\sigma }
  \Big[F^{+}_{r\sigma}(t-t_0) +F^{-}_{r\sigma}(t-t_0) \Big]
  \Big\{
    |\chi_\sigma)(Z_L| +  e^{-\Gamma_{\bar{\sigma}} (t-t_0)} |Z_R)(\chi_{\bar{\sigma}}| 
  \Big\}
  + \vartheta(t-t_0) |Z_R)(Z_L|
  ,
  \nonumber
\end{align}
\end{widetext}
where the first term is given by \Eq{Pibar0},
$F^{\pm}_{r\sigma}$ by \Eq{Fresult} and $\vartheta$ by \Eq{varthetaresult}.

\subsubsection{Time-dependent density operator}
Evaluating
\footnote{
Note that the stationary contributions to $\rho(t)$ are generated by terms in $\Pi(t,t_0)$ [\Eq{integral-solution-wbl}] of the form $|A(t))(Z_L|$ :
since by probability conservation any possible  $\rho(t_0)$ contains $\tfrac{1}{2}|Z_L)$ in its expansion [see \Eq{general-initial-dot} with $t=t_0$],
we have $|A(t))(Z_L| \rho(t_0)=\tfrac{1}{2} |A(t))$, independent of $\rho(t_0)$.
}
$\rho(t) = \Pi(t,t_0) \rho(t_0)$,
where the initial state $\rho(t_0)$ has the general
 form \eq{general-initial-dot}
 with $t=t_0$,
we obtain the exact time-dependent density operator for $U=0$ in the wideband limit:
\begin{widetext}
\begin{subequations}
  \label{exact-RDDO}
  \begin{align}
    \rho(t)
   =
   &
   \lim_{T\rightarrow \infty} \rho(t)
  +
  \sum_{r,\sigma}
  \frac{\Gamma_{r\sigma} }{\Gamma_\sigma }
      \Big[ F_{r\sigma}^{+}(t-t_0) +  F_{r\sigma}^{-}(t-t_0) \Big]
    |\chi_\sigma)  
    \label{exactPhi}
  \\
  &
  +
  \left(
     \sum_{r,\sigma}
    {e^{-\Gamma_{\bar{\sigma}} (t-t_0)}} \frac{2 \Gamma_{r\sigma} }{\Gamma_\sigma }
    \Big[F_{r\sigma}^{+}(t-t_0)+ F_{r\sigma}^{-}(t-t_0)\Big]
    \Phi_{\bar{\sigma}}(t_0)
    + \tfrac{1}{2} \vartheta(t-t_0)
  \right)
  |Z_R)
    \label{exactXi}
  ,
\end{align}
\end{subequations}
\end{widetext}
Here, the first term on the right hand side of \Eq{exactPhi} is the exact $T\rightarrow \infty$ result \eq{InfTrelax} discussed in \Sec{sec:infinite}.

\paragraph{Electron-pair and spin coherence.}
We first note the absence of corrections to the $T\rightarrow \infty$ decay as given by \Eq{InfTrelax}
of the electron-pair coherence [\Eq{Upsilon}] and transverse spin coherence coefficients [\Eq{Omega}] of the initial state $\rho(t_0)$.
For $U=0$ these therefore decay exponentially to zero with rates which are independent of temperature
and equal:
$\Upsilon_\eta (t)=e^{-\left(\Gamma/2\right)(t-t_0)} \Upsilon_\eta (t_0)$ and $\Omega_\sigma(t)=e^{-\left(\Gamma/2\right)(t-t_0)} \Omega_\sigma(t_0)$, respectively.
This is much like the fermion parity discussed below, but in contrast to the latter, here the decay is altered when $U \neq 0$.
The stationary values are zero by charge and spin-rotation symmetry, see \Sec{sec:stationary}.
\footnote{
  This can be seen from Eq. (151) in \Cite{Saptsov12a}.
}
The finite temperature gives rise to corrections to both the stationary values and to the decay, which we now discuss. 
\paragraph{Spin-orbital occupancies.}
The second term in \eq{exactPhi} modifies the decay of the level occupancies:
\begin{align}
  \Phi_\sigma(t)
  & =
  \langle n_\sigma \rangle(t) -  \tfrac{1}{2}
  \nonumber
  \\
  &
  = e^{-\Gamma_\sigma (t-t_0)} \Phi_\sigma(t_0)
  \label{Phit}
  \\
  &
  + \sum_{r}
  \frac{\Gamma_{r\sigma} }{\Gamma_\sigma }
  \Big[ F_{r\sigma}^{+}(t-t_0) +  F_{r\sigma}^{-}(t-t_0) \Big]
  .
  \nonumber
\end{align}
\refAB{Our result \eq{Phit} for a single spin
agrees with the one obtained in \Cite{Jin10} for spin-independent tunneling, i.e., $\Gamma_{r\sigma}=\tilde{\Gamma}_r$ and zero magnetic field, $B=0$, after calculating the $\omega$-integral expression left unevaluated in \Cite{Jin10}.
Additionally switching off the reservoir dependence of the tunnel coefficients, i.e., $\tilde{\Gamma}_{r}=\tilde{\Gamma}$, in \Eq{Phit} and assuming 
that initially
the dot is unoccupied, $\langle n_\sigma\rangle(t_0)=0$, we find agreement with the result of \Cite{Schmidt08} obtained within these assumptions. }
The stationary value, obtained using  \Eq{inf_F}, 
\begin{align}
  \Phi_\sigma(\infty)
  &
  :=
  \langle n_\sigma \rangle(\infty) -   \tfrac{1}{2}
  \nonumber
  \\
  &=
  \frac 1 {\pi}\sum_{r} 
  \frac{\Gamma_{r \sigma }}{\Gamma_\sigma}
  \Im \Psi\left(\frac 1 2 + \frac {\tfrac{1}{2}\Gamma_\sigma  -i\epsilon_{r\sigma}}{2\pi T}\right)
  ,
  \label{av_n_2}
\end{align}
\refAB{under corresponding simplifications also agrees with that obtained in \Cites{Schmidt08,Jin10}.}
Both the stationary value and the decay towards it depend on the spin $\sigma$:
for spin-independent tunneling $\Gamma_{r \sigma} := \tilde{\Gamma}_{r}$,
this is a consequence of the Zeeman splitting $B$ on the quantum dot.
For $B=0$, however, this is due to nonequilibrium spin accumulation on the quantum dot,
$\braket{S_z}(t)= \sum_\sigma \braket{\sigma n_\sigma}/2=\sum_\sigma \sigma \Phi_\sigma/2$,
caused by the spin-dependent tunneling.
Only when both $B=0$ and $\Gamma_{r \sigma} := \tilde{\Gamma}_{r}$ do
 we have full spin-rotation symmetry.
In this case, the occupancies are equal $\braket{n_\uparrow}=\braket{n_\downarrow}$,
and there is no correction to the longitudinal spin $\braket{S_z}(t)$ as given by the $T \rightarrow \infty$ value \eq{InfTrelax}:
$\braket{S_z}(t)$ decays exponentially to zero,
in agreement with
the spin-rotation symmetry ($z$-axis).
The decay rate, $\Gamma_\sigma=\sum_r \tilde{\Gamma}_r$, is
identical to the rate $\Gamma/2=\sum_r \tilde{\Gamma}_r$ of the spin- and electron-pair coherences $\Upsilon_\eta(t)$ and $\Omega_\sigma(t)$, respectively, all of which are temperature independent.
%\HL{DISCUSS how to measure all these quantities? Or to skip such a discussion?}
\paragraph{Fermion-parity and two-particle correlations.}
The full time-evolution of the fermion-parity operator expansion coefficient [\Eq{zr-phi}] thus reads as follows:
\begin{align}
  \Xi(t)
  = &
  e^{-\Gamma (t-t_0)}  \Xi(t_0)
  +  \sum_{\sigma}   e^{-\Gamma_{\sigma} (t-t_0)} \times
  \nonumber
  \\
  &
  \sum_{r}
  \frac{2 \Gamma_{r\sigma} }{\Gamma_{\bar{\sigma}} }
  \Big[
  F_{r{\bar{\sigma}}}^{+}(t-t_0)+ F_{r\bar{\sigma}}^{-}(t-t_0)
  \Big]
  \Phi_{{\sigma}}(t_0)
  \nonumber
  \\
  &+ \tfrac{1}{2} \vartheta(t-t_0).
  \label{Xit}
\end{align}
This coefficient takes account of the correlations of the occupancies
through the average of the fermion-parity operator:
$\Xi(t)
= \tfrac 1 2 \braket{ e^{i\pi n} }
= 2\braket{ \prod_\sigma (n_\sigma-1/2)}
= 2 \braket{ n_\uparrow n_\downarrow} -\braket{n} +1/2
$.
The result \eq{Xit} is valid for an arbitrary initial state of the quantum dot with two-particle correlations: 
$\langle n_\uparrow n_\downarrow\rangle(t_0)
\neq
\langle n_\uparrow\rangle(t_0) \langle n_\downarrow\rangle(t_0)$,
equivalent to
\begin{align}
  \Xi(t_0)\neq 2\prod_\sigma \Phi_\sigma (t_0)
  \label{neqXi}
  .
\end{align}
The first contribution to \Eq{Xit} is the exponential decay determined by the $T \rightarrow \infty$ limit.
As explained in \Sec{sec:generalWBL}, this part of the time-dependent-decay \emph{never} has any finite temperature corrections:
it is, in fact, independent of all parameters except the sum of all rates $\Gamma=\sum_{r\sigma} \Gamma_{r\sigma}$,
 even when the interaction is switched on ($U \neq 0$).
The second term describes the transient effect of the initial occupancies
on the correlations through $\Phi_\sigma(t_0) =\langle n_\sigma \rangle(t_0) -  1/2$.
This term decays to zero in an oscillatory fashion, for small temperatures with an exponential envelope
$
\propto
e^{-( {\Gamma_{\bar{\sigma}}/2 + \Gamma_\sigma } )(t-t_0)}
$.
Thus, besides the rates encountered so far,
there are two additional decay rates:
\begin{align}
  \tfrac{1}{2} \Gamma_{\bar{\sigma}} + \Gamma_\sigma \quad \text{for $\sigma=\uparrow, \downarrow$}
  .
  \label{3gamma}
\end{align}
When all tunnel rates are spin and reservoir independent and equal to $\Gamma_{r\sigma}=\tilde{\Gamma}$,
this rate reduces\footnote{For spin-independent tunneling $\Gamma_{r\sigma}=\tilde{\Gamma}$ 
this $3\tilde{\Gamma}$ decay rate already appeared as the imaginary part of one of the eigenvalues of the noninteracting Anderson model ($U=0$),
see Eq. (244) of \Cite{Saptsov12a}. This eigenvalue was found~\cite{Saptsov12a} to drop out of the calculation of the stationary current and the 
single-level
occupancies. Note that in \Cite{Saptsov12a} Eq. (244) does not follow from RG considerations, but rather directly from the renormalized perturbation 
theory, in particular Eq. (163) there. It also does not make use of the $T\rightarrow 0$ limit. See also \Sec{sec:oneloopself}.} 
to $\Gamma_\sigma+\tfrac 1 2 \Gamma_{\bar{\sigma}}=3\tilde{\Gamma}$,
in contrast to the other rates encountered so far,
$\Gamma=4\tilde{\Gamma}$,
$\Gamma_\sigma=2\tilde{\Gamma}$ 
or
$\tfrac{1}{2}\Gamma_\sigma=\tilde{\Gamma}$.
For this simple case this additional energy scale was noted in \Cite{Oguri13a}
and related to the rates of a virtually excited particle-hole pair and an incident particle. Our spin-resolved result \eq{3gamma}, written as
$( \Gamma_{\bar{\sigma}} + \Gamma_\sigma )/2 + \Gamma_\sigma/2$,
agrees with this.
Here, we find that this scale is related to decay of two-particle correlations arising from the initial spin-orbital occupations.

Finally, the coefficient $\vartheta(t-t_0)$ defines
the stationary value $\Xi(\infty)$.
It is defined exclusively by the stationary two-loop propagator \eq{4term}
and its nonzero stationary value \eq{varthetaresult}
factorizes into the stationary values $\Phi_\sigma(\infty)$ [see \Eq{av_n_2}]: 
\begin{align}
  \Xi(\infty) =  \tfrac{1}{2} \vartheta(\infty) = 2 \prod_{\sigma} \Phi_\sigma(\infty)
  .
  \label{xiphi}
\end{align}
This relation simply expresses that
the stationary nonequilibrium averages of the total fermion-parity operator factorize
into the averages of the fermion parity of the two spinorbitals:
\begin{align}
  \lim_{t\rightarrow \infty} \langle e^{i\pi n}\rangle (t) 
  =
  \lim_{t\rightarrow \infty} \prod_\sigma \langle e^{i\pi  n_\sigma }\rangle (t)
  .
  \label{spinparity}
\end{align}
Rewritten using $e^{i\pi  n_\sigma }= (1-2n_\sigma)$,
this is equivalent to the factorization of the correlator of the occupancies,
$  \lim_{t\rightarrow \infty} \langle n_\uparrow n_\downarrow \rangle (t)
 = \lim_{t\rightarrow \infty} \langle n_\uparrow \rangle(t) \cdot \langle n_\downarrow \rangle (t)$,
which is expected for the noninteracting limit ($U=0$):
each spin-$\sigma$ ``channel'' can be averaged independently, see \App{sec:factorize}.
This relation \emph{always} holds in the stationary limit:
the initial correlations between the orbital occupancies, expressed by the inequality \Eq{neqXi}, are ``forgotten'' in the long time limit.
For a two-particle correlated initial state on the quantum dot, satisfying \Eq{neqXi},
the relation \Eq{spinparity} is violated on time scales $t-t_0 \lesssim \Gamma^{-1}$.
This timescale is set by the fermion-parity protected eigenvalue, and is therefore independent of all other parameters in the problem,
even for nonzero \emph{interaction} $U$, as discussed in \Sec{sec:generalWBL}.
For an initially uncorrelated quantum dot 
the relation $\Xi(t)= 2\prod_\sigma \Phi_\sigma (t)$ or equivalently,
$\langle n_\uparrow\rangle(t) \langle n_\downarrow\rangle(t)=\langle n_\uparrow n_\downarrow\rangle(t)$
at $t=t_0$, continues to hold for all times $t \geq t_0$.

\refAB{Although the effects of the initial dot state for $U=0$ have been studied previously~\cite{Andergassen11a,Jin10,Komnik09,Schmidt08} all these works
 are based on spinless electrons or, equivalently, on a single spin-orbital model.
For the a single-spin orbital the electron-parity operator practically coincides with the level occupancy operator $n_\sigma$: $(-1)^{n_\sigma}\propto n_\sigma-1/2$.
Thus by the simplicity of this model the effect of the initial two-particle correlators $\langle n_\uparrow n_\downarrow \rangle$ cannot appear.
However, for electrons with spin it makes a crucial difference if the dot was initially prepared in the nonfactorizable form, \Eq{neqXi}, or not, as our result show.
We are not aware of any work presenting an exact analytical result for the nonfactorizing time evolution of this correlator in this limit.\cut{before.}
}

\subsection{Stationary limit \label{sec:stationary}}

We now illustrate how simplifications arise in Laplace space in the noninteracting limit
and directly calculate the exact $U=0$ \emph{stationary} density operator.
This independent calculation is also of more general interest since
it involves the direct, explicit calculation of the stationary self-energy $\Sigma(i0)$ 
and the effective Liouvillian $L(i0)=L+\Sigma(i0)$ for $U=0$
 using the renormalized perturbation theory but now formulated in Laplace space.
This independent result was used (but not derived) in \Cite{Saptsov12a}
to check that the real-time renormalization group explicitly recovers the noninteracting limit
for the stationary current and the corresponding self-energy parts [see \Eq{Sigma-r}].
Here, we provide the derivation
 of this important benchmark not only for the current, but also for the full density operator and the
self-energy.

\subsubsection{Frequency space perturbation expansion for the self-energy}
To keep the paper self-contained, we first briefly outline the renormalized perturbation theory in Laplace space which is also valid for $U \neq 0$.
Although this theory was formulated in \Cite{Saptsov12a}, it was not used to explicitly calculate the $U=0$ limit, but the RG approach was used instead
to analytically verify the current
and the relevant self-energy parts in this limit.
We proceed in close analogy to the above time-dependent formulation and start with the Laplace transform
of the (nonrenormalized) perturbation expansion \Eq{reduced-expansion-usual} for the time-evolution propagator.
Alternatively,~\cite{Saptsov12a} we transform the general solution \Eq{pit} and then expand  the resolvent $1/(z-L^{\mathrm{tot}})$ in powers of $L^V$ using $\Tr{R}L^R =0$:
\begin{align}
  \label{laplace_rhod}
  \Pi(z) 
  & := \int_{t_0}^{\infty} dt e^{i z t} \Pi(t,t_0) = \Tr R \frac {i}{z-L^{\mathrm{tot}}}
     \rho_R
    \\
  & = \sum_{k=0}^{\infty} \frac {i}{z-L} \Tr{R} \left( L^V \frac {1}{z-L^R-L}\right)^k
     \rho_R
     .
   \nonumber
\end{align}
Inserting \Eq{q-li} for $L^V$,
we have for the $m$th-order term of the expansion of the Laplace-space resolvent, $\Pi(z)=\sum_{m=0}^{\infty} \Pi_m(z)$
[cf. \Eq{mth-expansion}]:
\begin{align}
  &\Pi_m(z)
  =
  i 
  \Tr{R} \left (J^{q_m}_{\bar{m}}  ... J^{q_{1}}_{\bar{1}} \rho^{R} \right)
  \times
  \label{R_average}
  \\
  &\frac{1}{z-L}G^{q_m}_m \frac{1}{z-L-X_m} \ldots
  \frac{1}{z-L-X_{2}}G^{q_{1}}_{1} \frac{1}{z-L} 
  .
  \nonumber
\end{align}
Here, we have commuted all $J^q$ to the far right, using $[L^R,J^q_1]_-=\eta(\omega+\mu_r) J^q_1$
and collected all \emph{reservoir} energies $x_k= \eta_k (\omega_k+\mu_k)$
of the $G^{q_k}_k (J_{\bar{k}}^{q_k})$ originally standing to the left to the resolvent $i$
in $X_i=\sum_i^m x_k$.
Evaluating the reservoir average using the Wick theorem \eq{wick},
we obtain terms that can be represented by the same diagrams as in \Fig{fig:diagrams},
where the free propagators connecting vertices stand for $1/(z-L-X_i)$
and the contraction line connecting a pair of vertices $\bar{G}$-$\bar{G}$ and $\bar{G}$-$\tilde{G}$
 stands for the contraction function \eq{ret-contr} and \eq{keld-contr},
where the line is now assigned a frequency $x_k$.
The irreducible parts of these diagrams [cf. \Sec{sec:self-energy}]
 are collected into the superoperator self-energy $\Sigma (z)$,
which is just the Laplace transform of $\Sigma(t-t')$ in \Eq{dyson}:
\begin{align}
  \Sigma(z)=\int_0^{\infty} e^{i z\tau}\Sigma(\tau,0) d\tau
  .
\end{align}
After grouping the diagrams into $\Sigma(z)$ blocks, we can resum the resulting geometric series and obtain the Laplace-space solution of the Dyson equation \eq{dyson} / the kinetic equation \eq{effective-l}:
\begin{align}
  \label{rhoz}
  \rho(z)=\frac i {z-L(z)}\rho(t_0)
  ,
\end{align}
where the effective Liouvillian in the Laplace representation [the Laplace transform of \Eq{effective-bar}] is
decomposed as in \Eq{effective-bar}:~\cite{Schoeller09a}
\begin{align}
  L(z)=L+\Sigma(z)
  .
\end{align}
The required expansion for $\Sigma(z)$ thus has the form:
\begin{align}
  \label{sig_irr} 
  & \Sigma (z) = (-1)^{P} \left(\prod \gamma \right)_{\mathrm{irr}}
  \times
  \\
  & \bar{G}_m\frac{1}{z-X_m-L} ...
  ...{G}^{q_{2}}\frac{1}{z-X_{2}-L} G^{q_{1}}
  .
  \nonumber
\end{align}
The causal structure of this expansion [cf. discussion of \Eq{pim}]
enforces that the leftmost vertex in each contraction is always a \emph{creation} superoperator $\bar{G}_m$ as a consequence of \Eq{trace}.
We use the same conventions as in \Eqs{pim} and \eq{mth-expansion}, suppressing all sums and integrations over reservoir frequencies.
Note that unlike in the time representation, these integrals cannot be pulled into the contraction functions, i.e., 
the
 $\gamma^q_{i,j}$ are given by \Eqs{ret-contr} and \eq{keld-contr}.
This is because 
in
 the Laplace transform the frequencies are convoluted with the dot evolution in the propagators $(z-L+X)^{-1}$.
The main advantage of \Eq{sig_irr} is that we can now directly work in the stationary limit by taking the limit $z \rightarrow i0$
and then calculating the stationary state by finding the right zero eigenvector of $L(i0)=L+\Sigma(i0)$.

Similar to \Sec{sec:eliminate} we explicitly work out the wide-band limit by noting that diagrams with a vertex inside a $\tilde{\gamma}$ contraction can be neglected,~\cite{Saptsov12a}
and that one can integrate out 
the
 $\tilde{\gamma}$ contractions and the $\tilde{G}$ vertices, incorporating them into the Laplace transform $\tilde{\Sigma}$ of the 
 self-energy $\tilde{\Sigma}(t,t')$ [\Eq{tildesigma}],
which is simply the time-independent factor given by \Eq{Sigmatilde}.
What remains is to evaluate the perturbative expansion for $\bar{\Sigma}(z)=\Sigma(z)-\tilde{\Sigma}$, with the simplified diagram rules schematized by
\begin{align}
  \label{barSigma}
  \bar{\Sigma}(z)
   =
   &(-1)^P\left(\prod_i\bar{\gamma}_i\right)_{\mathrm{irr}}
  \times
  \\
  &
  \bar{G}_m \frac{1}{z-X_m-\bar{L}} ... \frac{1}{z-X_{2}-\bar{L}}\bar{G}_1,
  \nonumber
\end{align}
where we sum 
over
 irreducible $\bar{\gamma}$ contractions and $\bar{L}=L+\tilde{\Sigma}$ as before [\Eq{barl}].
We can now find the stationary state as the right zero eigenvector of $L(i0)=L+\Sigma(i0)=\bar{L}+
\bar{\Sigma}(i0)
$.
Notably, in \Eq{barSigma} we can \emph{set} $z=0$ since $\bar{L}$ is a \emph{dissipative} Liouvillian which automatically regularizes all propagators.
\footnote{
  That, indeed, no zero can appear in the denominator  is related to the causal structure \Eq{barSigma}
  and to the stationary properties of the infinite-temperature self-energy discussed in \Sec{sec:infinite}.
  By the causal property \Eq{trace}, the creation superoperators $\bar{G}_1$ have a left zero eigenvector $(Z_L|$
  which is \emph{unique}, as mentioned in \Sec{sec:generalWBL}.
  At the same time, $|Z_L)$, the stationary $T \rightarrow \infty$  density operator, is the \emph{unique} right zero eigenvector of $\bar{L}=L+\bar{\Sigma}$,
  even when accounting for the both the bosonic and the fermionic diagonal blocks.~\cite{Saptsov12a}
  Therefore, when inserting a complete set of superkets in the superoperator expressions $(z-\bar{L}+X_k)^{-1} \bar{G}_1$
  the only zero of $\bar{L}$, occurring for $|Z_L)$, is canceled by the vanishing of the projection $(Z_L|\bar{G}_1$.
  Finally, one could worry that insertion of $|Z_L)(Z_L|$ could generate a zero in the leftmost expression $\bar{G}_1 (z-\bar{L}+X_k)^{-1}$ in \Eq{barSigma},
  since for the $T \rightarrow \infty$ limit the superadjoint $|Z_L)$ is the corresponding right zero eigenvector of the effective Liouvillian $\bar{L}$
  [cf. discussion in \Sec{sec:infinite}].
  However, this zero is canceled as well by the adjacent creation superoperator on the right:
  $\bar{G}_1 |Z_L)(Z_L| (z-\bar{L}+X_k)^{-1} \bar{G}_2= \bar{G}_1 (z + X_k)^{-1}  |Z_L)(Z_L| \bar{G}_2= 0$.
}
We note how these technical properties neatly tie in with the physical meaning.
That the denominators in \Eq{barSigma} contain no zeros was a key point in setting up the real-time renormalization group flow in Laplace space for the calculation of effective Liouvillians,
referred to as 
the
``zero-eigenvector problem''.\cite{Korb07,Schoeller09a}

\subsubsection{Noninteracting limit and super-Pauli principle\label{sec:Pauli2}}
We now work out the simplifications that occur for the self-energy in the limit $U=0$, in full analogy with the time-representation.
The part $\tilde{\Sigma}$ is obtained by setting $U=0$ in \Eq{Sigmatilde} or \eq{sigmatilde_basis}.
In the remaining calculation of $\bar{\Sigma}$, [\Eq{barSigma}], using \Eq{barLbarGcom} we commute the $\bar{G}$ through the resolvents:
this gives the analog of \Eq{mth-expansion-bar-non}:
\begin{align}
  &
  \bar{\Sigma}(z) =(-1)^P\left(\prod_i\bar{\gamma}_i\right)_{\mathrm{irr}}
  \bar{G}_m ...\bar{G}_1\times
  \label{barG_left}
  \\
  &
  \frac{1}{z-X_m-E_m-\bar{L}} ... \frac{1}{z-X_{2}-E_{2} - \bar{L}},
  \nonumber
\end{align}
where
$\bar{L}$ is given by \Eq{barL_basis} with $U=0$
and
$E_i=\sum_{k=2}^i \epsilon_k$ is the sum over renormalized, single-particle \emph{quantum dot} energies
\begin{align}
  \epsilon_i = \eta_i\epsilon_{\sigma_i}-i\tfrac 1 2 \Gamma_{\sigma_i}.
\end{align}
These have acquired an imaginary part by the inclusion of broadening of the $T \rightarrow \infty$ limit through the self-energy term $\tilde{\Sigma}$ in the renormalized Liouvillian $\bar{L}$.
Thus, also in the Laplace representation, the super-Pauli principle \eq{superpauli} directly reveals   
that in the $U=0$ limit nonzero terms  of the renormalized self-energy of order $m>4$ [cf. \Eq{superpauli2}] 
cannot appear:
$\bar{\Sigma}(z)  =  \bar{\Sigma}_2(z) +\bar{\Sigma}_4(z)$, cf. \Eq{sigmatrunc}.

\paragraph{One loop self-energy: stationary occupancies and current.\label{sec:oneloopself}}
We first calculate the one-loop diagram for $\bar{\Sigma}_2$
by substituting \Eq{GG} into
\begin{align}
  &  \bar{\Sigma}_2(i0)
  =
  \sum_{\eta,\sigma,r}
  \bar{G}_2 \bar{G}_{\bar{2}}
  \int \frac {d\bar{\omega} \bar{\gamma}(\bar{\omega})}{i0-\bar{\omega}-\bar{\mu}_2-\epsilon_{\bar{2}}-\bar{L}}
  \nonumber
  \\
  & =
    \sum_{\eta,\sigma,r} \eta   \frac{\Gamma_{r \sigma}}{2\pi}
    \left(
      |\chi_\sigma)(Z_L| 
      \int \frac {d\bar{\omega} \tanh(\bar{\omega}) }{i0-\bar{\omega}-\bar{\mu}_2-\epsilon_{\bar{2}}-\bar{L}}
      +
    \right.
    \nonumber
    \\
    &
    \, \, \quad\quad\quad
    \left.
      |Z_R)(\chi_{\bar{\sigma}}|
      \int \frac {d\bar{\omega} \tanh (\bar{\omega}) }{i0-\bar{\omega}-\bar{\mu}_2-\epsilon_{\bar{2}}-\bar{L}}
    \right)  + F
    .
\end{align}
Here and in the following,
we abbreviate $\bar{\omega}_i\equiv\eta_i\omega_i=x$
and $\bar{\mu}_i := \eta_i \mu_{r_i}$
and we write this in the form (omitting the fermionic part)
\begin{align}
  \bar{\Sigma}_2(i0) =   -i \sum_\sigma \left\{ \psi_\sigma |\chi_\sigma)(Z_L|+\phi_\sigma |Z_R)(\chi_\sigma| \right\}
  .
\end{align}
Using $(Z_L|\bar{L}=0$ and summing over $\eta$ explicitly,
this gives after replacing $x\rightarrow -x$ in the second integral 
the stationary value of the first two coefficients 
(see \App{sec:digamma_app}):
\begin{align}
  \psi_\sigma  &= \HL{-i}\sum_{r}\frac {\Gamma_{r\sigma}}{2\pi}\int dx\left[ \frac {\tanh(x/2T)}{i\tfrac 1 2 \Gamma_\sigma-x+\epsilon_{r\sigma}}
    -\frac {\tanh(x/2T)}{i\tfrac 1 2 \Gamma_\sigma-x-\epsilon_{r\sigma}}
  \right]
  \nonumber
  \\ 
  &=-\sum_r \frac{2\Gamma_{r\sigma}}{\pi} \Im \Psi \left(\frac 1 2 +\frac {\tfrac 1 2 \Gamma_\sigma-i\epsilon_{r\sigma}}{2\pi T} \right)
  ,
  \label{psiio}
\end{align}
where, [cf. \Eq{ers}]
\begin{align}
  \epsilon_{r\sigma}=\epsilon+\sigma B/2 - \mu_r
  .
\end{align}
We will see that 
the coefficients \eq{psiio}
of the \emph{one-loop} self-energy $\bar{\Sigma}_2(i0)$
completely determine the stationary current.
As a result,
 the current only shows the usual broadening $\Gamma_\sigma/2 = \sum_r \Gamma_{r\sigma}/2$
due to the cumulative relaxation rate for spin $\sigma$.

Analogously, we calculate the remaining coefficients $\phi_\sigma$ $(\chi_\sigma|\bar{L}=-2i\Gamma(\chi_\sigma|$, cf. \Eq{barL_chi}:
\begin{align}
\phi_{\sigma}  & =\HL{-i}
  \sum_r\frac{\Gamma_{r\bar{\sigma}}}{2\pi}\int dx
  \left[
    \frac {\tanh(x/2T)}{i(\tfrac 1 2\Gamma_{\bar{\sigma}}+\Gamma_\sigma)-x+\epsilon_{r\bar{\sigma}}}
  \right.
  \nonumber
  \\
  &
  \quad\quad\quad\quad\quad
  \left.
    - \frac {\tanh(x/2T)}{i(\tfrac 1 2\Gamma_{\bar{\sigma}}+\Gamma_\sigma)-x-\epsilon_{r\bar{\sigma}}}
  \right]
  \nonumber
  \\ 
  &=
  - \sum_r \frac{2\Gamma_{r\bar{\sigma}}}{\pi} \Im \Psi \left(\frac{1}{2}+\frac{\tfrac 1 2 \Gamma_{\bar{\sigma}}+\Gamma_\sigma-i \epsilon_{r\bar{\sigma}}}{2\pi T}\right)
  .
  \label{phi-digamma}
\end{align}
This part of the self-energy involves a quite
different broadening, $\Gamma_\sigma+\tfrac 1 2 \Gamma_{\bar{\sigma}}$ instead of $\tfrac 1 2\Gamma_\sigma$,
an energy scale
 that we noted earlier in \Eq{3gamma}.
However, we will see that the exact $U=0$ stationary current is not sensitive to this quantity.
We note that for $U \neq 0$, it can be shown that this broadening, although modified by the interaction, does enter into the stationary current.
\footnote{
  That the rate $3\tilde{\Gamma}$ drops out for $U=0$ was shown in \Cite{Saptsov12a}, Sec. III.C.3,
  due to the vanishing of a propagator supermatrix element, Eq. (246) of that reference.
  For $U \neq 0$, this matrix element does not vanish,
  and as a result the renormalization group flow is affected by this decay rate.
}

\paragraph{Two loop self-energy: stationary average fermion-parity.}
The two-loop contribution to $\bar{\Sigma}$ is calculated in close analogy to the two-loop contributions to $\bar{\Pi}_4$ in \Eq{formal-2loop}:
\begin{widetext}
\begin{subequations}
\begin{align}
  &\bar{\Sigma}_{4}(i0)
   =
  \sum_{\sigma,r,r^{'}}\frac{\Gamma_{r\sigma}\Gamma_{r^{'}\bar{\sigma}}}{(2\pi)^2}\bar{G}_l\bar{G}_{\bar{l}}\bar{G}_{l'}\bar{G}_{\bar{l}^{'}}
  \int d\bar{\omega}_l\int d\bar{\omega}_{l'}
  \frac{\tanh(\bar{\omega}_l/2T) \tanh(\bar{\omega}_{l'}/2T) }{( \eta_l\epsilon_{r\sigma}+i(\frac{\Gamma_\sigma}{2}+\Gamma_{\bar{\sigma}})-\bar{\omega}_l )} 
  \times
  \label{2loopPT-Xst} 
  \\
  &
  \left(
  \frac{1}{
    ( \eta_{l'}\epsilon_{r' \bar{\sigma}}+i\frac{\Gamma_{\bar{\sigma}}}2-\bar{\omega}_{l'} )
  }
  \frac{1}{
    ( \eta_l\epsilon_{r\sigma}+\eta_{l'}\epsilon_{r' \bar{\sigma}}+i\frac{\Gamma}{2} -\bar{\omega}_l-\bar{\omega}_{l'} )
  }
  +
  \frac{1}{
    (\eta_l\epsilon_{r{\sigma}}+i\frac{\Gamma_\sigma}2-\bar{\omega}_l )
  }
  \frac{1}{
    ( \eta_l\epsilon_{r\sigma}+\eta_{l'}\epsilon_{r'\bar{\sigma}}+i\frac{\Gamma} 2 -\bar{\omega}_l-\bar{\omega}_{l'} )
  }
  \right)
  \nonumber
  \\
  &=
  \sum_{\sigma,r,r^{'}}\frac{\Gamma_{r\sigma}\Gamma_{r^{'}\bar{\sigma}}}{(2\pi)^2}\bar{G}_l\bar{G}_{\bar{l}}\bar{G}_{l'}\bar{G}_{\bar{l}^{'}}
  \int d\bar{\omega}_l
  \frac{\tanh(\bar{\omega}_l/2T)}{ ( \eta_l\epsilon_{r\sigma}+i(\frac{\Gamma_\sigma} 2 +\Gamma_{\bar{\sigma}})-\bar{\omega}_l )} 
  \frac{1}{(\eta_l\epsilon_{r{\sigma}}+i\frac{\Gamma_\sigma} 2-\bar{\omega}_l )} 
  \int d\bar{\omega}_{l'} 
  \frac{\tanh(\bar{\omega}_{l'}/2T)}{ ( \eta_{l'}\epsilon_{r' \bar{\sigma}}+i\frac{\Gamma_{\bar{\sigma}}} 2 -\bar{\omega}_{l'} )}
  \label{2loopPT-gamma} 
\end{align}
\end{subequations}
\end{widetext}
Here, we made
 use of the fact that for $\bar{\gamma}_{ij}$ the multi-indices are related as $i=\bar{j}$, and 
we relabeled 
$4\rightarrow l=\omega_l,\eta_l,r, \sigma$
and
$2\rightarrow l'=\omega_{l'},\eta_{l'},r', \bar{\sigma}$ for compactness.
We also explicitly made use of relation $\sigma_l=\sigma=\bar{\sigma}_{l'}$,
which expresses the fact that in the product $\bar{G}_l\bar{G}_{\bar{l}}\bar{G}_{l'}\bar{G}_{\bar{l}^{'}}$ all operators must be different
[otherwise the product is zero due to the super-Pauli principle \eq{superpauli}\HL{-\eq{superpauli2}}].
Equation \eq{2loopPT-Xst} has two types of contributions, represented by
the irreducible parts of the last two diagrams in \Fig{fig:diagramsren}.
To obtain the form \eq{2loopPT-Xst} we have anticommuted the field superoperators
to have the same order in each type of contribution.
This cancels the Wick sign,
similar to the rewriting of \Eq{formal-2loop} into \Eq{formal-2loop2}, cf. \Fig{fig:factorize} and \App{sec:twoloop_ap}.
Notably, when combining these two types of contributions in \Eq{2loopPT-gamma}, the double integral over frequencies $\bar{\omega}_{l}$ and $\bar{\omega}_{l'}$ becomes factorizable \emph{but only for zero} quantum-dot \emph{frequency $z=0$}.
Therefore the complete expression \eq{2loopPT-gamma} is a sum over factorizable integrals.
As in \Eqs{formal-2loop} and \eq{4term}, the superoperator structure of \eq{2loopPT-gamma} is very simple:
\begin{align}
  \bar{\Sigma}_4(i0) = \zeta |Z_R)(Z_L|
  .
  \label{4termsigma}
\end{align}
Using
$\bar{G}_l\bar{G}_{\bar{l}}\bar{G}_{l'}\bar{G}_{\bar{l}'}=\eta_l\eta_{l'}|Z_R)(Z_L|$
and \Eq{dig-sum} [\App{sec:digamma_app}] we obtain:
\begin{align}
  \label{zeta}
   &\zeta
    =
   \sum_{\sigma,r,r'}\frac{ 4\Gamma_{r\sigma} \Gamma_{r^{'}\bar{\sigma}}}{\pi^2}\times
   \\ 
   &\left[
     \Im \Psi\left(\frac 1 2 + \frac {\tfrac 1 2 \Gamma_\sigma+\Gamma_{\bar{\sigma}}-i\epsilon_{r\sigma}}{2\pi T} \right)
     \Im \Psi\left(\frac 1 2 + \frac {\tfrac 1 2 \Gamma_{\bar{\sigma}}-i\epsilon_{r' \bar{\sigma}}}{2\pi T} \right)
   \right.
   \nonumber
   \\
   &\left.
     -\Im \Psi\left(\frac 1 2 + \frac {\tfrac 1 2 \Gamma_\sigma-i\epsilon_{r\sigma}}{2\pi T} \right)
       \Im  \Psi\left(\frac 1 2 + \frac {\tfrac 1 2\Gamma_{\bar{\sigma}}-i\epsilon_{r' \bar{\sigma}}}{2\pi T} \right)
   \right]
   .
   \nonumber
 \end{align}

Summarizing, the exact zero-frequency effective dot Liouvillian in the wide-band limit
for the noninteracting Anderson model ($U=0$) is
\begin{align}
  &i L( i0)
  := i \big[ L + \tilde{\Sigma} + \bar{\Sigma}_2(i0) + \bar{\Sigma}_4(i0) \big]
  =
  \nonumber
  \\
  &
  \Gamma|Z_R)(Z_R|   +  \zeta|Z_R)(Z_L|
  +
  \label{Li0}
  \\
  &
  \sum_\sigma
  \Big[
    \phi_\sigma |Z_{R})(\chi_\sigma| +
    \psi_\sigma |\chi_\sigma)(Z_{L}|
  \Big]
  +
  \sum_\sigma\Gamma_\sigma|\chi_\sigma)(\chi_\sigma|
  +
  \nonumber
  \\
  &
  \sum_\sigma
  \Big[ i\sigma B    + \tfrac{1}{2} \Gamma \Big]
  |S_\sigma)(S_\sigma|
  +
  \sum_\eta  
  \Big[ i \eta ( 2\epsilon+U ) + \tfrac{1}{2} \Gamma \Big]
  |T_\eta)(T_\eta|
   ,
  \nonumber
\end{align}
 with $\psi_\sigma$ given by \Eq{psiio},
$\phi_\sigma$ given by \Eq{phi-digamma},
$\zeta$ given by \Eq{zeta} and not writing the irrelevant fermionic part.

\subsubsection{Stationary density operator}

The stationary state, the unique right zero eigenvector of  $L(i0)$,  can be determined conveniently using the form \eq{Li0} of the superoperator.
In fact,  the complete eigenspectrum of $L( i0)$, can be found in terms of the its coefficients:~\cite{Saptsov12a}
generally, the effective Liouvillian written in the basis \Eqs{basis-boson}
 must have the form (ignoring the fermionic part again):
\begin{align}
  \label{obliv}
  &i L(i0)  =  \Gamma |Z_{R})(Z_{R}|  + \zeta |Z_{R})(Z_{L}| +
  \\
  \nonumber
  &
  \sum_\sigma \Big[
  \phi_\sigma |Z_{R})(\chi_\sigma| +
  \psi_\sigma |\chi_\sigma)(Z_{L}|
   \Big]
  + \sum_{\sigma,\sigma} \xi_{\sigma,\sigma'} |\chi_\sigma)(\chi_{\sigma'}|
  +
  \nonumber
  \\
  &
  \sum_\sigma
  E_\sigma |S_\sigma)(S_\sigma|
  +
  \sum_\eta
  M_\eta |T_\eta)(T_\eta|
  .
  \nonumber
\end{align}
 This form follows from
the causal structure,
the wide-band limit (fixing the first coefficient to the constant, fermion-parity eigenvalue $-i \Gamma$, cf. \Sec{sec:generalWBL})
\footnote{
  The limit considered in \Cite{Saptsov12a} corresponds to the case: $\Gamma_{r\sigma}=\tilde{\Gamma}$ for all $r,\sigma$.
  The coefficient before the first term in \Eq{obliv} is then $\Gamma=\sum_{r,\sigma}\Gamma_{r\sigma}=4\tilde{\Gamma}$.
  This $\Gamma$ [from \eq{obliv}] should not be confused with one used in \Cite{Saptsov12a},
  the latter corresponds to the $\tilde{\Gamma}$ in the present notations.
}
and the symmetries of the Anderson model. (Notably, it does \emph{not} require $U=0$.)
As in \Cite{Saptsov12a}, we have conservation of the charge and of the spin along the direction of the magnetic field $B$,
even though we now also include spin-dependent tunneling
(we assume that the magnetic field $B$ and the polarization vectors are collinear,
 preserving the spin-rotation symmetry about this axes).
The stationary state can be expressed in terms of the effective Liouvillian
coefficients~\cite{Saptsov12a} and by comparing \Eq{Li0} with the general form \Eq{obliv},
we obtain in terms of the matrix $\xi_{\sigma,\sigma'} = \Gamma_\sigma \delta_{\sigma,\sigma'}$,
and the vectors [\Eq{psiio} and \Eq{phi-digamma}]:
\begin{subequations}
\begin{align}
  \rho(\infty)
  =&
  \tfrac{1}{2}|Z_{L})
  -\tfrac{1}{2}\sum_{\sigma,\sigma'} \xi^{-1}_{\sigma,\sigma'} \psi_{\sigma'} |\chi_\sigma)
  \nonumber
  \\
  & +\tfrac{1}{2\Gamma}
  ( \sum_{\sigma,\sigma'} \phi_\sigma \xi^{-1}_{\sigma,\sigma'} \psi_{\sigma'}  -\zeta )
  |Z_{R})
  \label{rho}
  \\
  =&
  \tfrac{1}{2}|Z_{L})
  +\sum_\sigma \Phi_\sigma(\infty) |\chi_\sigma)
  + \Xi(\infty) |Z_{R})
  ,
  \label{rho2}
\end{align}
\end{subequations}
where the expansion coefficients are
\begin{align}
  \Phi_\sigma(\infty)
  &=
  \frac{1}{\pi}\sum_{r}\frac{\Gamma_{r\sigma}}{\Gamma_\sigma} \Im
  \Psi \left(\frac 1 2 +\frac {\tfrac{1}{2} \Gamma_\sigma -i\epsilon_{r\sigma}}{2\pi T} \right),
  \label{Phiresult}
  \\
  \Xi(\infty)
  &=  2\Phi_\uparrow(\infty) \cdot \Phi_\downarrow(\infty)
  .
  \label{xiphi2}
\end{align}
We note that the additional broadening scale $\tfrac 1 2 \Gamma_\sigma+\Gamma_{\bar{\sigma}}$ that we noted already in \Eqs{3gamma} and \eq{phi-digamma}
is also  present in $\zeta$ [\Eq{zeta}]
but drops out in the calculation of $\Xi(\infty)$ when one sums over $\sigma$.
Equation \eq{Phiresult} reproduces the stationary spin-orbital occupancies \Eq{av_n_2} through $\braket{n_\sigma} = \tfrac{1}{2} + \Phi_\sigma$.
Moreover, \Eq{xiphi2} confirms 
the factorization \eq{xiphi} of the \emph{stationary} value $\Xi(\infty)$ into the coefficients $\Phi_\sigma(\infty)$.
On the superoperator level, the renormalized two-loop self-energy $\bar{\Sigma}_4$ does not factorize for \emph{any} frequency $z$,
not even at $z=0$, see \Eqs{4termsigma} and \eq{zeta}.
One can verify that to achieve such a factorization
a reducible term needs to be added to $\bar{\Sigma}_4(t,t_0)$.
This is precisely what happens in \Eq{barpi4} and produces essentially the Laplace transform of superoperator $\bar{\Pi}_4(t,t_0)$,  which indeed 
factorizes at any time $t$ by \Eq{formal-2loop2}.
\hl{This shows that for $\bar{\Sigma}_4(z)$ itself no such factorization is to be expected at any frequency.
There are thus advantages of working directly with full propagators in 
time-space in comparison with working with self-energies in Laplace-space,
when considering the noninteracting limit of the Anderson model and its generalizations,
Furthermore, \Eqs{rho2} and \eq{xiphi2} show the physical importance of the stationary two-loop self-energy superoperator $\bar{\Sigma}_4(i0)$, i.e.,
 the \emph{quartic} term \eq{2loopPT-gamma} appearing in the effective theory despite the absence of two-particle interactions ($U=0$).
To obtain the correct two-particle correlations for $U=0$ in the stationary limit
it is crucial to calculate both the one- and two-loop self-energies, i.e.,
 to work with the effective Liouvillian which is quartic in the fields.}

Finally, we note that in the expansion \eq{rho2} of the \emph{stationary} state, the terms containing superkets
$|S_\sigma)$ and $|T_\eta)$,
describing spin- and electron-pair coherence [cf. \Sec{sec:examples}],
do not appear since they are forbidden by charge and spin rotation symmetry.
If such coherences are prepared in the initial state, they must decay to zero in time,
in agreement with the central result \Eq{exact-RDDO}.
\subsection{Time-dependent current\label{sec:current-time}}
In this last section of the paper
we illustrate that in the calculation of \emph{observable} averages
very similar simplifications can be made using the causal field superoperators.
We focus on the example of the time-dependent charge current
in the noninteracting limit $U=0$.

\subsubsection{Current self-energy}
We first present considerations which apply generally, i.e., to the interacting Anderson model ($U\neq0$),
and, in fact, to multiorbital generalizations.
Generally, an observable $A$ that is not local to the dot requires the calculation of
an additional self-energy $\Sigma_A$ with its own real-time diagrammatic expansion.\cite{Schoeller09a}
However, for a quantity which is conserved in the tunneling, such as the current $I^r$ into reservoir $r$, this is not necessary:
it can be obtained by simply keeping track of the part of the self-energy that is related to reservoir $r$.
That is, we decompose $\Sigma = \sum_r \Sigma^r$ by
splitting up the interaction $L^V=\sum_r L^{V^r} $ into $r$-contributions $L^{V^r}=[V^r,\bullet]_-$ [cf. \Eq{Vdef}]
at the \emph{latest time} $t_m$ in each term of the perturbation series [cf. \Eq{pt_lv}].
Then, by rewriting $I^r=-i[H^{\mathrm{tot}},n^r]=-i[V^r,n^r]=-iL^{V^r}n^r$ with fixed $r$ one finds the relation~\cite{Saptsov12a}
\begin{subequations}
  \begin{align}
    \langle I^{r}(t)\rangle
    & = - \tfrac{i}{2} \Tr D L^{n+} \Tr R L^{V^r} \rho^{\mathrm{tot}} (t)
    \\
    & = -\tfrac{i}{2} \Tr D L^{n+} \int_{t_0}^t dt' \Sigma^r(t,t') \rho(t')
    ,
    \label{Ir}
  \end{align}
\end{subequations}
where $L^{n+}=[n,\bullet]_+$ is the \emph{anti}commutator with the dot particle number operator $n$.
Besides the computational simplification, extended here to the time-dependent case,
this result can be used to show very easily~\cite{Saptsov12a} nonperturbatively that the \emph{stationary} current at zero bias voltage is always zero
(as it should be), something which is not always obvious. 
\HL{This is an obvious physical requirement.
However, within the real-time approach (or its equivalent, the Nakajima-Zwanzig approach), designed to deal with strongly interacting models, it is not obvious 
how to verify explicitly that in general this is actually the case, in particular, when going to higher orders in the perturbation theory in $\Gamma$ or when making nonperturbative approximations in this framework, as, for instance, in \Cite{Saptsov12a}.
When properly done, concrete calculations of this type always seem to comply with zero current at zero bias, but why this is so in general has not been clarified before. We found that \Eq{Ir} provides this key step in explicitly verifying this physical requirement.}
We can now again take advantage of the causal structure by decomposing $\Sigma^r=\tilde{\Sigma}^r+\bar{\Sigma}^r$
into the $T \rightarrow \infty$ part  $\bar{\Sigma}^r$ and the finite temperature corrections $\bar{\Sigma}^r$.
The only difference with the analysis in \Sec{sec:infiniteT} is that one simply does not sum over the reservoir index $r$
of the contraction with the latest field superoperator $\bar{G}_m$ in the $m/2$-loop renormalized perturbation expansion for $\bar{\Sigma}$ discussed in \Sec{sec:self-energy}.
To make progress, we first expand the dual supervector $\tr_D \tfrac{1}{2} L^{n+}=\tfrac{1}{2} (Z_L|L^{n+}$,
appearing in \Eq{Ir},
  in the dual Liouville-Fock basis \eq{basis-boson}:
\begin{align}
  \Tr D \tfrac{1}{2} L^{n+} \bullet
  = \sum_\sigma \Tr D ( n_\sigma\bullet )
  = \sum_\sigma (\chi_\sigma| + 2(Z_L|
  .
\end{align}
When this is inserted into \Eq{Ir}, it follows from the causal structure of the perturbation theory, $(Z_L|\Sigma^r \propto \tr_D \Sigma^r= 0$ (\emph{not} from probability conservation)
\footnote{
  $\tr_D \Sigma^r= 0$ does \emph{not} follow from probability conservation,
  which only requires $\tr_{D} \sum_r \Sigma^r = 0$.
  The causal structure is a stronger constraint.
}
that the current is a sum of projections onto two \emph{doubly excited} superkets $|\chi_\sigma)$ [cf \Eq{chichi}]:
\begin{align}
  \langle I^r (t)\rangle
  &=
  -i \sum\limits_\sigma (\chi_\sigma| \Big[ \tilde{\Sigma}^r(t,t')+\bar{\Sigma}^r(t,t') \Big] \rho(t')
  \nonumber
  \\
  &=\tilde{I}^r(t)+ \bar{I}^r(t)
  .
  \label{cur-fin}
\end{align}

\subsubsection{Wideband limit and artifacts}
The expression for $\bar{I}^r(t)$ is, in general, rather complicated since it requires the nontrivial part of the self-energy $\bar{\Sigma}^r$.
However, in the wideband limit $\tilde{\Sigma}^r =-i \sum_{\sigma} \frac {1}{2} \Gamma_{r\sigma} \bar{G}_1\tilde{G}_{\bar{1}}\bar{\delta}(t-t')$ 
is just \Eq{tildesigma} without the sum over $r$. The superoperator expansion of $\tilde{\Sigma}^r$ is given then by \Eq{sigmatilde_basis},
where one has to replace $\Gamma_\sigma\rightarrow\Gamma_{r\sigma}$, $\Gamma\rightarrow \sum\limits_{\sigma}\Gamma_{r\sigma}.$
For $\tilde{I}^r$ this gives (compare with the Green's function result of \Cite{Jauho94} and Eq. (30) in \Cite{Schmidt08}):
\begin{subequations}
\label{tildeI}
  \begin{align}
    \tilde{I}^r(t) & 
    =-i\sum\limits_{\sigma}(\chi_\sigma|\int\limits_{t_0}^t dt' \tilde{\Sigma}^r(t,t')\rho(t')
    \label{Itildeint}
    \\
    & =-\sum_\sigma {\Gamma_{r\sigma}} \Phi_\sigma(t).
    \label{Itilde}
  \end{align}
\end{subequations}
This only depends on the coefficients $\Phi_\sigma(t)=\langle n_\sigma \rangle(t)-1/2$ of the density operator \eq{general-initial-dot},
i.e., the deviations of the average spin-orbital occupations from the stationary, $T \rightarrow \infty$ values.
In the $T \rightarrow \infty$ limit the current is given by the contribution \eq{Itilde} alone,
when substituting for $\Phi_\sigma(t)$ the value
 $\lim_{T \rightarrow \infty} \Phi_\sigma(t)=e^{-\Gamma_\sigma(t-t_0)}\Phi_\sigma(t_0)$ [cf. \Eq{InfTrelax}].
In the stationary limit, this gives a vanishing current,
which is as it should be since the bias voltage is dominated by thermal fluctuations.
Note that this holds generally for $U \neq 0$ and nonperturbatively in $\Gamma$, as in \Sec{sec:infinite}.

%
% Artifact: the problem
%
For finite temperature, the current requires the calculation of $\bar{I}^r(t)$
but also $\tilde{I}^r(t)$  changes since $\Phi_\sigma(t)$ takes another value for finite $T$ [cf. \Eq{Itilde}], also requiring a calculation.\footnote{
  Note that although $\tilde{\Sigma}^r(t,t')$ is the (reservoir-resolved) $T \rightarrow \infty$ self-energy,
  $\tilde{I}^r(t)$ is \emph{not} simply the $T \rightarrow \infty$ current.
}
Before we turn to this,
we note that the wideband limit result for \eq{Itilde} has the disconcerting property that at the initial time $t_0$
it can give rise to a nonzero total current for a general initial condition $\Phi_\sigma(t_0)$.
This is again clear in the $T \rightarrow \infty$ limit mentioned above:
$\lim_{T \rightarrow \infty} I^r(t_0) = \lim_{T \rightarrow \infty} \tilde{I}^r(t_0) = \Phi_\sigma(t_0)$
yields a nonzero value.
In the next section, we explicitly show that for the $U=0$ limit this also occurs at finite temperature.

This nonzero current is inconsistent with our initial assumption that the dot and reservoirs are decoupled at $t=t_0$ and is unphysical.
This is an artifact of the wideband limit and raises the question on which time-scale \eq{Itilde} is correct.
For the noninteracting\footnote{In \Cite{Schmidt08}, also the first-order expansion in $U$ as well as Monte Carlo simulations
for an arbitrary $U$ were discussed.} case $U=0$ this has  been discussed, e.g., in \Cites{Maciejko06, Schmidt08, Jin10} 
and it was shown by explicit calculation for a finite bandwidth $D$ that the current starts from zero at $t=t_0$ as it should,
but then on the time-scale set by the inverse band width, $1/D$, the result rapidly approaches the wideband limit result.

%
% Artifact: the solution
%
The use of causal field superoperators allows us to generalize this qualitative understanding to the interacting case ($U \neq 0$)
since the effect of the wideband limit can be traced explicitly on the level of superoperators.
The effect is twofold:
first, as explained in \Sec{sec:widebandlimit}, the $\delta$-function constraint on time integrations arising from the large bandwidth energy
forbids diagrams in which a retarded contraction crosses with any other contraction line.
Since experimentally the bandwidth $D$ is usually much larger than any characteristic energy of the dot, temperature or transport bias
this should be a good approximation.
Second,  the time-dependent retarded contraction \eq{time-ret},
retained in $\tilde{\Sigma}(t,t')$,
is basically the Fourier transform of the function $\Gamma_{1,\omega_1}$,
$
\tilde{\gamma}_{1,{2}}(t,t')
\propto
\int d \omega_{1} \Gamma_{1,\omega_1} e^{-i\eta_1 (\omega_{1}+\mu_{r_1}) (t-t')}
$.
For large but finite bandwidth $D$
this therefore has a characteristic finite time support of order $1/D$.
If we now correct for this in \Eq{Itildeint},
then the integral over $t'$  starts from zero at $t_0$, as it should,
and only on the time scale $1/D$ does it acquires the wideband limit value given by \Eq{Itilde}.
Note however, that the times on which this difference is noticeable is below femtoseconds for the typical energies $D \sim 1.0 $~eV,
which seems to be far below realistic time scales of switching on the tunnel coupling $\Gamma$ at $t_0$.

\subsubsection{Noninteracting current\label{sec:current-nonint}}

We now again return to the noninteracting limit $U=0$.
In this case, the first contribution $\tilde{I}^r(t)$ to the current \eq{cur-fin} is given by \Eq{Itilde}
 with the value of $\Phi_\sigma(t)$ given by the central result \eq{Phit}.
We see that $\tilde{I}^r(t)$ depends on the initial state through $\Phi_\sigma(t_0)$ and decays to a nonzero stationary value.
The second contribution $\bar{I}^r(t)$ can now be simplified for $U=0$ using the super-Pauli principle \eq{superpauli}\HL{-\eq{superpauli2}}:
in the renormalized perturbation expansion for the finite $T$ corrections only two terms survive,
$\bar{\Sigma}^r(t,t')=\bar{\Sigma}^r_2(t,t')+\bar{\Sigma}^r_4(t,t')$,
in full analogy to \Eq{sigmatrunc}. 
In addition, by the same principle
 $(\chi_\sigma|\bar{\Sigma}^r_4(t,t') = 0$ in \Eq{cur-fin}
[since $\bar{\Sigma}^r_4\propto \bar{G}_2\bar{G}_{\bar{2}}\bar{G}_1\bar{G}_{\bar{1}}\propto |Z_R)(Z_L|$],
and therefore only the two-loop self-energy $\bar{\Sigma}^r_2(t,t')$ contributes to the current in this limit (here there is no summation
over the $r$-component of the multi-index $2$):
\begin{align}
\label{Sigma-r}
  \bar{\Sigma}^r_2(t,t')
  & =
  \sum_{2}
  \bar{G}_2 e^{-i\bar{L}(t-t')}\bar{G}_{\bar{2}}\bar{\gamma}_{2,\bar{2}}(t,t')
  \\
  &=
  \sum_{\eta,\sigma}
  -\frac{\Gamma_{r\sigma} T e^{\left(i\eta\epsilon_{r\sigma} +\tfrac 1 2 \Gamma_\sigma\right)(t-t')}}{\sinh\left(\pi T(t-t')\right)}
  e^{-i\bar{L}(t-t')}\bar{G}_2\bar{G}_{\bar{2}}
  .
  \nonumber
\end{align}
Using \Eq{GG}
together with
$(\chi_\sigma|\bar{L}=(\chi_\sigma|\tilde{\Sigma}=-i\Gamma_\sigma(\chi_\sigma|$
[cf. \Eq{barL_chi}]
and summing over $\eta$ one obtains
\begin{align}
 & \bar{\Sigma}^r_2(t,t')
  =
  \\
  &
  -2i\Gamma_{r\sigma}T e^{- \tfrac 1 2 \Gamma_\sigma (t-t') }\frac{\sin\left(\epsilon_{r\sigma}(t-t')\right)}{\sinh\left(\pi T (t-t')\right)}
  |\chi_\sigma)(Z_L| + \ldots
  \nonumber
\end{align}
The terms not written out give no contribution when inserted in \Eq{cur-fin} for the current
(either proportional to $|Z_R)(\chi_{\bar{\sigma}}|$ or to fermionic projectors),
and the term shown gives a contribution independent of the initial dot state (since  $(Z_L|\rho(t')=1/2$).
The explicit result in terms of the function \eq{digamma-lerch-phi} reads as
\begin{align}
  \bar{I}^r&
  =
  -\sum_{\sigma} \Gamma_{r\sigma}T \int_{t_0}^t d t' e^{-\tfrac 1 2 \Gamma_\sigma(t-t')}\frac{\sin\left(\epsilon_{r\sigma}(t-t')\right)}{\sinh\left(\pi T (t-t')\right)}
  \nonumber\\ 
  &=\sum\limits_\sigma\Gamma_{r\sigma}F_{r\sigma}^{+}(\Delta )
  ,
\end{align}
\hl{where again $\Delta=t-t_0.$}
The total average current through reservoir $r$,
written for the case of two reservoirs $r=\pm$, $\mu_r=r V_b/2$ is then
\begin{align}
  \label{final-current}
  \langle I^r \rangle (t)
  &=  \sum\limits_{\sigma} \frac{\Gamma_{r\sigma}\Gamma_{\bar{r}\sigma}}{\Gamma_\sigma}\left({F_{r\sigma}^{+}(\Delta )-F_{\bar{r}\sigma}^{+}(\Delta )}\right)
  \\
  &- \sum\limits_{r,\sigma}  \frac {\Gamma_{r\sigma}^2}{\Gamma_\sigma} F_{r\sigma}^{-}(\Delta )
  - e^{-\Gamma_\sigma\Delta t}\sum\limits_\sigma\Gamma_{r\sigma}\Phi_\sigma(t_0)
  .
\nonumber
\end{align}
\hl{
For the noninteracting case the time-dependent current has already been explicitly calculated in the limit of one spin-orbital, e.g., in \Cite{Schmidt08}
(using the Keldysh Green's function approach), under the assumption that the dot was initially \HL{fully} unoccupied,
and for an arbitrary occupation in \Cite{Andergassen11a} (using the real-time renormalization group
for the interacting resonant-level model in the limit of $T\rightarrow 0$ and vanishing nonlocal interaction, corresponding the noninteracting Anderson model). We have verified that under the corresponding simplifications our result \Eq{final-current} agrees with these works.
% with one of \Cite{Schmidt08} and \Cite{Andergassen11a}, respectively.
The last two terms in \Eq{final-current} are not antisymmetric in \me{the reservoir index} $r$ and do not vanish as $V_b=0$. They originate from the current caused 
the change of the dot charge ($I_{\mathrm{dis}}=dn(t)/dt$), the displacement current.~\cite{Schmidt08}
The displacement current decays, as it should, to zero in the stationary limit $\Delta\rightarrow +\infty$, which follows from the asymptotic relation \Eq{inf_F}.
We note the deviation of the results of~\Cite{Jin10} from the above body of works.
\footnote{\HL{
    The results of \Cite{Jin10} seem to be at variance with other known solutions~\cite{Schmidt08,Andergassen11a} and with ours. 
    These authors used a Feynman path integral to derive a convolutionless master equation.
    We have carefully checked our result against theirs by using the identity 
    % \begin{align}
      ${T} / {\sinh(\pi T\tau)}=i\int_{-\infty}^{+\infty}  e^{-i\omega\tau}\tanh\left(\frac {\omega}{2T}\right) \frac{d\omega}{2\pi}$
    %\end{align}
    to rewrite our explicitly evaluated current \Eq{final-current} in the form of an unevaluated $\omega$ integral, as is done in \Cite{Jin10}.
    We find that our result, \Eq{final-current} (and therefore that of others~\cite{Schmidt08,Andergassen11a})
    can only be recovered by adding by hand a term to Eq.(35) of \Cite{Jin10} that seems to be missing:
    $
    \Gamma_{r\sigma}e^{-(\Gamma_\sigma/2) \Delta}\int\frac{d\omega}{2\pi}
    [
    \frac{(\Gamma_\sigma/2) \cos\left((\omega-\epsilon)\Delta\right)}{(\omega-\epsilon)^2+(\Gamma/2)^2}
    -\frac{(\omega-\epsilon)\sin\left((\omega-\epsilon)\Delta\right)}{(\omega-\epsilon)^2+(\Gamma/2)^2}
    ]
    $, considering only one spinorbital $\sigma$, as has been done in \Cite{Jin10}.
    The result for the initial current in \Cite{Jin10}, $-\Gamma_{r\sigma} n_\sigma(t_0)$,
    also deviates from the results of other works and our \Eq{displ},
    by missing the term $\Gamma_{r\sigma}/2$ 
    (resulting from the preceding equation for $\Delta=0$, and writing again fixed spin $\sigma$).
    As a consequence, the result of \Cite{Jin10} \me{also} does not seem to fit in the intuitive physical picture sketched in \Cite{Schmidt08} after Eq. (36) and  here after \Eq{displ}.
    It is puzzling why, despite these differences, the result of \Cite{Jin10} for the average dot occupation number does coincide with that of other works~\cite{Schmidt08,Andergassen11a} and with our \Eq{Phit}.
    The authors of \Cite{Jin10}, discussing their results for the current, Eqs. (38-39a) in \Cite{Jin10}, only mention that 
   that ``some of their results [for the  the current] were also obtained using nonequilibrium Green function technique~\cite{Schmidt08}''
   % (\Cite{Jin10}, Sec. 4.1, p. 15),
   without mentioning or investigating \me{the} discrepancy with \Cite{Schmidt08}.
   The results of the present paper and of \Cite{Andergassen11a} confirm the results of \Cite{Schmidt08}
   without using the $t_0\rightarrow-\infty$ limit
   which can thus not be used to explain the difference with the results \Cite{Jin10}.
   The use of this limit, commonly assumed in Keldysh Green's function techniques, was criticized in \Cite{Jin10}
   and distinguishes their method from that of \Cite{Schmidt08}.
  }
}
}

Close to the initial time, $D^{-1} \ll |t-t_0| \ll \Gamma^{-1}_{r\sigma}$ (cf. discussion above)
\begin{align}
\label{displ}
  \langle I^r \rangle(t)
  \approx
  \tilde{I}^r(t_0)
  =
  \sum_\sigma \Gamma_{r\sigma} [ \tfrac{1} 2- n_\sigma(t_0) ],
\end{align}
i.e., the total current is dominated by the last term in \Eq{final-current}, \HL{the part of the displacement current} coming from \Eq{Itilde}. 
\HL{
This is again in agreement with the zero-temperature results of \Cite{Andergassen11a} for the spinless, interacting resonant level model
in the limit of vanishing nonlocal interaction.
This also agrees with the result for the initial current in \Cite{Schmidt08}, however, in contrast to that work,
we take into account arbitrary initial dot level occupancies.
The physical picture behind the result \eq{displ} again nicely relates to the fundamental importance of the $T=\infty$ limit built into our causal superfermion technique.
Extending~\footnote{
\me{In \Cite{Schmidt08} the artificial instantaneous current due to the wideband limit
was physically related to the infinite \emph{bias} limit $V_b\rightarrow +\infty$.
In \Cite{Oguri02} it was shown that for symmetric tunnel coupling this is equivalent to the $T\rightarrow \infty$ limit.
Here we discuss spin- and reservoir-dependent tunneling for which no such equivalence seems be known.~\cite{Oguri13a}
Still, the $T\rightarrow \infty$ limit provides the essential physical starting point.
Moreover, our analysis simply extends to an arbitrary number of reservoirs.}
}
the discussions in \Cite{Schmidt08} [cf. Eq. (36) there] it is as follows.
The initial current \eq{displ} stems form the part \eq{tildeI}, which describes the current
in the $T\rightarrow+\infty$ limit
[see discussion following \Eq{tildeI}].
Due to the wideband limit, the processes described by $\tilde{\gamma}$ [cf. \Eq{time-ret}] are very fast, taking place 
on the times of order $D^{-1}$, and in the wideband limit giving a finite instantaneous current response at $t=t_0$.
In contrast, the temperature-induced processes described by $\bar{\gamma}$ [cf. \Eq{time-keldysh}] are much slower and do not contribute on such short time scales.
The current thus ``does not know yet'' about the actual temperature of the reservoirs on such time-scales and therefore behaves such as if $T$  would be infinite.
This is what the physical decomposition \eq{cur-fin} of the charge current expresses, which follows naturally on a general level from our formalism.
In the concrete result \eq{displ} the factors $\langle n_\sigma \rangle (t_0)\rangle-1/2$ show that a deviation of the initial dot charge
from the value $1/2$, the stationary value in the limit $T\rightarrow+\infty$, determines the response:
%Thus the current \eq{displ} is zero for $\langle n_\sigma \rangle (t_0)=1/2$ (stationary value), while 
the empty dot $\langle n_\sigma \rangle (t_0)=0$ will charge up, $I^r(t_0)=\sum_\sigma\Gamma_{r\sigma}/2$,
whereas the filled dot $\langle n_\sigma \rangle (t_0)\rangle=1$ will discharge, $I^r=-\sum_\sigma\Gamma_{r\sigma}/2$.
%The result of \Cite{Jin10} seems does not fit in this qualitative picture.
}

The stationary value of the \HL{current, attained at much later times $ |t-t_0| \gg \Gamma^{-1}_{r\sigma}$,} is determined by the first term of \Eq{final-current}, which is antisymmetric in the reservoirs 
(and thus vanishes at zero bias) [cf. \Eq{inf_F}]:
\begin{align}
  \label{stationary-current}
  \langle I^r \rangle (\infty)
  =
  \sum_{r',\sigma}\frac{\Gamma_{r\sigma}\Gamma_{\bar{r}\sigma}} {\pi \Gamma_\sigma}
   r' \Im \Psi \left( 
    \frac 1 2+\frac{ \tfrac{1}{2} \Gamma_\sigma - i\epsilon_{r'\sigma}}{2\pi T} \right)
  .
\end{align}
Expressed in the fermi-function $f(x)=\frac 1 {e^{x/T}+1} = \frac 1 2 -\frac 1 2 \tanh(x/2T)$
and using \Eq{dig-tanh}, this can be rewritten as the more familiar form
of a sum of current contributions from the independent spin-orbitals, each broadened by $\Gamma_\sigma=\sum_r \Gamma_{r\sigma}$:
\begin{align}
  \nonumber
  \langle I^r \rangle (\infty)
   =
  &
  \sum_\sigma
  \frac{\Gamma_{r\sigma}\Gamma_{\bar{r}\sigma}} {\pi \Gamma_\sigma}
  \int_{-\infty}^\infty \frac{ \Gamma_\sigma/2}{(x-\epsilon_\sigma)^2+(\Gamma_\sigma/2)^2}
  \\
  & \times \Big( f(x+\mu_r)-f(x+\mu_{\bar{r}})\Big)\HL{dx}.
\end{align}
\HL{This result coincides with either of \Cite{Schmidt08,Andergassen11a,Jin10} in the corresponding limits mentioned above.}
Finally, we note that the stationary current \eq{final-current},
calculated here by explicitly taking the long-time limit,
is recovered from our direct calculation of the stationary quantities \HL{in \Sec{sec:stationary}:
with} the help of \Eqs{Ir} and \eq{rho} one can show that the stationary current depends on just two stationary self-energy coefficients:~\cite{Saptsov12a}
\begin{align}
  I^r(\infty)
  =
  \sum\limits_\sigma
  \frac{\Gamma_{\bar{r}\sigma}\psi^r_\sigma-\Gamma_{{r}\sigma}{\psi}^{\bar{r}}_\sigma} {2\Gamma_\sigma}
.
  \label{Iinfty}
\end{align}
Inserting the $U=0$ result for $\psi^r_\sigma$ by leaving out the $r$-sum in \Eq{psiio} reproduces \Eq{stationary-current}.
This confirms the result for the current in the $U=0$ limit obtained in \Cite{Saptsov12a} by a real-time RG \HL{calculation of these coefficients.}
\HL{This is another way of seeing that}
the additional broadening scales $\Gamma_\sigma+\tfrac 1 2 \Gamma_{\bar{\sigma}}$ [\Eq{3gamma}] do not affect the stationary current for $U=0$
\HL{since} the coefficients $\phi_\sigma$ [\Eq{phi-digamma}] and $\zeta$ [\Eq{zeta}] do not appear in \Eq{Iinfty}.

\refB{The main objective of the above was to illustrate in a tractable example how the causal superfermion technique works for the calculation of an observable, in this case the current.
Although we were able to include all possible initial coherences and correlations in the initial density operator locally on the quantum dot,
the time-dependent current reduces to the sum over its spin-resolved components.
The current is not sensitive to the fermion-parity decay of the quantum-dot mixed state, which can be detected in ways discussed in the introduction [cf. \Eq{keyresult}].
In the noninteracting and wideband limit the effect of the spin-polarization of the ferromagnetic leads is to merely introduce different decay time-scales for different spin states ($\Gamma_{r\sigma}^{-1}$).
The situation becomes more interesting when Coulomb interaction is included since this generates
of the effective exchange magnetic field,~\cite{Koenig03,Martinek03a,Choi04,Braun04set,Martinek05} a nondissipative effect.
The time evolution of this field after switching on the tunnel processes between the dot and the ferromagnetic leads is of considerable interest.
The method presented in the present paper may serve as a starting point for conveniently addressing how such effects develop,
in particular even for small Coulomb interaction nonperturbatively but strong tunnel coupling ($\Gamma \gg U$).
It is advantageous that this can be done in the same formalism which can treat the complementary limit ($\Gamma \ll U$).
}

% \HL{We see that the time-dependent current is the sum over its spin-resolved components and the ferromagnetic leads do not bring any sufficient
% effect in the noninteracting limit. Although even small Coulomb interaction leads to an interesting effects in the such system, e.g., generation
% of the effective exchange magnetic field.~\cite{Koenig03,Martinek03a,Choi04,Braun04set,Martinek05} This effect and its development in time after switching 
% on the tunnel processes between the dot and the ferromagnetic leads can be conveniently described within our 
% approach, e.g., perturbatively in the Coulomb interaction
% using the ideas of \Sec{sec:limitations}. This problem, however, being interesting on its own requires much more detailed discussion
% and is beyond the scope of the present paper. It deserves a separate publication, which is in planes,~\cite{Saptsov14b} and for which
% the methods presented in the present paper will serve as a starting point.  }

\section{Discussion and outlook\label{sec:conclusion}}

\hl{As outlined in the introduction,
the time evolution of strongly interacting quantum dots is of great experimental interest,
but analytical theoretical methods struggle to deal with it.
Taking an Anderson model description as a starting point,
we focused on improving the real-time approach which has already been successfully applied to explain various experiments.
}
The goal of this paper was twofold: we wanted
(i)
% to systematically apply the causal superfermion technique to all aspects of the 
\HL{to set up from scratch the real-time approach to time-dependent decay in interacting transport problems,
systematically exploiting the causal superfermion technique (\Sec{sec:timeevolution})}
and (ii) to highlight \HL{its practical advantages} by a complete solution of the noninteracting Anderson model describing a quantum dot 
with spin-dependent tunneling rates $\Gamma_{r\sigma}$ \HL{and for an arbitrarily correlated initial mixed state (\Sec{sec:results}).}
We now  summarize \hl{these two} aspects separately, starting with the concrete results \HL{(ii)},
\HL{and then turning to the general framework (i).
In the process we generalize the concrete results to multiorbital models, in both the interacting and the noninteracting case.}
We also comment on the limitations imposed by the few assumptions that we made
and provide an outlook on possible further applications \HL{which have motivated this work all along.}

\subsection{Quantum-dot spin-valve:
  \newline
  $U=0$ and interaction corrections}
\HL{In \Sec{sec:time-evolution} we} calculated the exact time-evolution propagator of the complete two-fermion density operator in the
noninteracting limit ($U=0$).
The exact result, nonperturbative in the tunneling rates $\Gamma_{r\sigma}$,
is obtained from a simple \emph{second-order} renormalized perturbation theory,
expanding in the Keldysh reservoir correlation function $\bar{\gamma}(\omega)  \propto \Gamma_{r\sigma} \tanh(\omega/2T)$
instead of just $\Gamma_{r\sigma}$.
Our result  \eq{exact-RDDO} includes all possible coefficients of the density operator:
spin-orbital occupancies ($\braket{ n_\uparrow}$,  $\braket{ n_\downarrow}$),
 transverse spin coherences (e.g. $\braket{ d_{\uparrow}^\dag d_{\downarrow}}$)  and electron-pair coherences  (e.g., $\braket{d_{\downarrow}d_\uparrow}$),
but also the two-particle correlations 
quantified
by the nonequilibrium average of the fermion-parity operator $\braket{ (-1)^n}  \sim \braket{n_\downarrow n_\uparrow}(t)\HL{+...}$.
The last three arise only due to the initial preparation of the quantum-dot state.

% Factorization
Besides recovering known results for the one-particle quantities,
we noted that in general
the \emph{transient} two-particle correlator  does not  factorize
$ \braket{n_\downarrow n_\uparrow}(t)   \neq \braket{n_\downarrow}(t) \cdot  \braket{ n_\uparrow}(t)$
until stationarity is reached,
$ \braket{n_\downarrow n_\uparrow}(\infty) =  \braket{n_\downarrow}(\infty) \cdot  \braket{ n_\uparrow}(\infty)$.
  This happens when the quantum dot state is initially prepared in a two-particle correlated state.
  In the stationary state these correlations, however, die out.
\footnote{
   Formally, for $U=0$
  the stationary state has a form that allows for a Wick theorem on the \emph{quantum dot},
  although with \emph{unknown} nonequilibrium parameters $\Phi_\sigma(\infty) =\braket{n_\sigma}(\infty)-1/2$.
   This is especially clear in terms of superfermions:
  there is a formal analogy between
  the nonequilibrium dot stationary state \Eq{rho2}, which can be written as
    $\rho(\infty)=\tfrac 1 2 \exp\left(2\sum_\sigma \Phi_\sigma(\infty)\bar{G}_{+\sigma}\bar{G}_{-\sigma}\right)|Z_L)$,
  and the thermal-equilibrium state, written as 
  $\rho_{\rm eq}=\exp(-\epsilon_\sigma n_\sigma/T)/Z=\tfrac 1 2 \exp\left(-\sum_\sigma \tanh(\epsilon_\sigma/2T)\bar{G}_{+\sigma}\bar{G}_{-\sigma}\right)|Z_L)$.
    \HL{The nonequilibrium analog of the fluctuation-dissipation relation \Eq{FDT} for the dot then reads as 
$\tilde{G}_1\rho(\infty)=-2\eta\Phi_\sigma\bar{G}_1\rho(\infty)$. The Wick theorem for the dot in this case can be proven in full analogy with
the equilibrium reservoir Wick theorem,~\cite{Saptsov12a} although the dot Keldysh contraction functions are not, in general, expressible via 
$\tanh(\epsilon/2T)$:
$\bar{\gamma}_{\mathrm{dot}}=\langle \tilde{G}_1\tilde{G}_2\rangle\propto -\eta_1\Phi_{\sigma_1}\not\propto \tanh(\eta_1\epsilon/2T)$.}
Before stationarity is reached  ($t_0< t < \infty$) this analogy does not apply
  unless one starts at $t_0$ from a quantum-dot state without two-particle correlations.
   }
Another, more  striking aspect of the decay of these initial correlations on the quantum dot,
$\braket{ (-1)^n}(t) \sim e^{-\Gamma(t-t_0)} \braket{ (-1)^n}(t_0)+...$,
is that the strict exponential form and the decay rate
$\Gamma=\sum_{r\sigma} \Gamma_{r\sigma}$
is independent of any other parameter in the problem.
Within our superfermion formulation of the real-time approach it is immediately \HL{clear} that
no corrections to this simple ``universal'' behavior can appear,
 due to neither \me{finite} temperature $T$ (see \Sec{sec:generalWBL}), nor  bias voltage $V$, nor magnetic field $B$, nor \emph{interaction} $U$.
Notably, $\Gamma$ depends only on the sum of the spin-dependent rates, i.e.,  even the spin-polarization of the tunneling drops out,
an aspect not addressed in~\Cite{Saptsov12a}.
This generalizes an earlier conclusion based on perturbation theory:~\cite{Contreras12}
\me{this absence of corrections} holds \emph{nonperturbatively} in the tunnel coupling $\Gamma_{r \sigma}$ for the \emph{interacting} Anderson model
but also for the decay in multiorbital generalizations, recently studied in~\Cite{Schulenborg13thesis}.
The key point is that by the fundamental fermion-parity superselection rule, any local quantum-dot Hamiltonian must commute with the operator $(-1)^n$.
Therefore, the decay of the initial correlations $\braket{ (-1)^n}(t_0)$, appearing in the expansion of the density operator,
can only come from the tunnel coupling to the reservoirs
  and has the above mentioned form.

In addition, we found that the time evolution of $ \braket{ (-1)^n}(t) \sim \braket{n_\downarrow n_\uparrow}(t)\HL{+\ldots}$ contains
additional
oscillatory decaying terms coming from the initial occupations $\braket{n_\sigma}(t_0)$
with rate $\Gamma_\uparrow + (\Gamma_\downarrow/2)$ and $\Gamma_\downarrow + (\Gamma_\uparrow/2)$.
These unexpected rates were noted earlier for spin- and junction-independent tunnel rates $\Gamma_{r \sigma}=\tilde{\Gamma}$
as an additional broadening scale $3\tilde{\Gamma}$ in the stationary density operator~\cite{Saptsov12a} 
and in related self-energies.~\cite{Oguri13a}
Thus,
even in this simple limit
the time-dependent decay of the density operator of the noninteracting ($U=0$) Anderson model shows
four characteristic decay rates: $\tilde{\Gamma}$, $2\tilde{\Gamma}$, $3\tilde{\Gamma}$, $4\tilde{\Gamma}$.

\HL{Finally, in \Sec{sec:current-time}} we illustrated the application of superfermions to the calculation of observable quantities
\me{for} the time-dependent charge current. We showed that the \me{small-time} artifacts of the wideband limit in the transport current
can be discussed on the superoperator level.
\HL{Also, the $T\rightarrow \infty$ limit, built into the field superoperators,
naturally appears in the expressions for the displacement current.}
We furthermore confirmed the RG results for the stationary noninteracting limit in \Cite{Saptsov12a},
in particular, we related the observation made there
-- that only one-loop self-energy corrections matter for the current -- to the super-Pauli principle introduced here.

\subsection{Superfermions in the real-time approach}
The results summarized above served to illustrate three general aspects of superfermions -- announced in the title of the paper -- as applied to 
the real-time transport theory \HL{that we discussed in \Sec{sec:timeevolution}.}
\me{Therefore these can be also generalized to multiorbital Anderson quantum dots.}

\emph{(i) Causal structure of superfermions.}
Using various examples,
we illustrated 
that physical meaning can be assigned to formal objects
appearing in a Liouville-space theory of a strongly interacting, open fermionic system.
Although many concepts carry over from the usual Hilbert-Fock space, many others require careful reconsideration,
e.g., the role of the super-kets in the expansion of a mixed state [\Eq{general-initial-dot}] or the superfermion number [\Eq{Netasigma}].
The crucial feature distinguishing quantum fields in Liouville-Fock space from those in Hilbert-Fock space is what we refer to as the ``causal structure''.
On the one hand, this entails [\Eq{parity}]
\begin{align}
  \bar{G}_{\eta \sigma} |Z_R) = 0,
  \label{mostfilled}
\end{align}
where $|Z_R) \sim (-1)^n $ is the fermion-parity operator appearing in the corresponding superselection rule of quantum mechanics.
Roughly speaking, this imposes the constraint that ``fermions on different Keldysh contours anticommute''.~\cite{Saptsov12a} 
On the other hand, the identity [\Eq{trace}]
\begin{align}
  (Z_L|  \bar{G}_{\eta \sigma} = 0,
  \label{empty}
\end{align}
where the $(Z_L|= \text{Tr} $ represents the trace operation, is involved in the probability conservation of the density operator.
As we showed \HL{in \Sec{sec:widebandlimit}}, the causal structure implies much more than probability conservation of the dynamics.
Whereas the former has received much attention in Green's function formalism, 
in density operator approaches much less attention seems to have been given to this more fundamental structure.

We emphasized the central importance of the unit operator $|Z_L) \sim \unit$ as the Liouville-Fock space vacuum and its physical meaning as the $T \rightarrow \infty$ maximally mixed state. 
We used the  $T \rightarrow \infty$ limit as a point of reference,
not only in the construction of the Liouville-Fock space but also in the calculation of the time-evolution propagator and its self-energy.
This may be compared with the limit of infinite bias $V_b=\infty$, which also admits an exact analysis.
It has recently been studied by Oguri and Sakano~\cite{Oguri13a} and earlier by Gurvitz,~\cite{Gurvitz96,Gurvitz97}
while the relevance of renormalization corrections at finite bias were pointed out in \Cite{Wunsch05}.
In comparison with this,  we emphasize that our formulation using the $T\rightarrow \infty$ limit has the important technical advantage that it is provides a unique starting point
irrespective of the number of reservoirs. \me{Moreover, it applies irrespective of the asymmetry of the tunnel couplings: the latter spoils the relation between the $V_b \rightarrow \infty$ and $T \rightarrow \infty$ limits for two electrodes, discussed in \Cite{Oguri02}.}

Another interesting consequence of incorporating the $T \rightarrow \infty$ limit is that the unperturbed  evolution
(i.e., the reference problem for the renormalized time-dependent perturbation theory) is \emph{dissipative} and therefore damped as a function of time.
This may prove to be interesting for numerical schemes that aim to calculate memory kernels.~\cite{Cohen11, Cohen13a, Cohen13b, Weiss08,Segal10, Segal11, Segal13}
This damping depends only on the tunnel couplings,
in contrast to the broadening obtained by a recently proposed dressing scheme~\cite{Kern13}, \me{cf. also~\cite{Marthaler11},}
which depends on the quantum-dot energies
and is based on a partial resummation of real-time diagrams that serves a different purpose.

\HL{Finally, the $T \rightarrow \infty$ limit also aids the physical understanding of observables, such as the \me{displacement part of the current \eq{displ}.}}

\emph{(ii) Fermion-parity protected decay mode.}
\HL{As shown in \Sec{sec:infiniteT}}
the striking independence of \me{the key result \Eq{keyresult}},
$ \braket{ (-1)^n}(t) \sim e^{-\Gamma(t-t_0)} \braket{ (-1)^n}(t_0) \HL{+\ldots}$,
of all remaining parameters
including the \emph{interaction} $U$
relates to a formal property of the general theory.
Since the causal superfermion approach uses the $T \rightarrow \infty$ limit as a reference point,
it reveals that finite-temperature corrections to the time evolution only involve \emph{creation} superoperators $\bar{G}$.
Clearly then, the time evolution of the superket $|Z_R) \sim (-1)^n$ in Liouville space cannot have any such correction:
as expressed by \Eq{mostfilled}, it is the ``most filled'' superket and simply cannot accommodate more superfermions.
Moreover, it is readily seen that \emph{any interacting} $N$-spin-orbital Anderson model (orbitals $l=1,\ldots,N/2)$ with quadratic tunnel coupling exhibits 
exactly this purely exponential decay mode with rate $\Gamma =\sum_{r\sigma,l} \Gamma_{r\sigma,l}$.
Finally, it is interesting to note that half of the decay modes that we studied in the noninteracting case $(U=0)$ are, in fact, fixed completely by the $T=\infty$ calculation.

\emph{(iii) Super-Pauli principle.}
The super-Pauli principle \eq{superpauli} states that formal superkets cannot be ``doubly occupied'' :
\begin{align}
  ( \bar{G}_{\eta \sigma})^2 = 0
  .
\end{align}
This simple consequence of the causal Liouville-Fock space construction provides useful insights in two directions.
First, when applying the real-time approach to noninteracting problems, the super-Pauli principle is the key simplification that keeps the calculations 
completely tractable on the \emph{superoperator} level.
The renormalized perturbation theory is simple to set up,
and a \emph{finite-order} $N$  calculation gives the \emph{exact} result for $N$ spinorbitals, including all local $N$-particle nonequilibrium correlations.
The higher-order corrections vanish exactly, not by their scalar magnitude but by their superoperator \me{structure:
generalizing} \Eq{superpauli2} to the case for $N$ spinorbitals we have
\begin{align}
  \bar{G}_{m} ... \bar{G}_{1} = 0 \quad \text{for $m>2N$},
\end{align}
as a direct consequence of the super-Pauli principle.
The other major implication of taking the noninteracting limit, \Eq{formal-2loop2}, can also be generalized to this case
by extending the simple considerations in \Fig{fig:factorize} to even orders $m=4,\ldots, 2N$:
the $m/2$-loop time propagator factorizes into one-loop propagators,
\begin{align}
  &\bar{\Pi}_m(t,t_0)
  =
  \label{factorize}
  \\
  &
  \frac{1}{(m/2)!}
  \underset{\text{($m/2$ times)}}
  {
    \bar{\Pi}_2(t,t_0) e^{i\bar{L}(t-t_0)}
    \bar{\Pi}_2(t,t_0) 
    \ldots  e^{i\bar{L}(t-t_0)}
    \bar{\Pi}_2(t,t_0)
  }
  .
  \nonumber
\end{align}
The superoperator algebraic structure thus carries important physical information, which is naturally revealed by the causal superfermions.
We emphasize that these simple general features of the noninteracting limit remain hidden in the real-time approach
unless one starts from the renormalized the perturbation theory \eq{mth-expansion-bar-all}, incorporating the wideband limit.
We furthermore showed that certain observables, such as the charge current, turn out to be insensitive to corrections beyond the one-loop order.
This raises a question of practical importance: given a physical $M$-particle quantity, to which loop order does one need to calculate the self-energy in order to get the exact noninteracting result?

This leads to the second important insight
which is relevant to applications of the real-time renormalization group approach,~\cite{Schoeller09a,Pletyukhov10,Andergassen11a}
which aims to provide a good solution in both the strong and \emph{weak} interaction limits.
Our complementary frequency-space calculation of the stationary limit \HL{in \Sec{sec:stationary}} confirmed that the real-time renormalization group
in the one- plus two-loop approximation~\cite{Saptsov12a} correctly reproduces the exact noninteracting limit, in particular the self-energy part relevant to the current, \me{relating this to the super-Pauli principle.}
\hl{We additionally \me{calculated} the stationary state \me{in this limit,} obtaining the exact effective Liouvillian by a \emph{two-loop} order calculation.
For generalized Anderson models with $N$ spin-orbitals, 
we inferred above that at least a $N$-loop calculation in the renormalized perturbation theory will reproduce the exact noninteracting limit.
This implies that real-time renormalization group schemes  must include \emph{at least} a consistent $N$-loop RG flow  for the Liouvillian 
together with the corresponding vertex corrections
in order to capture the exact noninteracting limit.
In view of the complications encountered in \Cite{Saptsov12a}, already at the $N=2$-loop order for the Anderson model,
a question becomes practically relevant
under which conditions may higher loop orders be avoided,
 (e.g., for a given observable or specific density operator component)?}
\schoeller{
\me{Here we should point out that it can be shown that if one is interested in the evolution of single-particle quantities (expressible through two (super)fields) only one-loop diagrams are required for the time-dependent decay in the noninteracting limit.}
This carries over to multiparticle quantities  only under the condition that initial correlations on the dot are absent,
i.e, these factorize at the initial time \HL{(see main result \Eq{Xit}, cf. \Eqs{neqXi} and \eq{varthetaresult})}.
When initial correlations are present, however, higher loop evolution does matter.
\me{Note that even} when the initial density operator contains nonfactorizable correlators,
our key result \HL{\Eqs{formal-2loop2} and \eq{factorize}} shows that the time-evolution superoperator can still be factorized.
}

We also \me{found} that the noninteracting limit becomes most transparent
when considering the renormalized two-loop propagator $\bar{\Pi}_4(t,t_0)$ in time space \me{(rather than its Laplace transform),}
because it factorizes in the limit $U=0$.
\HL{This seems to have no equally simple counterpart in Laplace space for the two-loop renormalized self-energy $\bar{\Sigma}_4(z)$.}
The generalized \me{time-space} relations \eq{factorize}
allow for a convenient verification on the superoperator level
that a real-time RG scheme correctly reproduces the noninteracting limit in all nonvanishing loop orders $m/2=1,2,\ldots,N$.

\subsection{Limitations and further extensions\label{sec:limitations}}
Our considerations were quite general.
We now end with comments on the limitations imposed by our assumptions
and provide an outlook on how these may be overcome.

% Noninteracting limit
\emph{Noninteracting limit ($U=0$)}
Although we focused \HL{in \Sec{sec:results}} on the noninteracting limit for illustrative purposes,
the principles demonstrated here, can be applied to interacting problems, as we showed, e.g., in \Sec{sec:Lil}.
\me{This is what has motivated our exhaustive study all along.}
\HL{A more} advanced example is our RG study \Cite{Saptsov12a}, but other approaches may also be developed.
For instance, one may consider expanding the time-evolution propagator $\Pi$ or its self-energy $\bar{\Sigma}$ in the \emph{nonlinear part} of the  Coulomb interaction,
i.e., not simply in the parameter $U$, but in the term $U(-1)^n/2$ in the Anderson Hamiltonian,  cf. \Sec{sec:Lil}.
This \me{perturbative expansion} then yields an approximate result which is nonperturbative in both $U$ and $\Gamma_{r\sigma}$ \me{(the quadratic term \eq{quadratic} also depends on $U$, and this part is treated nonperturbatively).}
The coefficients for the $m$th-order term in this nonlinear interaction can be calculated using our causal superfermion approach through
an expansion in the Keldysh correlation function $\bar{\gamma}$, which is truncated as \me{in the noninteracting limit,} but now at the $(2+m)$-loop order.\footnote{This is 
because the superoperator expansion of the nonquadratic part of the Liouvillian, \Eq{quartic}, always contains a term \emph{decreasing}
the number of superparticles by two:
the first term in \Eq{quartic},
$\bar{G}_{\eta\sigma}\tilde{G}_{\bar{\eta}\sigma}\tilde{G}_{\eta\bar{\sigma}}\tilde{G}_{\bar{\eta}\bar{\sigma}}$, contains one creation superoperator
$\bar{G}_{\eta\sigma}$ and three destruction superoperators $\tilde{G}_{\bar{\eta}\sigma}\tilde{G}_{\eta\bar{\sigma}}\tilde{G}_{\bar{\eta}\bar{\sigma}}$
and thus decreases the total number of superfermions by two.
In contrast, each pair of Keldysh contracted creation superoperators always \emph{increases} this number by two. Since the total number of superfermions
cannot exceed four by super-Pauli principle, the terms with more than $(2+m)$ loops are identically equal to zero for given $m$.
}
Also, here the result simplifies due to the superoperator structure dictated by  the super-Pauli principle.

% Wide-band limit
\emph{Wide-band limit}
The wideband limit is another main simplifying assumption that we made \HL{in \Sec{sec:widebandlimit}.}
However, beyond this limit  the number of Keldysh contractions \me{($\bar{\gamma}$)} is still limited to \emph{two} 
by the super-Pauli principle \eq{superpauli}\HL{-\eq{superpauli2}}
in the non-interacting limit ($U=0$).
This expresses the general fact that retarded contractions \me{($\tilde{\gamma}$)} always connect a creation and an annihilation superoperator,
and thus its contribution does not change the total super particle number.
In contrast, Keldysh contractions $\bar{\gamma}$ always connect two \emph{creation} superoperators,
increasing the total superfermion number by two. Since the super-Pauli principle limits the total number of superfermions to four,
at most, two Keldysh contractions are allowed.
\me{This illustrates how common physical reasoning based on the usual second quantization can be transferred to nonequilibrium problems using our causal superfermions, aiding the solution of physical problems.}

\emph{Initial system-reservoir factorization}
The assumption of factorizing system-reservoir correlations at the initial time is not that restrictive either.
Much of the technical and physical conclusions presented can be generalized to apply also to \me{the case of nonfactorizing system-reservoir initial conditions}
and will be discussed in a forthcoming work.~\cite{Saptsov13b}

\refB{\emph{Time-dependent parameters}
So far, we have focused on the time evolution of the quantum dot to the new stationary state
 after the tunnel couplings $\Gamma_{r\sigma}$ experience a sudden change (quench)
at $t=t_0$ from $\Gamma_{r\sigma}=0$ to a set of finite values which further remain unchanged for $t>t_0$.
Although \Cites{Schmidt08,Jin10,Andergassen11a} also studied this problem,
the main motivation here was to illustrate the advantages of the causal superfermion approach in this most simple setting.
However, our formalism can be easily extended to deal with a time dependence of all the parameters involved, i.e., $\epsilon=\epsilon(t)$, $B=B(t)$, $V_b=V_b(t)$,
and $\Gamma_{r\sigma}=\Gamma_{r\sigma}(t)$ as we briefly outline. First, we note that once we consider the wideband limit,
\Eq{time-ret} remains valid
if the parameters vary much slower than the inverse band width, which always seems to be experimentally given.
This allows us to integrate out the retarded reservoir contractions also in this case and obtain an infinite-temperature kernel  $\tilde{\Sigma}$, cf. \Eq{full-tildesig}, but now with a time-dependent $\Gamma_{r\sigma}(t)$ and a corresponding time-dependent renormalized Liouvillian $\bar{L}(t)=L(t)+\tilde{\Sigma}(t)$, cf. \Eq{barl}.
The interaction-representation vertices \eq{WBL-intervertex} now include a time-ordering superoperator $\mathrm{T}$
and reduces in the noninteracting case to a result similar to \Eq{barG-evol}:
% can be calculated using time-ordered renormalized ``free'' 
%evolution $\bar{\Pi}_0(t,t_0)=\mathrm{T}e^{-i\int_{t_0}^t d\tau\bar{L}(\tau)}$:
\begin{align}
% \bar{G}_j(t)=\bar{\Pi}_0(t,t_0)\bar{G}_j\bar{\Pi}_0^{-1}(t,t_0).
\bar{G}_j(t) & = \mathrm{T}e^{-i\int_{t_0}^t d\tau\bar{L}(\tau)} \bar{G}_j \left(\mathrm{T}e^{-i\int_{t_0}^t d\tau\bar{L}(\tau)}\right)^{-1}
\\
             & =      e^{\int_{t_0}^t d\tau [ i\eta\epsilon_\sigma(\tau) + \tfrac{1}{2}\Gamma_\sigma(\tau) ] }   \bar{G}_1  \quad \text{for $U=0$}
\end{align}
The main ideas of our approach thus remain the same and apply also to multiorbital extensions without any crucial complications arising.
In particular, the super-Pauli principle \eq{superpauli} is also valid for the above field superoperators
 and causes the perturbation series to terminate at a finite order as in \Eq{pitrunc},
which is one of the central insights of this paper.
}

\acknowledgments
We acknowledge useful discussions with M. Hell, M. Pletyukhov, H. Schoeller, and J. Splettstoesser.
\appendix
\section{Field superoperators\label{app:superfields}}
In this Appendix, we provide further comments on the construction of field superoperators undertaken in \Sec{sec:superfermion}.
As discussed in \Sec{sec:parity}, we emphasize the importance of starting this construction from sets of fermionic operators  $d_1$ and $b_1$ (or $a_1$) {for the quantum dot and the reservoirs} respectively, which mutually \emph{commute}.
The crucial advantage of using such field operators is that \Eq{trace} holds for the \emph{partial} traces of the corresponding dot or reservoir superoperators $G^{q}_1$ [\Eq{superG}] and $J^{q}_1$ [\Eq{superJ}].
This allows one to obtain the reduced dynamics of the dot (by integrating out the reservoir degrees of freedom), while preserving the causal properties \Eq{trace},
which we have shown to bring many computational and physical insights.

If one uses in \Eq{superG} and \eq{superJ} instead of {$d_1$ and $b_1$} mutually anticommuting sets of fermion operators $d'_1$ and $b'_1=\sqrt{\Gamma_{r\sigma}/2\pi} a'_1$, {constructed} in \Sec{sec:parity}, then one obtains the same anticommutation relations \eq{anticommut_d}, \eq{anticommut_r}
{for the resulting field superoperators.}
However, in this case no definite commutation relations analogous to \Eq{anticommut_dr} are obtained (neither commutation, nor anticommutation relations),
which is a major disadvantage.
{In principle, one can introduce} other sets of field superoperators which are free of this problem even though one starts again from the anticommuting fields $d'_1$ and $b'_1=\sqrt{\Gamma_{r\sigma}/2\pi} a'_1$. Instead of using \Eq{superG} and \eq{superJ}, one defines
\begin{align}
  \label{superG_t}
  \mathfrak{G}^q_1\bullet=\frac 1{\sqrt{2}} \left\{ d'_1\bullet + q(-1)^{n+n^R}\bullet d'_1 (-1)^{n+n^R}
  \right\}
,
\end{align}
\begin{align}
\label{superJ_t}
 \mathfrak{J}^q_1\bullet=\frac 1{\sqrt{2}} \left\{ b'_1\bullet - q(-1)^{n+n^{R}}\bullet b'_1 (-1)^{n+n^{R}} \right\}
 .
\end{align}
Here, one uses, in contrast to \Eq{superG}-\eq{superJ}, the \emph{global} fermion parity operator $(-1)^{n+n^R}.$
The field superoperators $\mathfrak{G}_1$ {and} $\mathfrak{J}_1$ can be checked to satisfy the same anticommutation relations \Eq{anticommut_d}-\eq{anticommut_r}
as $G_1, J_1$ and the same super adjoint relation.
In contrast to \Eq{anticommut_dr}, they satisfy instead mutual \emph{anti}commutation relations \Eq{anticommut_dr}:
\begin{align}
  \label{anticommut_dr_t}
    [\tilde{\mathfrak{J}}_1, \tilde{\mathfrak{G}}_2]_+
  = [\bar{\mathfrak{J}}_1  , \bar{\mathfrak{G}}_2  ]_+
  = [\bar{\mathfrak{J}}_1  , \tilde{\mathfrak{G}}_2]_+
  = [\tilde{\mathfrak{J}}_1, \bar{\mathfrak{G}}_2  ]_+ = 0.
\end{align}
However, the disadvantage of this construction is that
instead of the causal property \Eq{trace} we now have
\begin{align}
  \Tr{}\bar{\mathfrak{G}}_1=0, ~~~ \Tr{}\tilde{\mathfrak{J}}_1=0,
\end{align}
where $\Tr{}=\Tr D \Tr R$ is a \emph{global} trace,
while the crucial \emph{local}-trace identities \Eq{trace} are not valid anymore:
\begin{align}
  \Tr D \bar{\mathfrak{G}}_1\not\equiv0, ~~~ \Tr R \tilde{\mathfrak{J}}_1\not \equiv 0.
\end{align}
This seems to drastically complicate\footnote{One can, in principle, formulate the problem in terms of $\mathfrak{J}$
and $\mathfrak{G}$ superoperators, separate $\mathfrak{G}$ and $\mathfrak{J}$, and then project the $\mathfrak{G}$ 
defined in the global reservoir-dot Liouville space onto the local superoperators $G$, \Eq{superG}.
This involves, however, unnecessary complications.} the calculation of the partial reservoir trace required in \Sec{sec:timeevolution}.
\section{Time representation for the Keldysh contraction \label{sec:keldysh}}
In the wide-band limit, the explicit form of the time-dependent Keldysh correlation function $\bar{\gamma}_{2,{1}}(t_2-t_1)$ [\Eq{keld-contr}] can 
be obtained using the partial fraction expansion for the meromorphic function 
\begin{align}
\label{meromorphic_tanh}
\tanh (z)=\sum\limits_{n=-\infty}^{+\infty} \left(z+i\pi(n+1/2)\right)^{-1}.
\end{align}
Closing the contour of integration over $\omega$ in the lower half of the complex plane and making use of the residual theorem, we obtain \Eq{time-keldysh} of the main text as follows:
\begin{align}
  \bar{\gamma}_{2,{1}}(t)
  & =\frac \Gamma {2\pi} \int d\omega e^{-i\eta(\omega+\mu)t} \tanh(\eta\omega/2T)\delta_{2,\bar{1}}
  \nonumber
  \\
  &= -i 2 T e^{-i\eta \mu t} \Gamma\sum_{n=0}^{+\infty}  e^{-\pi T(2n+1)t} \delta_{2,\bar{1}}
  \nonumber
  \\
  & =
  \frac {-i\Gamma T}{\sinh\left(\pi T t\right)}
  e^{-i \eta \mu t}
  \delta_{2,\bar{1}}
  .
\end{align}
Here, as in \Eq{time-ret} the multi-indices ${2,\bar{1}}$ in the $\delta$-function do not contain the reservoir frequencies.
\section{Integrals of Keldysh contraction -  digamma function\label{sec:digamma_app}}
In \Eq{psiio}, \eq{zeta} and \eq{stationary-current}, we used the following result, obtained from
$
\Im \Psi(z) = - \Im \sum_{n=0}^{+\infty} 1/(n+z)
$,
the expansion \eq{meromorphic_tanh} of $\tanh(z)$,
and application of the residual theorem (closing the integration contour in the lower half of the complex plane):
\begin{align}
  &  \tfrac{1}{2} \int_{-\infty}^{+\infty} dx 
  \frac { \Gamma \tanh(x/2T)}{\Gamma^2+(x-\epsilon)^2}
  \nonumber
  \\
  & =
  - \tfrac{1}{2}\Im
  \int_{-\infty}^{+\infty}dx \frac {\tanh(x/2T)}{i\Gamma+\epsilon-x}
  \label{dig-sum}
  \\
  & =
  - \tfrac{1}{2}
  \Im \sum_{n=-\infty}^{+\infty}
  \int_{-\infty}^{+\infty}dx
  \frac{1}{x/2T+i\pi(n+1/2)}\frac 1 {i\Gamma+\epsilon-x}
  \nonumber \\
  & =
  \tfrac{1}{2} \Im \sum\limits_{n=0}^{+\infty}\frac{2}{n+1/2+\frac\Gamma {2\pi T}-\frac{i\epsilon}{2\pi T}}=
  \nonumber \\
  & = 
  - \Im \Psi\left( \frac 1 2 + \frac {\Gamma - i\epsilon }{2\pi T}\right)
  .
  \label{dig-tanh}
\end{align}
\section{Evaluation of the functions $F^{+}_{r\sigma}(t)$ and $F^{-}_{r\sigma}(t)$\label{sec:Ffunc_ap}}
Here, we present the calculation of the function $F^{+}_{r\sigma}(\Delta t)$ given in \Eq{digamma-lerch-phi}.
We use the expansion
\begin{align}
  \frac 1 {\sinh(x)}
  =
  \frac {2e^{-x}}{1-e^{-2x}}=2\sum_{n=0}^{\infty} e^{-(2n+1)x},
\end{align}
which holds for any positive $x$ since $e^{-2x} < 1$ lies inside the convergence radius.
We obtain \Eq{digamma-lerch-phi} as:
\begin{align}
  & F^{+}_{r\sigma}(\Delta t):=
  -\int_0^{\Delta t} d \tau \frac {T \sin(\epsilon_{r\sigma}\tau)}{\sinh\left(\pi T\tau\right)}  e^{-\Gamma \tau}
  \label{f1-ap}
  \\
  =&
  -2T  \Im \sum_{n=0}^{\infty} \int_0^{\Delta t} d \tau e^{\left(i\epsilon_{r\sigma}-\Gamma\right)\tau-\pi T \tau (2n+1)}
  \nonumber
  \\ 
  =&
  2T \Im \sum_{n=0}^{\infty}
    \frac{1- e^{\left(i\epsilon_{r\sigma}-\Gamma\right)\Delta t-\pi T \Delta t (2n+1)}}{i\epsilon_{r\sigma}-\Gamma-\pi T  (2n+1)}
  \nonumber
  \\
  = &
  \frac 1 {\pi} \Im \Psi\left(\frac 1 2 +\frac {\Gamma - i\epsilon_{r\sigma}}{2\pi T} \right)
  \\
  & 
  +
  \Im
  \left\{
    \frac{    e^{( i\epsilon_{r\sigma} - \Gamma - \pi T ) \Delta t} } {\pi}
    \Phi\left(e^{-2\pi T\Delta t};1;\frac 1 2 + \frac {\Gamma - i\epsilon_{r\sigma}}{2 \pi T} \right)
  \right\}
  ,
  \nonumber
\end{align}
where $\Psi(z)=-\gamma - \sum_{k=0}^\infty [1/(z+k) - 1/(1+k)] $ is the digamma-function, and $\Phi(z;s;\nu)= \sum_{n=0}^{\infty}  {z^n}/{(n+\nu)^s}$ is the \emph{Lerch transcendent} (see, e.g., \Cite{Erdely}).
Analogously, we obtain for the function $F^{-}_{r\sigma}(t)$ [\Eq{Frelation}]:
\begin{subequations}
  \begin{align}
    &F^{-}_{r\sigma}(\Delta t)
    :=
    e^{-2\Gamma\Delta t} \int_0^{\Delta t} d\tau
    \frac{T\sin(\epsilon_{r\sigma}\tau)} {\sinh(\pi T \tau)}
    e^{\Gamma\tau}
    \\
    & =
    -e^{-2\Gamma\Delta t} 2T \sum_{n=0}^{\infty} \Im \int\limits_0^{\Delta t} d \tau e^{\left(i\epsilon_{r\sigma}+\Gamma\right)\tau-\pi T \tau (2n+1)}
    \label{pp_pole}
    \\  \nonumber
    & =-e^{-2\Gamma\Delta t}\left({F^{+}_{r\sigma}(\Delta t)}|_{\Gamma\rightarrow -\Gamma}\right)\\ 
    & = \Im \left\{
      \frac {e^{(i\epsilon_{r\sigma}-\Gamma-\pi T)\Delta t}}{\pi}\Phi\left(e^{-2\pi T\Delta t};1;\frac 1 2 -\frac{\Gamma + i\epsilon_{r\sigma}}{2\pi T} \right)
    \right.
    \nonumber
    \\
    &\phantom{=}
    \left. +
      \frac{e^{-2\Gamma\Delta t}}{\pi}\Psi\left(\frac 1 2 +\frac{-\Gamma+i\epsilon_{r\sigma}}{2\pi T} \right)
    \right\}.
    \nonumber
  \end{align}
\end{subequations}
Note that the results satisfy the formal relation  \Eq{Frelation}. One can think that the function $F^{-}_{r\sigma}(\Delta t)$
can have a pole at $\Gamma=2\pi T k$ ($k=1,2,...$) for $\epsilon_{r\sigma}=0$, since both $\Psi\left(\frac 1 2 +\frac{-\Gamma+i\epsilon_{r\sigma}}{2\pi T} \right)$ 
and $\Phi\left(e^{-2\pi T\Delta t};1;\frac 1 2 -\frac{\Gamma + i\epsilon_{r\sigma}}{2\pi T} \right)$ have it. However, the pole of the $\Psi$ function
exactly {compensates} the pole of the $\Phi$ function giving zero in that case.
That this should be the case is already clear from \Eq{pp_pole} by taking $\epsilon_{r\sigma}=0$.
\section{Two-loop contributions to the time-evolution \label{sec:twoloop_ap}}

In this Appendix, we write out the proof of the factorization \eq{formal-2loop2},
presented diagrammatically in \Fig{fig:factorize}.
The manipulations that we apply are analogous to those of \Cite{Koller10} for two-loop calculations.
That reference, however, deals with self-energy diagrams for interacting systems in the zero frequency limit.
We start from \Eq{formal-2loop}, repeated here for convenience:
 \begin{align}
  \bar{\Pi}_4(t,t_0)
  &=
    e^{-i\bar{L}(t-t_0)}
  \nonumber
  \\
  &
  \times 
  \timeintfour{t}{t_4}{t_3}{t_2}{t_1}{t_0}
  \bar{G}'_4(t_4) \bar{G}'_3(t_3)\bar{G}'_2(t_2)\bar{G}'_1(t_1)
  \nonumber
  \\
  &
  \times 
  \sum_{\langle i,j,k,l\rangle} (-1)^P \bar{\gamma}_{i j}(t_i-t_j) \bar{\gamma}_{k l}(t_k - t_j)
  .
\end{align}
Here, we sum over the following possible contractions:
$i,j,k,l=4,3,2,1$ (reducible) and $4,2,3,1$ and $4,1,3,2$ (both irreducible).
First, we relabel the times and multi-indices as indicated in  \Fig{fig:factorize}
such that the contraction function is the same for all terms, equal to
$\bar{\gamma}_{43}(t_4-t_3) \bar{\gamma}_{21}(t_2 - t_1)$.
In the irreducible contractions, this changes the order of the vertices from
$\bar{G}'_4(t_4) \bar{G}'_{3}(t_3)\bar{G}'_2(t_2)\bar{G}'_{1}(t_1)$,
but for each of these we can restore this order by anticommuting
the creation superoperators [\Eq{Gprimeanticom}].
{This puts} the superoperators connected by a $\bar{\gamma}$-contraction
adjacent to each other, {i.e., one} disentangles the contractions:
therefore the sign appearing from anticommutation of the creation super operators
 precisely cancels the fermionic Wick sign $(-1)^P$.
We are left with
\begin{align}
  &\bar{\Pi}_4(t,t_0)
  =
  e^{-i\bar{L}(t-t_0)} \times 
  \\
  & \Big[
  \timeintfour{t}{t_4}{t_3}{t_2}{t_1}{t_0}
  +\timeintfour{t}{t_4}{t_2}{t_3}{t_1}{t_0}
  +\timeintfour{t}{t_2}{t_4}{t_3}{t_1}{t_0} \Big]
  \nonumber
  \\
  &
  \times
  \sum_{4321}
  \bar{G}'_4(t_4) \bar{G}'_{3}(t_3)\bar{G}'_2(t_2)\bar{G}'_{1}(t_1) 
  \bar{\gamma}_{4 3}(t_4-t_3) \bar{\gamma}_{2 1}(t_2 - t_1)
  .
  \nonumber
\end{align}
By duplicating these terms, while compensating by a factor 1/2,
and {interchanging the dummy variables $t_1,t_2 \leftrightarrow t_3,t_4$ in the duplicates,}
we obtain a sum of integrals {which} can be factorized as \Eq{formal-2loop2}
by comparing with the definition of $\bar{\Pi}_2 (t,t_0)$ [\Eq{1loopdef}]:
\begin{subequations}
\begin{align}
  &\bar{\Pi}_4(t,t_0)
  =
  \tfrac{1}{2} e^{-i\bar{L}(t-t_0)} \times 
  \\
  & \Big[
  \timeintfour{t}{t_4}{t_3}{t_2}{t_1}{t_0}
  +\timeintfour{t}{t_4}{t_2}{t_3}{t_1}{t_0}
  +\timeintfour{t}{t_2}{t_4}{t_3}{t_1}{t_0} 
  \nonumber
  \\
  & +
  \timeintfour{t}{t_2}{t_1}{t_4}{t_3}{t_0}
  +\timeintfour{t}{t_2}{t_4}{t_1}{t_3}{t_0}
  +\timeintfour{t}{t_4}{t_2}{t_1}{t_3}{t_0} \Big]
  \nonumber
  \\
  &
  \times
  \sum_{4321}
  \bar{G}'_4(t_4) \bar{G}'_{3}(t_3)\bar{G}'_2(t_2)\bar{G}'_{1}(t_1)
  \bar{\gamma}_{4 3}(t_4-t_3) \bar{\gamma}_{2 1}(t_2 - t_1)
  \nonumber
  \\
  &
  = \tfrac{1}{2} e^{-i\bar{L}(t-t_0)} \times
  \nonumber
  \\
  &
  \quad
  \timeint{t}{t_4}{t_3}{t_0}
  \sum_{43}\bar{G}'_4(t_4) \bar{G}'_{3}(t_3)
  \timeint{t}{t_2}{t_1}{t_0}
  \sum_{21}\bar{G}'_2(t_2)\bar{G}'_{1}(t_1)
  \nonumber
  \\
  &
  =
  \tfrac 1 2 \bar{\Pi}_2(t,t_0) e^{i\bar{L}(t-t_0)} \bar{\Pi}_2 (t,t_0)
  .
\end{align}
\end{subequations}
\section{Spin-channel decomposition\label{sec:factorize}}

In this Appendix we outline the calculation of the propagator
by using the noninteracting limit \emph{in the first step} and then setting up the perturbation theory.
(In contrast to this, the calculations given in \Sec{sec:results} {first set up the perturbation theory and  make} use of the assumption $U=0$ \emph{in the last step},
and rather {aim} to show how in a framework applicable to interacting systems, this limit is achieved.
The approach we now outline, although shorter, does not make that clear.)
In particular, we use that for $U=0$ the quantum dot Liouvillian decomposes into single-spin species, $L=\sum_\sigma L_\sigma$.
Due to this special property, the total Liouvillian decomposes into commuting spin-resolved parts,
$L^{\mathrm{tot}} = \sum_\sigma L^{\mathrm{tot}}_\sigma$
with
$L^{\mathrm{tot}}_\sigma=
L_\sigma +L^R_\sigma + L^V_\sigma$,
since the reservoirs are noninteracting, the tunnel coupling \eq{Vrdef} is quadratic,
and all spin-dependencies (due the junctions and the magnetic field) are considered to be collinear.
Here $L_\sigma$, $L^R_\sigma$, and $L^V_\sigma$ are obtained from
 \Eqs{L2ndquant}, \eq{LR}, and \eq{q-li}, respectively,
by leaving out the sum over the spin-index $\sigma$.
One now splits the propagator \eq{pit} into commuting factors relating to different spins:
\begin{subequations}
  \begin{align}
    \Pi (t,t_0)
    &= \Tr{R}\left( e^{-i L^\mathrm{tot} (t-t_0)}\rho^{R} \right) \bullet \\
    &=\Tr{R}\left( e^{-i L^\mathrm{tot}_\uparrow (t-t_0)}e^{-i L^\mathrm{tot}_\downarrow (t-t_0)}\rho^{R}_\uparrow\rho^{R}_\downarrow  \right) \bullet
    \label{factorspinliv}
    \\
    &=\Pi^\uparrow(t,t_0)\Pi^\downarrow(t,t_0)
    ,
    \label{factorspinpi}
  \end{align}
\end{subequations}
where $\rho^R_\sigma=\prod_{r} e^{-\frac{1}{T} (H^{r}_\sigma-\mu^r {n}^r_\sigma)}{Z^r_\sigma}$
and
\begin{align}
  \label{pisigmat}
  \Pi^\sigma(t,t_0)=\Tr{R \sigma}\left( e^{-i L^\mathrm{tot}_\sigma (t-t_0)}\rho^{R}_\sigma \right) \bullet.
\end{align}
where the trace runs over one spin-degree of freedom.
The superoperator $\Pi^\sigma(t,t_0)$
can be again calculated using the renormalized perturbation series, \Eq{pitrunc},
which the super-Pauli principle now truncates at the {one-loop} order:
\begin{align}
  \Pi^\sigma(t,t_0)=\bar{\Pi}_0^\sigma(t,t_0) + \bar{\Pi}^\sigma_2(t,t_0).
\end{align}
Here,
$\bar{\Pi}_0^\sigma(t,t_0)=e^{-i\bar{L}_\sigma(t-t_0)}$
{with}
$\bar{L}_\sigma=L_\sigma+\tilde{\Sigma}_\sigma$.
{In turn,}
$\tilde{\Sigma}_\sigma$
and $\bar{\Pi}_2^\sigma(t,t_0)$ are defined by \Eqs{Sigmatilde} and \eq{1loopftpt}, respectively,
by leaving out the summation over the spin-index, fixing it to the value $\sigma$,
and in \Eq{1loopftpt} replacing $\bar{L} \rightarrow \bar{L}_\sigma$ in the exponential prefactor.
Inserting \Eq{pisigmat} into \Eq{factorspinpi} and comparing order by order with the expansion \Eq{pitrunc},
we obtain:
\begin{subequations}
\begin{align}
  \bar{\Pi}_0(t,t_0)
  & =  \bar{\Pi}_0^\uparrow(t,t_0)\bar{\Pi}_0^\downarrow(t,t_0)=e^{-i\sum_\sigma \bar{L}_\sigma(t-t_0)}
  ,
  \\
  \bar{\Pi}_2(t,t_0)
  & = \sum_\sigma \bar{\Pi}_0^\sigma(t,t_0) \bar{\Pi}_2^{\bar{\sigma}}(t,t_0)
  ,
  \\
  \bar{\Pi}_4(t,t_0)
  & = \bar{\Pi}_2^{\uparrow}(t,t_0)\bar{\Pi}_2^{\downarrow}(t,t_0)
  \nonumber
  \\
  & = \tfrac 1 2 \sum_\sigma\bar{\Pi}_2^{\sigma}(t,t_0)\bar{\Pi}_2^{\bar{\sigma}}(t,t_0)
  \nonumber
  \\ 
  & = \tfrac 1 2 \bar{\Pi}_2(t,t_0) e^{i\bar{L}(t-t_0)} \bar{\Pi}_2 (t,t_0)
  .
\end{align}
\end{subequations}
{where we used in the last step that $[\bar{\Pi}_2^{\sigma}(t,t_0)]^2=0$ by the super-Pauli principle \eq{superpauli}.}
The last equation is the factorization relation \eq{formal-2loop2} obtained in the main text.

\bibliographystyle{apsrev4-1}
\end{document}